\documentclass[aps,prl,twocolumn,secnumarabic,balancelastpage,amsmath,amssymb,floatfix,superscriptaddress]{revtex4-1}
\usepackage{amssymb}
\usepackage{amsmath}
\usepackage{amsfonts}
\usepackage{braket}
\usepackage{graphicx}
\usepackage{epstopdf}
\usepackage{epsfig}
\usepackage[toc,page,title,titletoc,header]{appendix}
\usepackage{dsfont,amsthm,amsbsy}
\usepackage{todonotes}
\usepackage{hyperref}
\usepackage{cancel}

\usepackage[normalem]{ulem}

\newcommand{\tr}{\operatorname{Tr}}
\newcommand{\sgn}{\operatorname{sgn}}
\newcommand{\eff}{\mathrm{eff}}
\newcommand{\Hstatic}{H_{\text{static}}}

\newcommand{\figOne}{
  \begin{figure}[h!]
    \includegraphics[width=3.2in]{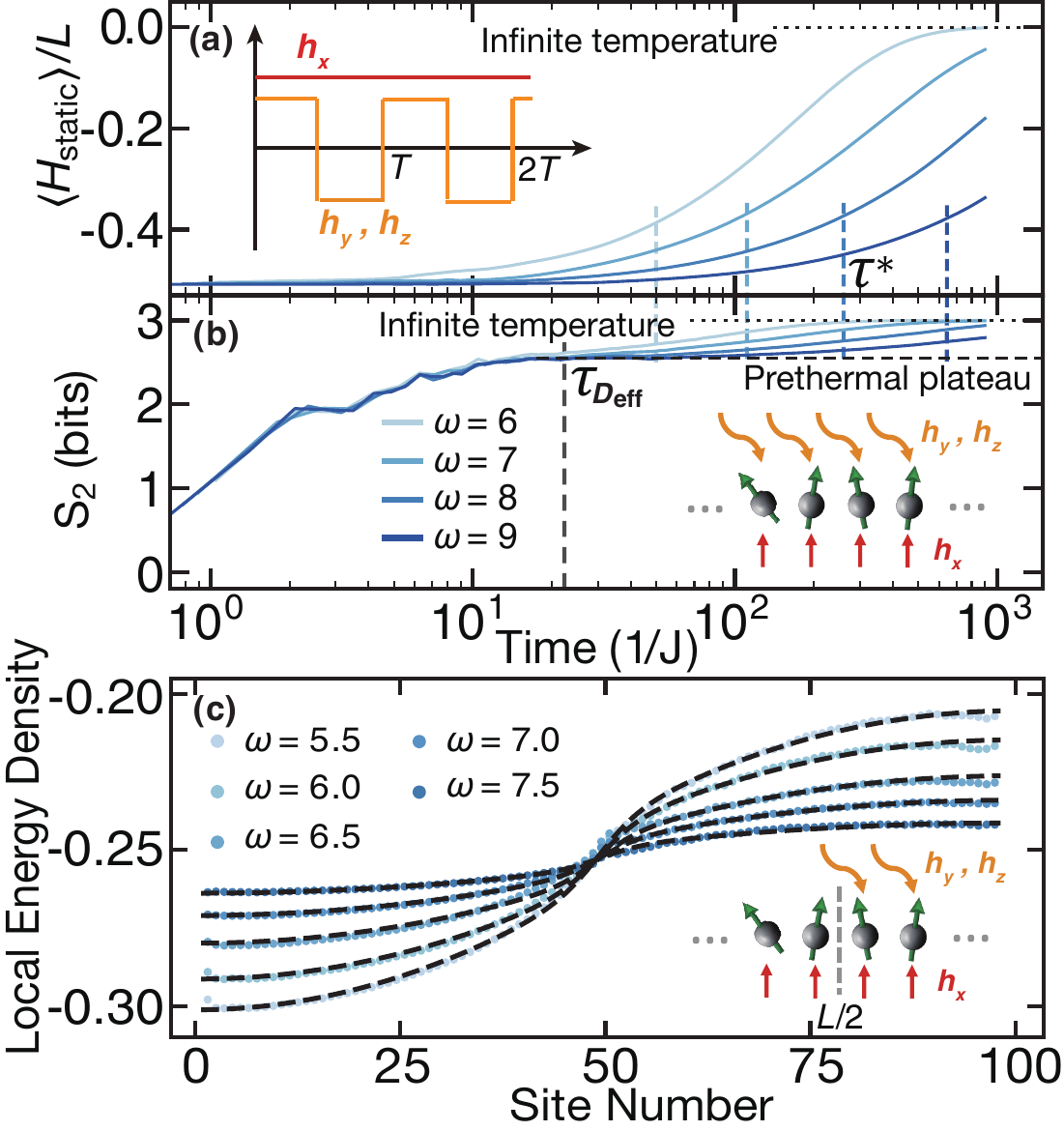}
    \centering
    \caption{
      Floquet thermalization of an $L=100$ spin chain. 
      (a) Average energy density measured with respect to $\Hstatic\approx D_\eff $ under a global drive. 
      The heating timescale $\tau^*$ is extracted from the energy's exponential approach to its infinite-temperature value, and depends exponentially on the driving frequency (for explicit scaling, see Fig.~\ref{fig4}a.) 
      (b) The second R\'{e}nyi entropy of the leftmost three sites. The dashed lines are computed using the prethermal Gibbs ensemble \cite{SM}. 
      (c) Spatial profiles of energy density under a half-system drive with $\langle \Hstatic\rangle/L=-0.25$.
      Insets: the drive's time dependence (a) and schematics of the global drive (b) and the half-system drive (c).
    }
    \label{fig1}
  \end{figure}
  
}

\newcommand{\figTwo}{
  \begin{figure}[t]
    \includegraphics[width=3.2in]{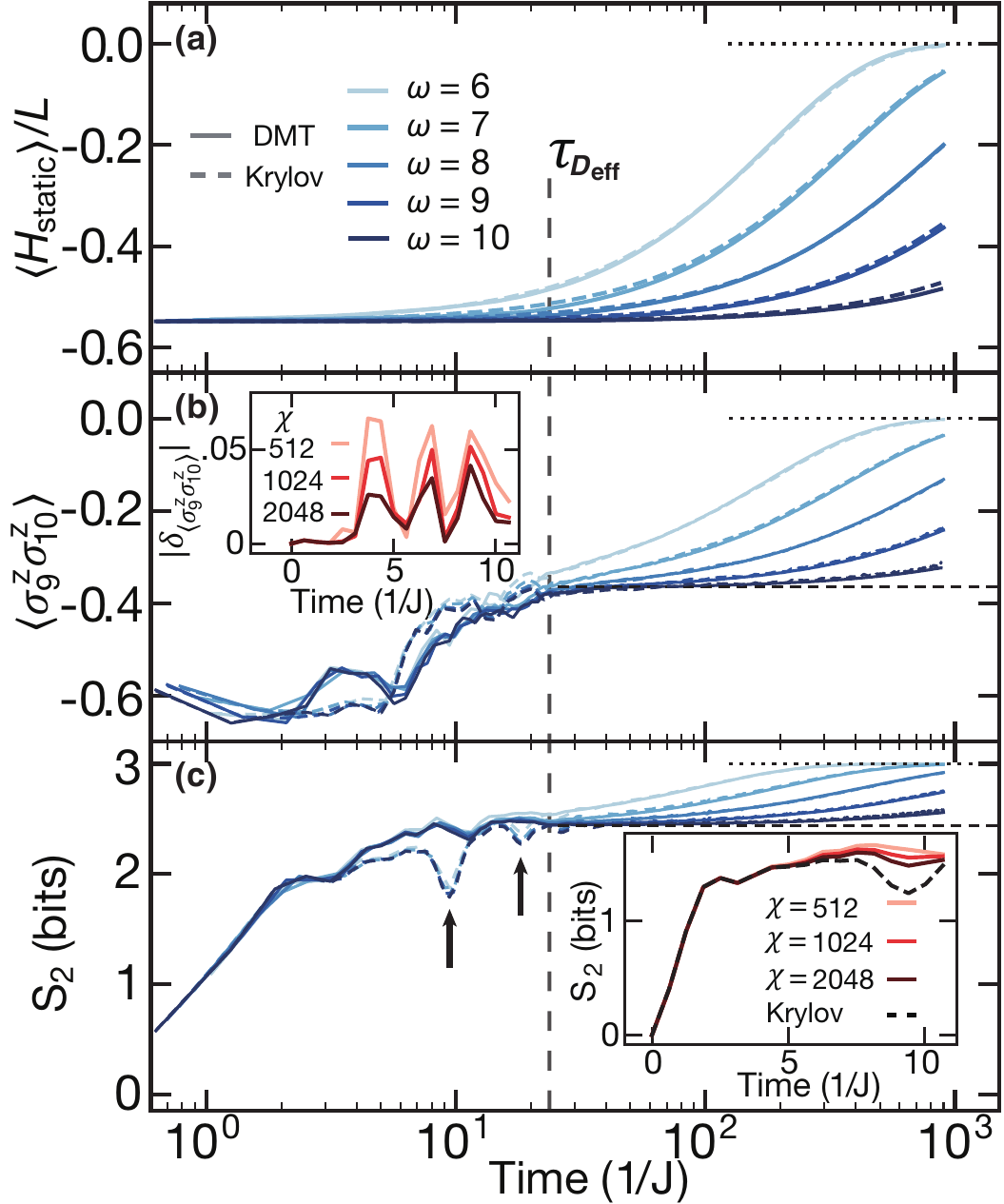}
    \centering
    \caption{
      Comparison between DMT and Krylov of the time evolution of an $L=20$ spin chain under a global drive (at fixed bond dimension $\chi = 64$).
      (a) Average energy density $\langle \Hstatic \rangle/L$.
      (b) A typical local observable $\sigma^z_9\sigma^z_{10}$.
      (c) The second R\'{e}nyi entropy $S_2$ of the leftmost three sites. The arrows mark resonance-like dips, which DMT fails to capture \cite{SM}. 
      The dashed lines are computed using the prethermal Gibbs ensemble. 
      Insets (early-time behavior at frequency $\omega = 10$): (b) errors in the local observable $\delta_{\langle \sigma^z_9\sigma^z_{10} \rangle}=\langle\sigma^z_9\sigma^z_{10}\rangle_{\mathrm{DMT}}-\langle\sigma^z_9\sigma^z_{10}\rangle_{\mathrm{Kry}}$, (c) errors in $S_2$. 
      }
    \label{fig2}
  \end{figure}
}

\newcommand{\figThree}{
    \begin{figure}[t]
    \includegraphics[width=3.3in]{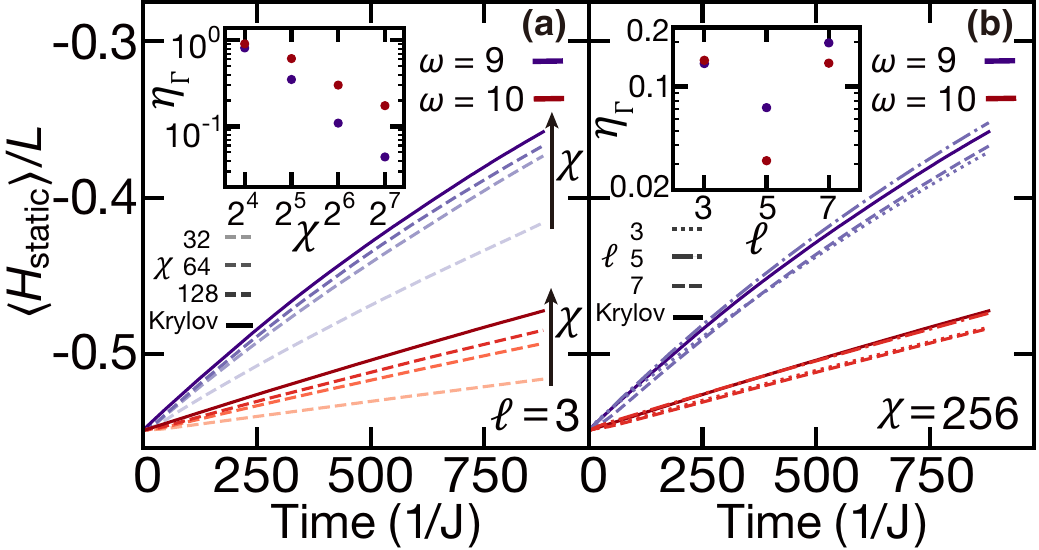}
    \centering
    \caption{Energy density at (a) bond dimension $\chi = 32,64,128$ and (b) the preservation diameter $\ell = 3,5,7$.
      Insets: relative error in the heating rate $\eta_\Gamma = |\Gamma_{\mathrm{DMT}} - \Gamma_{\mathrm{Krylov}}|/\Gamma_{\mathrm{Krylov}}$, where $\Gamma=1/\tau^*_E$ is defined by $\frac{d}{dt}\langle \Hstatic \rangle = -\Gamma \langle \Hstatic \rangle. $
    }
    \label{fig3}
    \end{figure} 
}

\newcommand{\figFour}{
  \begin{figure}[t]
    \includegraphics[width=3.2in]{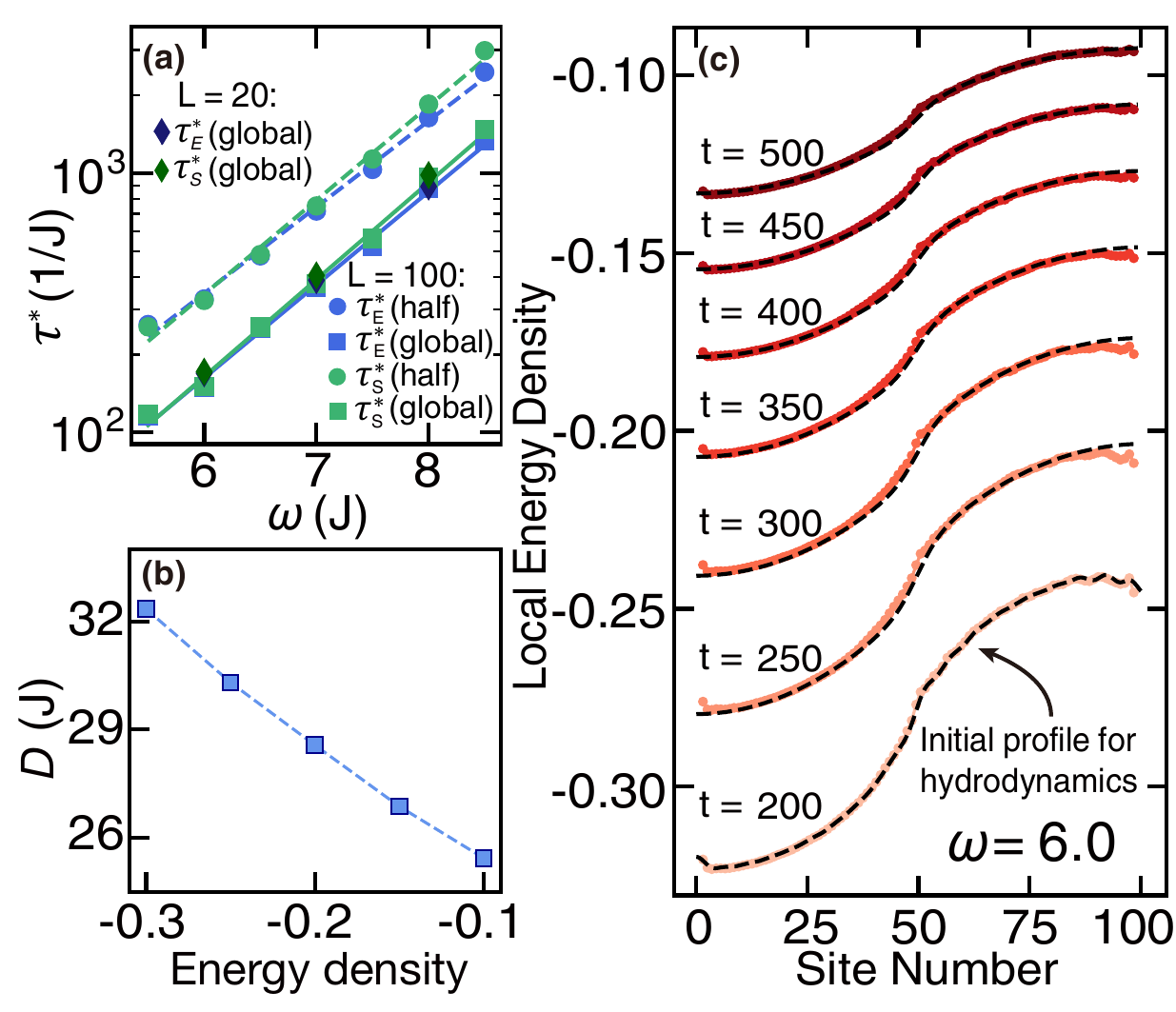}
    \centering
    \caption{(a) Heating timescale, $\tau^*$, extracted in energy density ($E$) and subsystem entropy ($S_2$) for $L=20,100$. In agreement with theoretical prediction, $\tau^*$ depends exponentially on $\omega$. For both driving protocols, we extract the same local energy scale $J_{local}^E\approx 1.21$. However, the half-system drive exhibits a heating timescale twice as large as the global drive. 
      (b) Energy dependence of the diffusion coefficient in the undriven spin chain. 
      (c) Dynamics of the energy density with half-system drive. Solid curves are computed using DMT. Dashed black curves are computed using a hydrodynamical equation, Eq.~\ref{eq:diff}, where one feeds in the DMT-calculated energy-density profile at time $t=200$. 
      Subsequent time evolution under the differential equation quantitatively reproduces the exact results from DMT. 
    }
    \label{fig4}
  \end{figure}
}

\begin{document}
\title{Emergent hydrodynamics in non-equilibrium quantum systems}
\author{Bingtian Ye}
\thanks{These two authors contribute equally to this work. }
\affiliation{Department of Physics, University of California, Berkeley, CA 94720, USA}
\author{Francisco Machado}
\thanks{These two authors contribute equally to this work. }
\affiliation{Department of Physics, University of California, Berkeley, CA 94720, USA}
\author{Christopher David White}
\affiliation{Institute for Quantum Information and Matter, Caltech, Pasadena, CA 91125, USA}
\author{Roger S. K. Mong}
\affiliation{Department of Physics and Astronomy, University of Pittsburgh, Pittsburgh, PA 15260, USA}
\affiliation{Pittsburgh Quantum Institute, Pittsburgh, PA 15260, USA}
\author{Norman Y. Yao}
\affiliation{Department of Physics, University of California, Berkeley, CA 94720, USA}
\affiliation{Materials Science Division, Lawrence Berkeley National Laboratory, Berkeley, CA 94720, USA}

\date{\today}
\begin{abstract}
A tremendous amount of recent attention has focused on characterizing the dynamical properties of periodically driven many-body systems.
  Here, we use a novel numerical tool termed `density matrix truncation' (DMT) to investigate the late-time dynamics of large-scale Floquet systems.
  We find that DMT accurately captures two essential pieces of Floquet physics, namely, prethermalization and late-time heating to infinite temperature.
  Moreover, by implementing a spatially inhomogeneous drive, we demonstrate that an interplay between Floquet heating and diffusive transport is crucial to understanding the system's dynamics.
  Finally, we show that DMT also provides a powerful method for quantitatively capturing the emergence of hydrodynamics in static (un-driven) Hamiltonians; in particular, by simulating the dynamics of generic, large-scale quantum spin chains (up to $L=100$), we are able to directly extract the energy diffusion coefficient. 
  
\end{abstract}

\maketitle

Understanding the non-equilibrium dynamics of strongly correlated quantum systems represents a central challenge at the interface of condensed matter, atomic physics and quantum information science.
This challenge stems in part from the fact that such systems can be taken out of equilibrium in a multitude of different ways, each with its own set of expectations and guiding intuition. 

For example, under a quench, one typically expects a many-body system to quickly evolve toward local thermal equilibrium \cite{deutsch_quantum_1991,srednicki_chaos_1994,rigol_thermalization_2008,dalessio_quantum_2016,Calabrese_2016,Gogolin_2016}. At first sight, this suggests a simple description. However, capturing both the microscopic details of short-time thermalization as well as the cross-over to late-time hydrodynamics remains an open challenge \cite{potter_universal_2015,vosk_theory_2015,agarwal_anomalous_2015,znidaric_diffusive_2016,Luitz_2016,Khait_2016,Luitz_2016_2,sahu_scrambling_2018,Bohrdt_2017}.
Indeed, despite nearly a century of progress, no general framework exists for perhaps the simplest question: How does one derive a classical diffusion coefficient from a quantum many-body Hamiltonian?

Alternatively, a many-body system can also be taken out of equilibrium via periodic (Floquet) driving --- a strategy which has received a tremendous amount of recent attention in the context of novel Floquet phases of matter \cite{else_floquet_2016,khemani_phase_2016,von_keyserlingk_phase_2016,von_keyserlingk_1d_2016,von_keyserlingk_absolute_2016,lindner_floquet_2011,kitagawa_topological_2010,yao_timecrystaltheory_2017,choi_timecrystal_2017,zhang_timecrystal_2017}.
In this case, the non-equilibrium system is generically expected to absorb energy from the driving field (so-called Floquet heating) until it approaches a featureless infinite temperature state \cite{prosen_1998, prosen_1999,lazarides_2014,d'alessio_drivenlongtime_2014,bukov_highfreq_2015,machado_exponentially_2017}.

\figOne

While these questions are naturally unified under the umbrella of non-equilibrium dynamics \cite{fn1}, understanding the interplay between Floquet heating, emergent hydrodynamics and microscopic thermalization represents a crucial step toward the characterization and control of non-equilibrium many-body systems \cite{abanin2015exponentially,Mori2016,abanin2017effective,Kuwahara2016,abanin2017rigorous,else2017prethermal, Bukov_2015,Bohrdt_2017_1}.
That one expects such connections can already been seen in certain limits; for example, in the limit of a high-frequency Floquet drive, energy absorption is set by an extremely slow heating rate. Thus, one anticipates a relatively long timescale where the system's stroboscopic dynamics can be captured by an effective static \textit{prethermal Hamiltonian}.
These expectations immediately lead to the following question: How do the late-time dynamics of driven quantum systems account for both the prethermal Hamiltonian's hydrodynamics and the energy absorption associated with Floquet heating?

Until now, such questions have remained largely unexplored owing to the fact that they sit in a region of phase space where neither theoretical techniques nor numerical methods easily apply. 
However, a number of recently proposed numerical methods \cite{White_2017,Leviatan_2017,Wurtz_2018.1,Wurtz_2018.2,Mike_2019} promise to bridge this gap and directly connect microscopic models to  emergent macroscopic hydrodynamics.
Here, we will focus on one such method --- density matrix truncation (DMT) \cite{White_2017} --- which modifies time-evolving block decimation (TEBD) by representing states as matrix product density operators (MPDOs) and  prioritizing short-range (over long-range) correlations.

Working with a generic, one-dimensional spin model, in this Letter, we use DMT to investigate a broad range of non-equilibrium phenomena ranging from Floquet heating to emergent hydrodynamics.  
Our main results are three fold. 
First, we find that DMT accurately captures two essential pieces of Floquet physics: prethermalization and heating to infinite temperature (Fig.~\ref{fig1}).
Crucially, the truncation step intrinsic to DMT enables us to efficiently explore the \emph{late-time} dynamics of large-scale quantum systems (up to $L=100$), at the cost of imperfectly simulating the system's  \emph{early-time} dynamics.

This trade-off hinges on DMT's efficient representation of local thermal states, making it a natural tool for studying emergent hydrodynamics. 
Our latter two results illustrate this in two distinct contexts: 1) directly measuring the energy diffusion coefficient for a static Hamiltonian, and 2) demonstrating the interplay between Floquet heating and diffusion in an inhomogeneously driven spin chain.
We hasten to emphasize that such calculations are fundamentally impossible for either exact diagonalization based methods (owing to the size of the Hilbert space) or conventional TEBD methods (owing to the large amount of entanglement at late times). 

\emph{Model and phenomenology}---We study the dynamics of a one-dimensional spin-$1/2$ chain whose evolution is governed by a time periodic Hamiltonian $H(t) = \Hstatic + H_{\mathrm{drive}}(t)$, where
\begin{align}
  \label{eq:Hamil}
  \begin{split}
    \Hstatic &= \sum^{L-1}_{i=1}[J\sigma^z_i \sigma^z_{i+1}+J_x \sigma^x_i \sigma^x_{i+1}] + h_x\sum_{i=1}^L\sigma^x_i~,
  \end{split}
\end{align}
with $\sigma^\alpha_i$ being the Pauli operators acting on site $i$ \cite{fn2}.
The drive, $H_{\mathrm{drive}}(t) = H_{\mathrm{drive}}(t+T)$, exhibits a period $T=2\pi/\omega$ and corresponds to an oscillating field in the  $\hat{y}$ and $\hat{z}$ directions:
\begin{align}
  H_{\mathrm{drive}}(t) = \sum_{i=1}^L v_i(t)~( h_y\sigma^y_i + h_z\sigma^z_i)~.
\end{align}
In this work we will consider two different driving protocols (Fig.~\ref{fig1} insets)
a \textbf{global drive}, with all spins driven [$v_i(t) = \sgn\cos(\omega t)$], and a \textbf{half-system drive}, with only the right half driven [$v_{i\le L/2}(t) = 0$ and $v_{i>L/2}(t) =  \sgn\cos(\omega t)$]. 
Throughout the letter, we work in the high-frequency regime with $\omega\ge 5J$, and choose the parameters to be $\{J, J_x, h_x, h_y, h_z\} = \{1, 0.75, 0.21, 0.17, 0.13\}$. 
We expect our choice of the model and parameters to be generic as we observe the same phenomenology upon varying both the parameters and the interaction Hamiltonian \cite{SM}. 

\figTwo
The quenched dynamics of a high-frequency driven system is characterized by two timescales.
The heating timescale, $\tau^*$ (Fig.~\ref{fig1}a), determines the rate of energy absorption from the drive 
and is proven to be at least exponential in the frequency of the drive,
$\tau^* \ge  \mathcal{O}(e^{\omega/J_{\mathrm{local}}})$, where $J_{\text{local}}$ is a local energy scale~
\cite{abanin2015exponentially,Mori2016,Kuwahara2016,abanin2017effective, abanin2017rigorous,else2017prethermal}.
Up until $\tau^*$, the  stroboscopic dynamics of the system is well described by the \emph{static} prethermal Hamiltonian $D_\eff = H_{\mathrm{static}} + \mathcal{O}(\omega^{-1})$, which can be obtained as the truncation of the Floquet-Magnus expansion of the evolution operator \cite{Kuwahara2016,abanin2017effective,abanin2017rigorous}.
The prethermalization timescale, $\tau_{D_\eff}$ (Fig.~\ref{fig1}a,b), determines the time at which the system approaches an equilibrium state with respect to $D_\eff$.
When $ \tau_{D_\eff} \ll \tau^*$, the system exhibits a well defined, long-lived prethermal regime.

In Figs.~\ref{fig1}a,b, we illustrate these two timescales by computing the dynamics of an $L = 100$ Floquet spin chain using DMT \cite{fn3}.
The average energy density $\langle \Hstatic(t)\rangle/L$ exhibits the expected phenomenology (Fig.~\ref{fig1}a): it remains constant (up to $\omega^{-1}$ corrections) until $\tau^*$, after which it begins to approach  its infinite temperature value $\langle \Hstatic \rangle_{T=\infty} = 0$.

To probe the prethermalization timescale $\tau_{D_\eff}$, a different diagnostic is needed. 
In particular, we compute the second R\'{e}nyi entropy, $S_2 = -\log_2 \tr [\rho^2_{\mathrm{s}}]$, where $\rho_{\mathrm{s}}$ is the reduced density matrix of the three leftmost spins. 
While the system begins in a product state with $S_2=0$, its entropy quickly approaches a \emph{prethermal plateau}, consistent with the Gibbs state of $D_\eff$ at a temperature that matches the initial energy density (Fig.~\ref{fig1}b) \cite{SM}.
The timescale at which this occurs corresponds to $\tau_{D_\eff}$ and, indeed, we observe $\tau_{D_\eff} \sim 1/J_{\mathrm{local}}$ independent of the driving frequency $\omega$.
Similar to the energy density, at late times $t > \tau^*$, $S_2$ begins to approach its infinite temperature value, $S_2^{T=\infty} = 3$ bits.


\emph{Benchmarking DMT}---To confirm the reliability of DMT in the simulation of Floquet dynamics, we compare it with Krylov subspace methods \cite{fn8,slepc1, slepc2, petsc1}. 
%
This analysis not only gauges the applicability of DMT, but also leads to insights into the nature of the Floquet heating process.

Time evolution with DMT proceeds via two repeating steps: a TEBD-like approximation of the time evolution unitary and a truncation of the MPDO.
In the TEBD-like step, we Trotter decompose the time evolution operator into a series of local gates which we then apply to the MPDO \cite{White_2017}.
Because each local gate application increases the bond dimension of the corresponding tensors, we must truncate them back to a fixed maximum bond dimension, which we call $\chi$. 
During this truncation step, a conventional TEBD method will discard the terms which contribute the least to the entanglement  \cite{Schollwock,Orus}. As a result, this truncation is agnostic to the locality of the discarded correlations.
By contrast, in DMT we explicitly prioritize the preservation of short-range correlations \cite{White_2017}. 
To this end, DMT separates $\chi$ into two contributions: $\chi = \chi^{\mathrm{preserve}} + \chi^{\mathrm{extra}}$.
$\chi^{\mathrm{preserve}} = 2^{\ell}$ is used to store the information of all observables on $\ell$ contiguous sites around the truncated tensor---we call $\ell$ the \textit{preservation diameter} \cite{SM,fn4}. 
$\chi^{\mathrm{extra}}$ is then used to preserve the remaining correlations with largest magnitude. 
Crucially, the preservation of short-range correlations allows DMT to conserve (up to Trotter errors) the instantaneous energy density \cite{White_2017}. 
To be more specific, in Floquet systems, DMT conserves $H(t)$ at each instant, but it \emph{does not} explicitly conserve $D_\eff$. 
%

We utilize three diagnostics to compare the time evolution between DMT and Krylov: the average energy density (Fig.~\ref{fig2}a), local two-point correlation functions (Fig.~\ref{fig2}b), and the second R\'{e}nyi entropy (Fig.~\ref{fig2}c).

At early times ($t<\tau_{D_\eff}$), one observes substantial disagreements between DMT and Krylov (Fig.~\ref{fig2}b,c).
This is to be expected.  
Indeed, the accurate description of early-time thermalization dynamics depends sensitively on the details of long-range correlations which DMT does not capture.
An exception to this is the energy density, whose changes are expected to be exponentially small in frequency \cite{abanin2015exponentially,Mori2016,Kuwahara2016,abanin2017effective}.
This is indeed born out by the numerics where one finds that $\langle H_{\mathrm{static}}\rangle/L$ remains quasi-conserved and in excellent agreement with Krylov (Fig.~\ref{fig2}a).

\figThree

One might naively expect the early-time disagreements to lead to equally large intermediate-time ($\tau_{D_\eff} < t < \tau^*$) deviations.
This is not what we observe.
Indeed, all \emph{three} diagnostics show excellent agreement between DMT and Krylov (Fig.~\ref{fig2}).
This arises from a confluence of two factors. 
First, as aforementioned, DMT accurately captures the system's energy density, which in turn, \emph{fully} determines the prethermal Gibbs state; second, DMT can efficiently represent such a Gibbs state.
Thus, although DMT fails to capture the \emph{approach} to the prethermal Gibbs state, it nevertheless reaches the same equilibrium state at $t \sim \tau_{D_\eff}$.
Afterwards (for $t>\tau_{D_\eff}$), the system is simply evolving between different Gibbs states of $D_\eff$, wherein one expects agreement between DMT and Krylov even at relatively low bond dimension (Fig.~\ref{fig2}).

Small disagreements between DMT and Krylov, however, re-emerge at very late times ($t > \tau^*$) and large frequencies, reflecting the physical nature of Floquet heating (Fig.~\ref{fig2}a). 
In particular, as the frequency increases, absorbing an energy quantum from the drive requires the correlated rearrangement of a greater number of spins~\cite{abanin2015exponentially,Mori2016,abanin2017effective}.
However, these longer-ranged correlations are \emph{not} strictly preserved by DMT, leading to an artificial (truncation-induced) suppression of heating at large frequencies (Fig.~\ref{fig3}).

This raises the question: How does the accuracy of DMT converge with both bond dimension and  preservation diameter?
As expected, increasing $\chi$ at fixed $\ell$ improves the accuracy of DMT since the amount of information preserved during each truncation step is greater, Fig.~\ref{fig3}a.
Curiously, tuning $\ell$ at fixed $\chi$ can \emph{also} affect the accuracy, despite not changing the amount of  information preserved, Fig.~\ref{fig3}b. 
This suggests the tantalizing possibility that one can achieve high accuracy at relatively low bond dimension by carefully choosing the operators which are preserved.

\emph{Floquet heating dynamics}---As a first demonstration of DMT's potential for extracting quantitative information about the Floquet dynamics, we directly measure the heating rate.
We find that both $\langle H_{\text{static}}\rangle/L$ and $S_2$ exhibit an exponential approach toward their infinite-temperature values: $|\langle H_{\text{static}}\rangle/L|\propto e^{-t/\tau^*_E}$ and $(S_2^{T=\infty} - S_2) \propto e^{-2t/\tau^*_S}$.
To this end, we extract $\tau^*_E$ and $\tau^*_S$ as independent measures of the Floquet heating timescale \cite{SM}. 
Crucially, they agree with one another across all system sizes studied ($ L =20$--$ 100$), as shown in Fig.~\ref{fig4}a.
Varying the frequency of the drive further allows us to extract the effective local energy scale which controls the heating dynamics: $J_{\mathrm{local}}^{E} = 1.21\pm 0.04$ and $J_{\text{local}}^{S} = 1.16\pm 0.04$. This is consistent with the microscopic onsite energy scale, $\|\Hstatic \| /L \simeq 1.26$ \cite{fn5}. 

\emph{Observing emergent hydrodynamics}---Having established that DMT accurately captures the late-time thermalization of Floquet systems, we now apply it to the study of a much broader question: the emergent hydrodynamics of large (undriven) quantum spin chains ($L = 100$).
In particular, our main goal here is to measure the diffusion coefficient as a function of temperature. 

%
Our setup is the following. 
On top of an initial thermal state with respect to $\Hstatic$, we add a small spatial inhomogeneity in the energy density (taken to be a Fourier mode) \cite{SM}.
As the system evolves under $\Hstatic$, one finds that the amplitude of this spatial variation decays exponentially, with a rate that scales as $q^2$, where $q$ is the wave-vector of the Fourier mode. 
This quadratic scaling is characteristic of diffusion and confirms the emergence of hydrodynamics from our microscopic quantum Hamiltonian \cite{SM}.
By further varying the temperature of the initial Gibbs ensemble, one can also study the diffusion coefficient, $D(\epsilon)$, as a function of the energy density $\epsilon$ (Fig.~\ref{fig4}b) \cite{fn6}. 

We emphasize that such a numerical observation of emergent hydrodynamics is well beyond the reach of conventional numerics and fundamentally leverages DMT's ability to prepare and evolve highly-entangled states near thermal equilibrium. 
Moreover, we note that our procedure can also be applied to the study of \emph{integrable} systems, where different types of anomalous transport can occur \cite{Prosen,Moore,Gopalakrishnan,Yoshimura,Fagotti,Misguich,Gobert}.
We highlight this by computing spin transport in the XXZ model  and observing ballistic, super-diffusive and diffusive exponents as a function of the Ising anisotropy (for details see \cite{SM}).

\figFour
\emph{Interplay between driving and hydrodynamics}---Taking things one step further, we now combine the two previous  settings and explore a situation where the interplay between Floquet heating and diffusive transport is crucial for understanding the system's thermalization dynamics. 
In particular, let us consider the time evolution of a spin chain where only the right half of the system is periodically driven (inset, Fig.~\ref{fig1}c).
At time $t=0$, the system is initialized in a N\'{e}el state with a domain wall every four spins \cite{fn3}. 

After an initial period of local equilibration, the combination of inhomogenous driving and interactions leads to three distinct features in the dynamics of the local energy density, as illustrated in Fig.~\ref{fig4}c. 
First, the local energy density on the right half of the spin chain is larger, reflecting the location where driving, and thus Floquet heating, is occurring. 
Second, the energy density across the entire chain gradually increases in time as energy from the right half is transported toward the left half.
Third, as the system approaches its infinite temperature state, the overall energy-density inhomogeneity between the left and right halves of the system is reduced.

Leveraging our previous characterizations of both heating and transport, we combine them into such a single hydrodynamical description.
The only missing element is a small correction to the transport due to the inhomogeneity of the drive, whose strength we characterize by a small, frequency dependent parameter $\eta$.

We now ask the following question: Can all three of these behaviors be \emph{quantitatively} captured using a simple hydrodynamical equation?
If so, one might naturally posit the following  modified diffusion equation \cite{SM}:
\begin{equation}
  \label{eq:diff}
    \partial_t \epsilon(x,t)= D(\epsilon)\partial^2_x\Big([1+\eta g(x)]\epsilon(x,t) \Big) - g(x)\frac{\epsilon(x,t)}{\tau^*_E}.
\end{equation}
Here, $g(x)$ is a step-like spatial profile which accounts for the fact that only half the spin chain is being driven \cite{fn7}.
The term proportional to $\eta$ corresponds to the aforementioned correction to the transport owing to the inhomogeneity of the drive, while the final term in the equation captures the Floquet heating.
Note that for the heating rate and the diffusion coefficient, we utilize the previously (and independently) determined values $1/\tau^*_E$ and $D(\epsilon)$, respectively (Fig.~\ref{fig4}a,b).

In order to test our hydrodynamical description, we feed in the energy density profile computed using DMT (at time $t=200$) into Eq.~\ref{eq:diff} and check whether the differential equation can quantitatively reproduce the remaining time dynamics (Fig.~\ref{fig4}c).
Our only fitting parameter is $\eta$, and we take it to be constant across the entire evolution. 
We find that $\eta \ll 1$ and decreases as frequency increases, consistent with our expectation that for larger driving frequencies, $D_\eff$ is more homogenous across the chain \cite{SM}. 
Remarkably, we observe excellent agreement for the remaining time evolution across all frequencies tested (Fig.~\ref{fig1}c and \ref{fig4}c)!
To this end, our results confirm that only a few coarse-grained observables are relevant to the late-time evolution of an interacting quantum system, even under a periodic drive \cite{SM}.

\begin{acknowledgments}
 \emph{Acknowledgments}---We would like to thank Yuval Baum, Soonwon Choi, Bryce Kobrin, Gregory D. Meyer, Mark Rudner, Michael Zaletel, Canxun Zhang, and Chong Zu for helpful conversations.
  Krylov-space simulations were performed using Meyer's software package Dynamite, a wrapper for the PETSc/SLEPc libraries~\cite{slepc1, slepc2, petsc1, fn9}.
This work is supported by the DARPA DRINQS program, the DOE (GeoFlow grant: DE-SC0019380), the Sloan foundation, the Packard foundation, the NSF (DMR-1848336), and the W. M. Keck Foundation.
  CDW gratefully acknowledges the support of the Caltech Institute for Quantum Information and Matter, an NSF Physics Frontiers Center supported by the Gordon and Betty Moore Foundation, and the National Science Foundation Graduate Research Fellowship under Grant No.~DGE‐1745301.
\end{acknowledgments}

\bibliography{references}

\begin{thebibliography}{67}%
\makeatletter
\providecommand \@ifxundefined [1]{%
 \@ifx{#1\undefined}
}%
\providecommand \@ifnum [1]{%
 \ifnum #1\expandafter \@firstoftwo
 \else \expandafter \@secondoftwo
 \fi
}%
\providecommand \@ifx [1]{%
 \ifx #1\expandafter \@firstoftwo
 \else \expandafter \@secondoftwo
 \fi
}%
\providecommand \natexlab [1]{#1}%
\providecommand \enquote  [1]{``#1''}%
\providecommand \bibnamefont  [1]{#1}%
\providecommand \bibfnamefont [1]{#1}%
\providecommand \citenamefont [1]{#1}%
\providecommand \href@noop [0]{\@secondoftwo}%
\providecommand \href [0]{\begingroup \@sanitize@url \@href}%
\providecommand \@href[1]{\@@startlink{#1}\@@href}%
\providecommand \@@href[1]{\endgroup#1\@@endlink}%
\providecommand \@sanitize@url [0]{\catcode `\\12\catcode `\$12\catcode
  `\&12\catcode `\#12\catcode `\^12\catcode `\_12\catcode `\%12\relax}%
\providecommand \@@startlink[1]{}%
\providecommand \@@endlink[0]{}%
\providecommand \url  [0]{\begingroup\@sanitize@url \@url }%
\providecommand \@url [1]{\endgroup\@href {#1}{\urlprefix }}%
\providecommand \urlprefix  [0]{URL }%
\providecommand \Eprint [0]{\href }%
\providecommand \doibase [0]{http://dx.doi.org/}%
\providecommand \selectlanguage [0]{\@gobble}%
\providecommand \bibinfo  [0]{\@secondoftwo}%
\providecommand \bibfield  [0]{\@secondoftwo}%
\providecommand \translation [1]{[#1]}%
\providecommand \BibitemOpen [0]{}%
\providecommand \bibitemStop [0]{}%
\providecommand \bibitemNoStop [0]{.\EOS\space}%
\providecommand \EOS [0]{\spacefactor3000\relax}%
\providecommand \BibitemShut  [1]{\csname bibitem#1\endcsname}%
\let\auto@bib@innerbib\@empty
\bibitem [{\citenamefont {Deutsch}(1991)}]{deutsch_quantum_1991}%
  \BibitemOpen
  \bibfield  {author} {\bibinfo {author} {\bibfnamefont {J.~M.}\ \bibnamefont
  {Deutsch}},\ }\href {\doibase 10.1103/PhysRevA.43.2046} {\bibfield  {journal}
  {\bibinfo  {journal} {Phys. Rev. A}\ }\textbf {\bibinfo {volume} {43}},\
  \bibinfo {pages} {2046} (\bibinfo {year} {1991})}\BibitemShut {NoStop}%
\bibitem [{\citenamefont {Srednicki}(1994)}]{srednicki_chaos_1994}%
  \BibitemOpen
  \bibfield  {author} {\bibinfo {author} {\bibfnamefont {M.}~\bibnamefont
  {Srednicki}},\ }\href {\doibase 10.1103/PhysRevE.50.888} {\bibfield
  {journal} {\bibinfo  {journal} {Phys. Rev. E}\ }\textbf {\bibinfo {volume}
  {50}},\ \bibinfo {pages} {888} (\bibinfo {year} {1994})}\BibitemShut
  {NoStop}%
\bibitem [{\citenamefont {{Rigol}}\ \emph {et~al.}(2008)\citenamefont
  {{Rigol}}, \citenamefont {{Dunjko}},\ and\ \citenamefont
  {{Olshanii}}}]{rigol_thermalization_2008}%
  \BibitemOpen
  \bibfield  {author} {\bibinfo {author} {\bibfnamefont {M.}~\bibnamefont
  {{Rigol}}}, \bibinfo {author} {\bibfnamefont {V.}~\bibnamefont {{Dunjko}}}, \
  and\ \bibinfo {author} {\bibfnamefont {M.}~\bibnamefont {{Olshanii}}},\
  }\href {\doibase 10.1038/nature06838} {\bibfield  {journal} {\bibinfo
  {journal} {Nature}\ }\textbf {\bibinfo {volume} {452}},\ \bibinfo {pages}
  {854} (\bibinfo {year} {2008})}\BibitemShut {NoStop}%
\bibitem [{\citenamefont {D'Alessio}\ \emph {et~al.}(2016)\citenamefont
  {D'Alessio}, \citenamefont {Kafri}, \citenamefont {Polkovnikov},\ and\
  \citenamefont {Rigol}}]{dalessio_quantum_2016}%
  \BibitemOpen
  \bibfield  {author} {\bibinfo {author} {\bibfnamefont {L.}~\bibnamefont
  {D'Alessio}}, \bibinfo {author} {\bibfnamefont {Y.}~\bibnamefont {Kafri}},
  \bibinfo {author} {\bibfnamefont {A.}~\bibnamefont {Polkovnikov}}, \ and\
  \bibinfo {author} {\bibfnamefont {M.}~\bibnamefont {Rigol}},\ }\href
  {\doibase 10.1080/00018732.2016.1198134} {\bibfield  {journal} {\bibinfo
  {journal} {Adv. Phys.}\ }\textbf {\bibinfo {volume} {65}},\ \bibinfo {pages}
  {239} (\bibinfo {year} {2016})}\BibitemShut {NoStop}%
\bibitem [{\citenamefont {Calabrese}\ \emph {et~al.}()\citenamefont
  {Calabrese}, \citenamefont {Essler},\ and\ \citenamefont
  {Mussardo}}]{Calabrese_2016}%
  \BibitemOpen
  \bibfield  {author} {\bibinfo {author} {\bibfnamefont {P.}~\bibnamefont
  {Calabrese}}, \bibinfo {author} {\bibfnamefont {F.~H.~L.}\ \bibnamefont
  {Essler}}, \ and\ \bibinfo {author} {\bibfnamefont {G.}~\bibnamefont
  {Mussardo}},\ }\href {\doibase 10.1088/1742-5468/2016/06/064001} {\bibfield
  {journal} {\bibinfo  {journal} {J. Stat. Mech.}\ }\textbf {\bibinfo {volume}
  {(2016)}},\ \bibinfo {pages} {064001}}\BibitemShut {NoStop}%
\bibitem [{\citenamefont {Gogolin}\ and\ \citenamefont
  {Eisert}(2016)}]{Gogolin_2016}%
  \BibitemOpen
  \bibfield  {author} {\bibinfo {author} {\bibfnamefont {C.}~\bibnamefont
  {Gogolin}}\ and\ \bibinfo {author} {\bibfnamefont {J.}~\bibnamefont
  {Eisert}},\ }\href {\doibase 10.1088/0034-4885/79/5/056001} {\bibfield
  {journal} {\bibinfo  {journal} {Rep. Prog. Phys}\ }\textbf {\bibinfo {volume}
  {79}},\ \bibinfo {pages} {056001} (\bibinfo {year} {2016})}\BibitemShut
  {NoStop}%
\bibitem [{\citenamefont {Potter}\ \emph {et~al.}(2015)\citenamefont {Potter},
  \citenamefont {Vasseur},\ and\ \citenamefont
  {Parameswaran}}]{potter_universal_2015}%
  \BibitemOpen
  \bibfield  {author} {\bibinfo {author} {\bibfnamefont {A.~C.}\ \bibnamefont
  {Potter}}, \bibinfo {author} {\bibfnamefont {R.}~\bibnamefont {Vasseur}}, \
  and\ \bibinfo {author} {\bibfnamefont {S.~A.}\ \bibnamefont {Parameswaran}},\
  }\href {\doibase 10.1103/PhysRevX.5.031033} {\bibfield  {journal} {\bibinfo
  {journal} {Phys. Rev. X}\ }\textbf {\bibinfo {volume} {5}},\ \bibinfo {pages}
  {031033} (\bibinfo {year} {2015})}\BibitemShut {NoStop}%
\bibitem [{\citenamefont {Vosk}\ \emph {et~al.}(2015)\citenamefont {Vosk},
  \citenamefont {Huse},\ and\ \citenamefont {Altman}}]{vosk_theory_2015}%
  \BibitemOpen
  \bibfield  {author} {\bibinfo {author} {\bibfnamefont {R.}~\bibnamefont
  {Vosk}}, \bibinfo {author} {\bibfnamefont {D.~A.}\ \bibnamefont {Huse}}, \
  and\ \bibinfo {author} {\bibfnamefont {E.}~\bibnamefont {Altman}},\ }\href
  {\doibase 10.1103/PhysRevX.5.031032} {\bibfield  {journal} {\bibinfo
  {journal} {Phys. Rev. X}\ }\textbf {\bibinfo {volume} {5}},\ \bibinfo {pages}
  {031032} (\bibinfo {year} {2015})}\BibitemShut {NoStop}%
\bibitem [{\citenamefont {Agarwal}\ \emph {et~al.}(2015)\citenamefont
  {Agarwal}, \citenamefont {Gopalakrishnan}, \citenamefont {Knap},
  \citenamefont {M\"uller},\ and\ \citenamefont
  {Demler}}]{agarwal_anomalous_2015}%
  \BibitemOpen
  \bibfield  {author} {\bibinfo {author} {\bibfnamefont {K.}~\bibnamefont
  {Agarwal}}, \bibinfo {author} {\bibfnamefont {S.}~\bibnamefont
  {Gopalakrishnan}}, \bibinfo {author} {\bibfnamefont {M.}~\bibnamefont
  {Knap}}, \bibinfo {author} {\bibfnamefont {M.}~\bibnamefont {M\"uller}}, \
  and\ \bibinfo {author} {\bibfnamefont {E.}~\bibnamefont {Demler}},\ }\href
  {\doibase 10.1103/PhysRevLett.114.160401} {\bibfield  {journal} {\bibinfo
  {journal} {Phys. Rev. Lett.}\ }\textbf {\bibinfo {volume} {114}},\ \bibinfo
  {pages} {160401} (\bibinfo {year} {2015})}\BibitemShut {NoStop}%
\bibitem [{\citenamefont {\ifmmode \check{Z}\else
  \v{Z}\fi{}nidari\ifmmode~\check{c}\else \v{c}\fi{}}\ \emph
  {et~al.}(2016)\citenamefont {\ifmmode \check{Z}\else
  \v{Z}\fi{}nidari\ifmmode~\check{c}\else \v{c}\fi{}}, \citenamefont
  {Scardicchio},\ and\ \citenamefont {Varma}}]{znidaric_diffusive_2016}%
  \BibitemOpen
  \bibfield  {author} {\bibinfo {author} {\bibfnamefont {M.}~\bibnamefont
  {\ifmmode \check{Z}\else \v{Z}\fi{}nidari\ifmmode~\check{c}\else
  \v{c}\fi{}}}, \bibinfo {author} {\bibfnamefont {A.}~\bibnamefont
  {Scardicchio}}, \ and\ \bibinfo {author} {\bibfnamefont {V.~K.}\ \bibnamefont
  {Varma}},\ }\href {\doibase 10.1103/PhysRevLett.117.040601} {\bibfield
  {journal} {\bibinfo  {journal} {Phys. Rev. Lett.}\ }\textbf {\bibinfo
  {volume} {117}},\ \bibinfo {pages} {040601} (\bibinfo {year}
  {2016})}\BibitemShut {NoStop}%
\bibitem [{\citenamefont {Luitz}\ \emph {et~al.}(2016)\citenamefont {Luitz},
  \citenamefont {Laflorencie},\ and\ \citenamefont {Alet}}]{Luitz_2016}%
  \BibitemOpen
  \bibfield  {author} {\bibinfo {author} {\bibfnamefont {D.~J.}\ \bibnamefont
  {Luitz}}, \bibinfo {author} {\bibfnamefont {N.}~\bibnamefont {Laflorencie}},
  \ and\ \bibinfo {author} {\bibfnamefont {F.}~\bibnamefont {Alet}},\ }\href
  {\doibase 10.1103/PhysRevB.93.060201} {\bibfield  {journal} {\bibinfo
  {journal} {Phys. Rev. B}\ }\textbf {\bibinfo {volume} {93}},\ \bibinfo
  {pages} {060201(R)} (\bibinfo {year} {2016})}\BibitemShut {NoStop}%
\bibitem [{\citenamefont {Khait}\ \emph {et~al.}(2016)\citenamefont {Khait},
  \citenamefont {Gazit}, \citenamefont {Yao},\ and\ \citenamefont
  {Auerbach}}]{Khait_2016}%
  \BibitemOpen
  \bibfield  {author} {\bibinfo {author} {\bibfnamefont {I.}~\bibnamefont
  {Khait}}, \bibinfo {author} {\bibfnamefont {S.}~\bibnamefont {Gazit}},
  \bibinfo {author} {\bibfnamefont {N.~Y.}\ \bibnamefont {Yao}}, \ and\
  \bibinfo {author} {\bibfnamefont {A.}~\bibnamefont {Auerbach}},\ }\href
  {\doibase 10.1103/PhysRevB.93.224205} {\bibfield  {journal} {\bibinfo
  {journal} {Phys. Rev. B}\ }\textbf {\bibinfo {volume} {93}},\ \bibinfo
  {pages} {224205} (\bibinfo {year} {2016})}\BibitemShut {NoStop}%
\bibitem [{\citenamefont {Luitz}\ and\ \citenamefont
  {Bar~Lev}(2016)}]{Luitz_2016_2}%
  \BibitemOpen
  \bibfield  {author} {\bibinfo {author} {\bibfnamefont {D.~J.}\ \bibnamefont
  {Luitz}}\ and\ \bibinfo {author} {\bibfnamefont {Y.}~\bibnamefont
  {Bar~Lev}},\ }\href {\doibase 10.1103/PhysRevLett.117.170404} {\bibfield
  {journal} {\bibinfo  {journal} {Phys. Rev. Lett.}\ }\textbf {\bibinfo
  {volume} {117}},\ \bibinfo {pages} {170404} (\bibinfo {year}
  {2016})}\BibitemShut {NoStop}%
\bibitem [{\citenamefont {Sahu}\ \emph {et~al.}(2018)\citenamefont {Sahu},
  \citenamefont {Xu},\ and\ \citenamefont {Swingle}}]{sahu_scrambling_2018}%
  \BibitemOpen
  \bibfield  {author} {\bibinfo {author} {\bibfnamefont {S.}~\bibnamefont
  {Sahu}}, \bibinfo {author} {\bibfnamefont {S.}~\bibnamefont {Xu}}, \ and\
  \bibinfo {author} {\bibfnamefont {B.}~\bibnamefont {Swingle}},\ }\href
  {http://arxiv.org/abs/1807.06086} {\bibfield  {journal} {\bibinfo  {journal}
  {arXiv:1807.06086}\ } (\bibinfo {year} {2018})}\BibitemShut {NoStop}%
\bibitem [{\citenamefont {Bohrdt}\ \emph {et~al.}(2017)\citenamefont {Bohrdt},
  \citenamefont {Mendl}, \citenamefont {Endres},\ and\ \citenamefont
  {Knap}}]{Bohrdt_2017}%
  \BibitemOpen
  \bibfield  {author} {\bibinfo {author} {\bibfnamefont {A.}~\bibnamefont
  {Bohrdt}}, \bibinfo {author} {\bibfnamefont {C.~B.}\ \bibnamefont {Mendl}},
  \bibinfo {author} {\bibfnamefont {M.}~\bibnamefont {Endres}}, \ and\ \bibinfo
  {author} {\bibfnamefont {M.}~\bibnamefont {Knap}},\ }\href {\doibase
  10.1088/1367-2630/aa719b} {\bibfield  {journal} {\bibinfo  {journal} {New J.
  Phys.}\ }\textbf {\bibinfo {volume} {19}},\ \bibinfo {pages} {063001}
  (\bibinfo {year} {2017})}\BibitemShut {NoStop}%
\bibitem [{\citenamefont {Else}\ \emph {et~al.}(2016)\citenamefont {Else},
  \citenamefont {Bauer},\ and\ \citenamefont {Nayak}}]{else_floquet_2016}%
  \BibitemOpen
  \bibfield  {author} {\bibinfo {author} {\bibfnamefont {D.~V.}\ \bibnamefont
  {Else}}, \bibinfo {author} {\bibfnamefont {B.}~\bibnamefont {Bauer}}, \ and\
  \bibinfo {author} {\bibfnamefont {C.}~\bibnamefont {Nayak}},\ }\href
  {\doibase 10.1103/PhysRevLett.117.090402} {\bibfield  {journal} {\bibinfo
  {journal} {Phys. Rev. Lett.}\ }\textbf {\bibinfo {volume} {117}},\ \bibinfo
  {pages} {090402} (\bibinfo {year} {2016})}\BibitemShut {NoStop}%
\bibitem [{\citenamefont {Khemani}\ \emph {et~al.}(2016)\citenamefont
  {Khemani}, \citenamefont {Lazarides}, \citenamefont {Moessner},\ and\
  \citenamefont {Sondhi}}]{khemani_phase_2016}%
  \BibitemOpen
  \bibfield  {author} {\bibinfo {author} {\bibfnamefont {V.}~\bibnamefont
  {Khemani}}, \bibinfo {author} {\bibfnamefont {A.}~\bibnamefont {Lazarides}},
  \bibinfo {author} {\bibfnamefont {R.}~\bibnamefont {Moessner}}, \ and\
  \bibinfo {author} {\bibfnamefont {S.~L.}\ \bibnamefont {Sondhi}},\ }\href
  {\doibase 10.1103/PhysRevLett.116.250401} {\bibfield  {journal} {\bibinfo
  {journal} {Phys. Rev. Lett.}\ }\textbf {\bibinfo {volume} {116}},\ \bibinfo
  {pages} {250401} (\bibinfo {year} {2016})}\BibitemShut {NoStop}%
\bibitem [{\citenamefont {von Keyserlingk}\ and\ \citenamefont
  {Sondhi}(2016{\natexlab{a}})}]{von_keyserlingk_phase_2016}%
  \BibitemOpen
  \bibfield  {author} {\bibinfo {author} {\bibfnamefont {C.~W.}\ \bibnamefont
  {von Keyserlingk}}\ and\ \bibinfo {author} {\bibfnamefont {S.~L.}\
  \bibnamefont {Sondhi}},\ }\href {\doibase 10.1103/PhysRevB.93.245145}
  {\bibfield  {journal} {\bibinfo  {journal} {Phys. Rev. B}\ }\textbf {\bibinfo
  {volume} {93}},\ \bibinfo {pages} {245145} (\bibinfo {year}
  {2016}{\natexlab{a}})}\BibitemShut {NoStop}%
\bibitem [{\citenamefont {von Keyserlingk}\ and\ \citenamefont
  {Sondhi}(2016{\natexlab{b}})}]{von_keyserlingk_1d_2016}%
  \BibitemOpen
  \bibfield  {author} {\bibinfo {author} {\bibfnamefont {C.~W.}\ \bibnamefont
  {von Keyserlingk}}\ and\ \bibinfo {author} {\bibfnamefont {S.~L.}\
  \bibnamefont {Sondhi}},\ }\href {\doibase 10.1103/PhysRevB.93.245146}
  {\bibfield  {journal} {\bibinfo  {journal} {Phys. Rev. B}\ }\textbf {\bibinfo
  {volume} {93}},\ \bibinfo {pages} {245146} (\bibinfo {year}
  {2016}{\natexlab{b}})}\BibitemShut {NoStop}%
\bibitem [{\citenamefont {von Keyserlingk}\ \emph {et~al.}(2016)\citenamefont
  {von Keyserlingk}, \citenamefont {Khemani},\ and\ \citenamefont
  {Sondhi}}]{von_keyserlingk_absolute_2016}%
  \BibitemOpen
  \bibfield  {author} {\bibinfo {author} {\bibfnamefont {C.~W.}\ \bibnamefont
  {von Keyserlingk}}, \bibinfo {author} {\bibfnamefont {V.}~\bibnamefont
  {Khemani}}, \ and\ \bibinfo {author} {\bibfnamefont {S.~L.}\ \bibnamefont
  {Sondhi}},\ }\href {\doibase 10.1103/PhysRevB.94.085112} {\bibfield
  {journal} {\bibinfo  {journal} {Phys. Rev. B}\ }\textbf {\bibinfo {volume}
  {94}},\ \bibinfo {pages} {085112} (\bibinfo {year} {2016})}\BibitemShut
  {NoStop}%
\bibitem [{\citenamefont {Lindner}\ \emph {et~al.}(2011)\citenamefont
  {Lindner}, \citenamefont {Refael},\ and\ \citenamefont
  {Galitski}}]{lindner_floquet_2011}%
  \BibitemOpen
  \bibfield  {author} {\bibinfo {author} {\bibfnamefont {N.~H.}\ \bibnamefont
  {Lindner}}, \bibinfo {author} {\bibfnamefont {G.}~\bibnamefont {Refael}}, \
  and\ \bibinfo {author} {\bibfnamefont {V.}~\bibnamefont {Galitski}},\ }\href
  {\doibase 10.1038/nphys1926} {\bibfield  {journal} {\bibinfo  {journal} {Nat.
  Phys.}\ }\textbf {\bibinfo {volume} {7}},\ \bibinfo {pages} {490} (\bibinfo
  {year} {2011})}\BibitemShut {NoStop}%
\bibitem [{\citenamefont {Kitagawa}\ \emph {et~al.}(2010)\citenamefont
  {Kitagawa}, \citenamefont {Berg}, \citenamefont {Rudner},\ and\ \citenamefont
  {Demler}}]{kitagawa_topological_2010}%
  \BibitemOpen
  \bibfield  {author} {\bibinfo {author} {\bibfnamefont {T.}~\bibnamefont
  {Kitagawa}}, \bibinfo {author} {\bibfnamefont {E.}~\bibnamefont {Berg}},
  \bibinfo {author} {\bibfnamefont {M.}~\bibnamefont {Rudner}}, \ and\ \bibinfo
  {author} {\bibfnamefont {E.}~\bibnamefont {Demler}},\ }\href {\doibase
  10.1103/PhysRevB.82.235114} {\bibfield  {journal} {\bibinfo  {journal} {Phys.
  Rev. B}\ }\textbf {\bibinfo {volume} {82}},\ \bibinfo {pages} {235114}
  (\bibinfo {year} {2010})}\BibitemShut {NoStop}%
\bibitem [{\citenamefont {Yao}\ \emph {et~al.}(2017)\citenamefont {Yao},
  \citenamefont {Potter}, \citenamefont {Potirniche},\ and\ \citenamefont
  {Vishwanath}}]{yao_timecrystaltheory_2017}%
  \BibitemOpen
  \bibfield  {author} {\bibinfo {author} {\bibfnamefont {N.~Y.}\ \bibnamefont
  {Yao}}, \bibinfo {author} {\bibfnamefont {A.~C.}\ \bibnamefont {Potter}},
  \bibinfo {author} {\bibfnamefont {I.-D.}\ \bibnamefont {Potirniche}}, \ and\
  \bibinfo {author} {\bibfnamefont {A.}~\bibnamefont {Vishwanath}},\ }\href
  {\doibase 10.1103/PhysRevLett.118.030401} {\bibfield  {journal} {\bibinfo
  {journal} {Phys. Rev. Lett.}\ }\textbf {\bibinfo {volume} {118}},\ \bibinfo
  {pages} {030401} (\bibinfo {year} {2017})}\BibitemShut {NoStop}%
\bibitem [{\citenamefont {Choi}\ \emph {et~al.}(2017)\citenamefont {Choi},
  \citenamefont {Choi}, \citenamefont {Landig}, \citenamefont {Kucsko},
  \citenamefont {Zhou}, \citenamefont {Isoya}, \citenamefont {Jelezko},
  \citenamefont {Onoda}, \citenamefont {Sumiya}, \citenamefont {Khemani},
  \citenamefont {von Keyserlingk}, \citenamefont {Yao}, \citenamefont
  {Demler},\ and\ \citenamefont {Lukin}}]{choi_timecrystal_2017}%
  \BibitemOpen
  \bibfield  {author} {\bibinfo {author} {\bibfnamefont {S.}~\bibnamefont
  {Choi}}, \bibinfo {author} {\bibfnamefont {J.}~\bibnamefont {Choi}}, \bibinfo
  {author} {\bibfnamefont {R.}~\bibnamefont {Landig}}, \bibinfo {author}
  {\bibfnamefont {G.}~\bibnamefont {Kucsko}}, \bibinfo {author} {\bibfnamefont
  {H.}~\bibnamefont {Zhou}}, \bibinfo {author} {\bibfnamefont {J.}~\bibnamefont
  {Isoya}}, \bibinfo {author} {\bibfnamefont {F.}~\bibnamefont {Jelezko}},
  \bibinfo {author} {\bibfnamefont {S.}~\bibnamefont {Onoda}}, \bibinfo
  {author} {\bibfnamefont {H.}~\bibnamefont {Sumiya}}, \bibinfo {author}
  {\bibfnamefont {V.}~\bibnamefont {Khemani}}, \bibinfo {author} {\bibfnamefont
  {C.}~\bibnamefont {von Keyserlingk}}, \bibinfo {author} {\bibfnamefont
  {N.~Y.}\ \bibnamefont {Yao}}, \bibinfo {author} {\bibfnamefont
  {E.}~\bibnamefont {Demler}}, \ and\ \bibinfo {author} {\bibfnamefont {M.~D.}\
  \bibnamefont {Lukin}},\ }\href {http://dx.doi.org/10.1038/nature21426}
  {\bibfield  {journal} {\bibinfo  {journal} {Nature}\ }\textbf {\bibinfo
  {volume} {543}},\ \bibinfo {pages} {221} (\bibinfo {year}
  {2017})}\BibitemShut {NoStop}%
\bibitem [{\citenamefont {Zhang}\ \emph {et~al.}(2017)\citenamefont {Zhang},
  \citenamefont {Hess}, \citenamefont {Kyprianidis}, \citenamefont {Becker},
  \citenamefont {Lee}, \citenamefont {Smith}, \citenamefont {Pagano},
  \citenamefont {Potirniche}, \citenamefont {Potter}, \citenamefont
  {Vishwanath}, \citenamefont {Yao},\ and\ \citenamefont
  {Monroe}}]{zhang_timecrystal_2017}%
  \BibitemOpen
  \bibfield  {author} {\bibinfo {author} {\bibfnamefont {J.}~\bibnamefont
  {Zhang}}, \bibinfo {author} {\bibfnamefont {P.~W.}\ \bibnamefont {Hess}},
  \bibinfo {author} {\bibfnamefont {A.}~\bibnamefont {Kyprianidis}}, \bibinfo
  {author} {\bibfnamefont {P.}~\bibnamefont {Becker}}, \bibinfo {author}
  {\bibfnamefont {A.}~\bibnamefont {Lee}}, \bibinfo {author} {\bibfnamefont
  {J.}~\bibnamefont {Smith}}, \bibinfo {author} {\bibfnamefont
  {G.}~\bibnamefont {Pagano}}, \bibinfo {author} {\bibfnamefont {I.-D.}\
  \bibnamefont {Potirniche}}, \bibinfo {author} {\bibfnamefont {A.~C.}\
  \bibnamefont {Potter}}, \bibinfo {author} {\bibfnamefont {A.}~\bibnamefont
  {Vishwanath}}, \bibinfo {author} {\bibfnamefont {N.~Y.}\ \bibnamefont {Yao}},
  \ and\ \bibinfo {author} {\bibfnamefont {C.}~\bibnamefont {Monroe}},\ }\href
  {http://dx.doi.org/10.1038/nature21413} {\bibfield  {journal} {\bibinfo
  {journal} {Nature}\ }\textbf {\bibinfo {volume} {543}},\ \bibinfo {pages}
  {217} (\bibinfo {year} {2017})}\BibitemShut {NoStop}%
\bibitem [{\citenamefont {Prosen}(1998)}]{prosen_1998}%
  \BibitemOpen
  \bibfield  {author} {\bibinfo {author} {\bibfnamefont {T.}~\bibnamefont
  {Prosen}},\ }\href {\doibase 10.1103/PhysRevLett.80.1808} {\bibfield
  {journal} {\bibinfo  {journal} {Phys. Rev. Lett.}\ }\textbf {\bibinfo
  {volume} {80}},\ \bibinfo {pages} {1808} (\bibinfo {year}
  {1998})}\BibitemShut {NoStop}%
\bibitem [{\citenamefont {Prosen}(1999)}]{prosen_1999}%
  \BibitemOpen
  \bibfield  {author} {\bibinfo {author} {\bibfnamefont {T.}~\bibnamefont
  {Prosen}},\ }\href {\doibase 10.1103/PhysRevE.60.3949} {\bibfield  {journal}
  {\bibinfo  {journal} {Phys. Rev. E}\ }\textbf {\bibinfo {volume} {60}},\
  \bibinfo {pages} {3949} (\bibinfo {year} {1999})}\BibitemShut {NoStop}%
\bibitem [{\citenamefont {Lazarides}\ \emph {et~al.}(2014)\citenamefont
  {Lazarides}, \citenamefont {Das},\ and\ \citenamefont
  {Moessner}}]{lazarides_2014}%
  \BibitemOpen
  \bibfield  {author} {\bibinfo {author} {\bibfnamefont {A.}~\bibnamefont
  {Lazarides}}, \bibinfo {author} {\bibfnamefont {A.}~\bibnamefont {Das}}, \
  and\ \bibinfo {author} {\bibfnamefont {R.}~\bibnamefont {Moessner}},\ }\href
  {\doibase 10.1103/PhysRevE.90.012110} {\bibfield  {journal} {\bibinfo
  {journal} {Phys. Rev. E}\ }\textbf {\bibinfo {volume} {90}},\ \bibinfo
  {pages} {012110} (\bibinfo {year} {2014})}\BibitemShut {NoStop}%
\bibitem [{\citenamefont {D'Alessio}\ and\ \citenamefont
  {Rigol}(2014)}]{d'alessio_drivenlongtime_2014}%
  \BibitemOpen
  \bibfield  {author} {\bibinfo {author} {\bibfnamefont {L.}~\bibnamefont
  {D'Alessio}}\ and\ \bibinfo {author} {\bibfnamefont {M.}~\bibnamefont
  {Rigol}},\ }\href {\doibase 10.1103/PhysRevX.4.041048} {\bibfield  {journal}
  {\bibinfo  {journal} {Phys. Rev. X}\ }\textbf {\bibinfo {volume} {4}},\
  \bibinfo {pages} {041048} (\bibinfo {year} {2014})}\BibitemShut {NoStop}%
\bibitem [{\citenamefont {Bukov}\ \emph
  {et~al.}(2015{\natexlab{a}})\citenamefont {Bukov}, \citenamefont
  {D'Alessio},\ and\ \citenamefont {Polkovnikov}}]{bukov_highfreq_2015}%
  \BibitemOpen
  \bibfield  {author} {\bibinfo {author} {\bibfnamefont {M.}~\bibnamefont
  {Bukov}}, \bibinfo {author} {\bibfnamefont {L.}~\bibnamefont {D'Alessio}}, \
  and\ \bibinfo {author} {\bibfnamefont {A.}~\bibnamefont {Polkovnikov}},\
  }\href {\doibase 10.1080/00018732.2015.1055918} {\bibfield  {journal}
  {\bibinfo  {journal} {Adv. Phys.}\ }\textbf {\bibinfo {volume} {64}},\
  \bibinfo {pages} {139} (\bibinfo {year} {2015}{\natexlab{a}})}\BibitemShut
  {NoStop}%
\bibitem [{\citenamefont {Machado}\ \emph {et~al.}(2019)\citenamefont
  {Machado}, \citenamefont {Kahanamoku-Meyer}, \citenamefont {Else},
  \citenamefont {Nayak},\ and\ \citenamefont
  {Yao}}]{machado_exponentially_2017}%
  \BibitemOpen
  \bibfield  {author} {\bibinfo {author} {\bibfnamefont {F.}~\bibnamefont
  {Machado}}, \bibinfo {author} {\bibfnamefont {G.~D.}\ \bibnamefont
  {Kahanamoku-Meyer}}, \bibinfo {author} {\bibfnamefont {D.~V.}\ \bibnamefont
  {Else}}, \bibinfo {author} {\bibfnamefont {C.}~\bibnamefont {Nayak}}, \ and\
  \bibinfo {author} {\bibfnamefont {N.~Y.}\ \bibnamefont {Yao}},\ }\href
  {\doibase 10.1103/PhysRevResearch.1.033202} {\bibfield  {journal} {\bibinfo
  {journal} {Phys. Rev. Research}\ }\textbf {\bibinfo {volume} {1}},\ \bibinfo
  {pages} {033202} (\bibinfo {year} {2019})}\BibitemShut {NoStop}%
\bibitem [{SM()}]{SM}%
  \BibitemOpen
  \href@noop {} {\ }\bibinfo {note} {\hspace{-1mm}See supplementary information
  for details.}\BibitemShut {Stop}%
\bibitem [{fn1()}]{fn1}%
  \BibitemOpen
  \href@noop {} {\bibinfo  {journal} {Many other phenomena also fall under this
  umbrella including dynamics in localized and open stochastic systems}\
  }\BibitemShut {NoStop}%
\bibitem [{\citenamefont {Abanin}\ \emph {et~al.}(2015)\citenamefont {Abanin},
  \citenamefont {De~Roeck},\ and\ \citenamefont
  {Huveneers}}]{abanin2015exponentially}%
  \BibitemOpen
\bibfield  {journal} {  }\bibfield  {author} {\bibinfo {author} {\bibfnamefont
  {D.~A.}\ \bibnamefont {Abanin}}, \bibinfo {author} {\bibfnamefont
  {W.}~\bibnamefont {De~Roeck}}, \ and\ \bibinfo {author} {\bibfnamefont
  {F.}~\bibnamefont {Huveneers}},\ }\href {\doibase
  10.1103/PhysRevLett.115.256803} {\bibfield  {journal} {\bibinfo  {journal}
  {Phys. Rev. Lett.}\ }\textbf {\bibinfo {volume} {115}},\ \bibinfo {pages}
  {256803} (\bibinfo {year} {2015})}\BibitemShut {NoStop}%
\bibitem [{\citenamefont {Mori}\ \emph {et~al.}(2016)\citenamefont {Mori},
  \citenamefont {Kuwahara},\ and\ \citenamefont {Saito}}]{Mori2016}%
  \BibitemOpen
  \bibfield  {author} {\bibinfo {author} {\bibfnamefont {T.}~\bibnamefont
  {Mori}}, \bibinfo {author} {\bibfnamefont {T.}~\bibnamefont {Kuwahara}}, \
  and\ \bibinfo {author} {\bibfnamefont {K.}~\bibnamefont {Saito}},\ }\href
  {\doibase 10.1103/PhysRevLett.116.120401} {\bibfield  {journal} {\bibinfo
  {journal} {Phys. Rev. Lett.}\ }\textbf {\bibinfo {volume} {116}},\ \bibinfo
  {pages} {120401} (\bibinfo {year} {2016})}\BibitemShut {NoStop}%
\bibitem [{\citenamefont {Abanin}\ \emph
  {et~al.}(2017{\natexlab{a}})\citenamefont {Abanin}, \citenamefont {De~Roeck},
  \citenamefont {Ho},\ and\ \citenamefont {Huveneers}}]{abanin2017effective}%
  \BibitemOpen
  \bibfield  {author} {\bibinfo {author} {\bibfnamefont {D.~A.}\ \bibnamefont
  {Abanin}}, \bibinfo {author} {\bibfnamefont {W.}~\bibnamefont {De~Roeck}},
  \bibinfo {author} {\bibfnamefont {W.~W.}\ \bibnamefont {Ho}}, \ and\ \bibinfo
  {author} {\bibfnamefont {F.}~\bibnamefont {Huveneers}},\ }\href {\doibase
  10.1103/PhysRevB.95.014112} {\bibfield  {journal} {\bibinfo  {journal} {Phys.
  Rev. B}\ }\textbf {\bibinfo {volume} {95}},\ \bibinfo {pages} {014112}
  (\bibinfo {year} {2017}{\natexlab{a}})}\BibitemShut {NoStop}%
\bibitem [{\citenamefont {Kuwahara}\ \emph {et~al.}(2016)\citenamefont
  {Kuwahara}, \citenamefont {Mori},\ and\ \citenamefont
  {Saito}}]{Kuwahara2016}%
  \BibitemOpen
  \bibfield  {author} {\bibinfo {author} {\bibfnamefont {T.}~\bibnamefont
  {Kuwahara}}, \bibinfo {author} {\bibfnamefont {T.}~\bibnamefont {Mori}}, \
  and\ \bibinfo {author} {\bibfnamefont {K.}~\bibnamefont {Saito}},\ }\href
  {\doibase http://dx.doi.org/10.1016/j.aop.2016.01.012} {\bibfield  {journal}
  {\bibinfo  {journal} {Ann. Phys.}\ }\textbf {\bibinfo {volume} {367}},\
  \bibinfo {pages} {96 } (\bibinfo {year} {2016})}\BibitemShut {NoStop}%
\bibitem [{\citenamefont {Abanin}\ \emph
  {et~al.}(2017{\natexlab{b}})\citenamefont {Abanin}, \citenamefont {De~Roeck},
  \citenamefont {Ho},\ and\ \citenamefont {Huveneers}}]{abanin2017rigorous}%
  \BibitemOpen
  \bibfield  {author} {\bibinfo {author} {\bibfnamefont {D.}~\bibnamefont
  {Abanin}}, \bibinfo {author} {\bibfnamefont {W.}~\bibnamefont {De~Roeck}},
  \bibinfo {author} {\bibfnamefont {W.~W.}\ \bibnamefont {Ho}}, \ and\ \bibinfo
  {author} {\bibfnamefont {F.}~\bibnamefont {Huveneers}},\ }\href {\doibase
  10.1007/s00220-017-2930-x} {\bibfield  {journal} {\bibinfo  {journal}
  {Commun. Math. Phys.}\ }\textbf {\bibinfo {volume} {354}},\ \bibinfo {pages}
  {809} (\bibinfo {year} {2017}{\natexlab{b}})}\BibitemShut {NoStop}%
\bibitem [{\citenamefont {Else}\ \emph {et~al.}(2017)\citenamefont {Else},
  \citenamefont {Bauer},\ and\ \citenamefont {Nayak}}]{else2017prethermal}%
  \BibitemOpen
  \bibfield  {author} {\bibinfo {author} {\bibfnamefont {D.~V.}\ \bibnamefont
  {Else}}, \bibinfo {author} {\bibfnamefont {B.}~\bibnamefont {Bauer}}, \ and\
  \bibinfo {author} {\bibfnamefont {C.}~\bibnamefont {Nayak}},\ }\href
  {\doibase 10.1103/PhysRevX.7.011026} {\bibfield  {journal} {\bibinfo
  {journal} {Phys. Rev. X}\ }\textbf {\bibinfo {volume} {7}},\ \bibinfo {pages}
  {011026} (\bibinfo {year} {2017})}\BibitemShut {NoStop}%
\bibitem [{\citenamefont {Bukov}\ \emph
  {et~al.}(2015{\natexlab{b}})\citenamefont {Bukov}, \citenamefont
  {Gopalakrishnan}, \citenamefont {Knap},\ and\ \citenamefont
  {Demler}}]{Bukov_2015}%
  \BibitemOpen
  \bibfield  {author} {\bibinfo {author} {\bibfnamefont {M.}~\bibnamefont
  {Bukov}}, \bibinfo {author} {\bibfnamefont {S.}~\bibnamefont
  {Gopalakrishnan}}, \bibinfo {author} {\bibfnamefont {M.}~\bibnamefont
  {Knap}}, \ and\ \bibinfo {author} {\bibfnamefont {E.}~\bibnamefont
  {Demler}},\ }\href {\doibase 10.1103/PhysRevLett.115.205301} {\bibfield
  {journal} {\bibinfo  {journal} {Phys. Rev. Lett.}\ }\textbf {\bibinfo
  {volume} {115}},\ \bibinfo {pages} {205301} (\bibinfo {year}
  {2015}{\natexlab{b}})}\BibitemShut {NoStop}%
\bibitem [{\citenamefont {Weidinger}\ and\ \citenamefont
  {Knap}(2017)}]{Bohrdt_2017_1}%
  \BibitemOpen
  \bibfield  {author} {\bibinfo {author} {\bibfnamefont {S.~A.}\ \bibnamefont
  {Weidinger}}\ and\ \bibinfo {author} {\bibfnamefont {M.}~\bibnamefont
  {Knap}},\ }\href {\doibase 10.1038/srep45382} {\bibfield  {journal} {\bibinfo
   {journal} {Sci. Rep.}\ }\textbf {\bibinfo {volume} {7}},\ \bibinfo {pages}
  {45382} (\bibinfo {year} {2017})}\BibitemShut {NoStop}%
\bibitem [{\citenamefont {White}\ \emph {et~al.}(2018)\citenamefont {White},
  \citenamefont {Zaletel}, \citenamefont {Mong},\ and\ \citenamefont
  {Refael}}]{White_2017}%
  \BibitemOpen
  \bibfield  {author} {\bibinfo {author} {\bibfnamefont {C.~D.}\ \bibnamefont
  {White}}, \bibinfo {author} {\bibfnamefont {M.}~\bibnamefont {Zaletel}},
  \bibinfo {author} {\bibfnamefont {R.~S.~K.}\ \bibnamefont {Mong}}, \ and\
  \bibinfo {author} {\bibfnamefont {G.}~\bibnamefont {Refael}},\ }\href
  {\doibase 10.1103/PhysRevB.97.035127} {\bibfield  {journal} {\bibinfo
  {journal} {Phys. Rev. B}\ }\textbf {\bibinfo {volume} {97}},\ \bibinfo
  {pages} {035127} (\bibinfo {year} {2018})}\BibitemShut {NoStop}%
\bibitem [{\citenamefont {Leviatan}\ \emph {et~al.}()\citenamefont {Leviatan},
  \citenamefont {Pollmann}, \citenamefont {Bardarson}, \citenamefont {Huse},\
  and\ \citenamefont {Altman}}]{Leviatan_2017}%
  \BibitemOpen
  \bibfield  {author} {\bibinfo {author} {\bibfnamefont {E.}~\bibnamefont
  {Leviatan}}, \bibinfo {author} {\bibfnamefont {F.}~\bibnamefont {Pollmann}},
  \bibinfo {author} {\bibfnamefont {J.~H.}\ \bibnamefont {Bardarson}}, \bibinfo
  {author} {\bibfnamefont {D.~A.}\ \bibnamefont {Huse}}, \ and\ \bibinfo
  {author} {\bibfnamefont {E.}~\bibnamefont {Altman}},\ }\href
  {http://arxiv.org/abs/1702.08894} {\bibinfo  {journal} {arXiv:1702.08894}\
  }\BibitemShut {NoStop}%
\bibitem [{\citenamefont {Wurtz}\ \emph {et~al.}(2018)\citenamefont {Wurtz},
  \citenamefont {Polkovnikov},\ and\ \citenamefont {Sels}}]{Wurtz_2018.1}%
  \BibitemOpen
\bibfield  {journal} {  }\bibfield  {author} {\bibinfo {author} {\bibfnamefont
  {J.}~\bibnamefont {Wurtz}}, \bibinfo {author} {\bibfnamefont
  {A.}~\bibnamefont {Polkovnikov}}, \ and\ \bibinfo {author} {\bibfnamefont
  {D.}~\bibnamefont {Sels}},\ }\href {\doibase 10.1016/j.aop.2018.06.001}
  {\bibfield  {journal} {\bibinfo  {journal} {Ann. Phys.}\ }\textbf {\bibinfo
  {volume} {395}},\ \bibinfo {pages} {341} (\bibinfo {year}
  {2018})}\BibitemShut {NoStop}%
\bibitem [{\citenamefont {Wurtz}\ and\ \citenamefont
  {Polkovnikov}()}]{Wurtz_2018.2}%
  \BibitemOpen
  \bibfield  {author} {\bibinfo {author} {\bibfnamefont {J.}~\bibnamefont
  {Wurtz}}\ and\ \bibinfo {author} {\bibfnamefont {A.}~\bibnamefont
  {Polkovnikov}},\ }\href {https://arxiv.org/abs/1808.08977} {\bibinfo
  {journal} {arXiv:1808.08977}\ }\BibitemShut {NoStop}%
\bibitem [{\citenamefont {Zaletel}\ and\ \citenamefont
  {Pollmann}()}]{Mike_2019}%
  \BibitemOpen
\bibfield  {journal} {  }\bibfield  {author} {\bibinfo {author} {\bibfnamefont
  {M.~P.}\ \bibnamefont {Zaletel}}\ and\ \bibinfo {author} {\bibfnamefont
  {F.}~\bibnamefont {Pollmann}},\ }\href {https://arxiv.org/abs/1902.05100}
  {\bibinfo  {journal} {arXiv:1902.05100}\ }\BibitemShut {NoStop}%
\bibitem [{fn2()}]{fn2}%
  \BibitemOpen
\bibfield  {journal} {  }\href@noop {} {\bibinfo  {journal} {While the bond
  terms can be mapped to a free-fermion integrable model, the additional field
  term breaks this integrability \cite{SM}}\ }\BibitemShut {NoStop}%
\bibitem [{fn3()}]{fn3}%
  \BibitemOpen
\bibfield  {journal} {  }\href@noop {} {\bibinfo  {journal} {In our
  calculations, we consider a generic initial state, typically taken to be a
  N\'{e}el state with a domain wall every four spins. We have checked that our
  observations are independent of initial state. Based on previous studies we
  expect this choice of initial state to be generic and to capture the main
  features of Floquet heating \cite{machado_exponentially_2017}}\ }\BibitemShut
  {NoStop}%
\bibitem [{fn8()}]{fn8}%
  \BibitemOpen
\bibfield  {journal} {  }\href@noop {} {\bibinfo  {journal} {Krylov methods
  compute the time evolution of the state by first constructing an appropriate
  subspace and then using it to build a suitable rational approximation to the
  exponential action of the Hamiltonian}\ }\BibitemShut {NoStop}%
\bibitem [{\citenamefont {Hernandez}\ \emph {et~al.}(2005)\citenamefont
  {Hernandez}, \citenamefont {Roman},\ and\ \citenamefont {Vidal}}]{slepc1}%
  \BibitemOpen
\bibfield  {journal} {  }\bibfield  {author} {\bibinfo {author} {\bibfnamefont
  {V.}~\bibnamefont {Hernandez}}, \bibinfo {author} {\bibfnamefont {J.~E.}\
  \bibnamefont {Roman}}, \ and\ \bibinfo {author} {\bibfnamefont
  {V.}~\bibnamefont {Vidal}},\ }\href {\doibase
  http://dx.doi.org/10.1145/1089014.1089019} {\bibfield  {journal} {\bibinfo
  {journal} {{ACM} Trans. Math. Software}\ }\textbf {\bibinfo {volume} {31}},\
  \bibinfo {pages} {351} (\bibinfo {year} {2005})}\BibitemShut {NoStop}%
\bibitem [{\citenamefont {Roman}\ \emph {et~al.}(2016)\citenamefont {Roman},
  \citenamefont {Campos}, \citenamefont {Romero},\ and\ \citenamefont
  {Tomas}}]{slepc2}%
  \BibitemOpen
  \bibfield  {author} {\bibinfo {author} {\bibfnamefont {J.~E.}\ \bibnamefont
  {Roman}}, \bibinfo {author} {\bibfnamefont {C.}~\bibnamefont {Campos}},
  \bibinfo {author} {\bibfnamefont {E.}~\bibnamefont {Romero}}, \ and\ \bibinfo
  {author} {\bibfnamefont {A.}~\bibnamefont {Tomas}},\ }\href@noop {} {\emph
  {\bibinfo {title} {{SLEPc} Users Manual}}},\ \bibinfo {type} {Tech. Rep.}\
  \bibinfo {number} {DSIC-II/24/02 - Revision 3.7}\ (\bibinfo  {institution}
  {D. Sistemes Inform\`atics i Computaci\'o, Universitat Polit\`ecnica de
  Val\`encia},\ \bibinfo {year} {2016})\BibitemShut {NoStop}%
\bibitem [{\citenamefont {Balay}\ \emph {et~al.}(1997)\citenamefont {Balay},
  \citenamefont {Gropp}, \citenamefont {McInnes},\ and\ \citenamefont
  {Smith}}]{petsc1}%
  \BibitemOpen
  \bibfield  {author} {\bibinfo {author} {\bibfnamefont {S.}~\bibnamefont
  {Balay}}, \bibinfo {author} {\bibfnamefont {W.~D.}\ \bibnamefont {Gropp}},
  \bibinfo {author} {\bibfnamefont {L.~C.}\ \bibnamefont {McInnes}}, \ and\
  \bibinfo {author} {\bibfnamefont {B.~F.}\ \bibnamefont {Smith}},\ }in\
  \href@noop {} {\emph {\bibinfo {booktitle} {Modern Software Tools in
  Scientific Computing}}},\ \bibinfo {editor} {edited by\ \bibinfo {editor}
  {\bibfnamefont {E.}~\bibnamefont {Arge}}, \bibinfo {editor} {\bibfnamefont
  {A.~M.}\ \bibnamefont {Bruaset}}, \ and\ \bibinfo {editor} {\bibfnamefont
  {H.~P.}\ \bibnamefont {Langtangen}}}\ (\bibinfo  {publisher}
  {Birkh{\"{a}}user Press},\ \bibinfo {year} {1997})\ pp.\ \bibinfo {pages}
  {163--202}\BibitemShut {NoStop}%
\bibitem [{\citenamefont {Schollw\"ock}(2005)}]{Schollwock}%
  \BibitemOpen
  \bibfield  {author} {\bibinfo {author} {\bibfnamefont {U.}~\bibnamefont
  {Schollw\"ock}},\ }\href {\doibase 10.1103/RevModPhys.77.259} {\bibfield
  {journal} {\bibinfo  {journal} {Rev. Mod. Phys.}\ }\textbf {\bibinfo {volume}
  {77}},\ \bibinfo {pages} {259} (\bibinfo {year} {2005})}\BibitemShut
  {NoStop}%
\bibitem [{\citenamefont {Or\'us}\ and\ \citenamefont {Vidal}(2008)}]{Orus}%
  \BibitemOpen
  \bibfield  {author} {\bibinfo {author} {\bibfnamefont {R.}~\bibnamefont
  {Or\'us}}\ and\ \bibinfo {author} {\bibfnamefont {G.}~\bibnamefont {Vidal}},\
  }\href {\doibase 10.1103/PhysRevB.78.155117} {\bibfield  {journal} {\bibinfo
  {journal} {Phys. Rev. B}\ }\textbf {\bibinfo {volume} {78}},\ \bibinfo
  {pages} {155117} (\bibinfo {year} {2008})}\BibitemShut {NoStop}%
\bibitem [{fn4()}]{fn4}%
  \BibitemOpen
  \href@noop {} {\bibinfo  {journal} {Although truncation does not directly
  affect $\ell$-sized operators, their dynamics is affected by the truncation
  of larger sized operators via the evolution of the system}\ }\BibitemShut
  {NoStop}%
\bibitem [{fn5()}]{fn5}%
  \BibitemOpen
\bibfield  {journal} {  }\href@noop {} {\bibinfo  {journal} {We define the
  microscopic onsite energy scale as the norm of the local Hamiltonian on each
  bond $|\mkern-1.4mu| J\sigma^z_i \sigma^z_{i+1}+J_x \sigma^x_i \sigma^x_{i+1}
  + h_x (\sigma^x_i+\sigma^x_{i+1})/2 |\mkern-1.4mu|$; this differs by a
  (subextensive) boundary term from $|\mkern-1.4mu| H_{\text{static}}
  |\mkern-1.4mu|/L$}\ }\BibitemShut {NoStop}%
\bibitem [{fn6()}]{fn6}%
  \BibitemOpen
\bibfield  {journal} {  }\href@noop {} {\bibinfo  {journal} {In this setup, we
  also confirmed that DMT gives dynamics consistent with Krylov at small system
  sizes ($L=20$) \cite{SM}. Moreover, our method, near infinite temperature
  ($\epsilon=0$), matches independent calculations of the diffusion
  \cite{Parker_1812}}\ }\BibitemShut {NoStop}%
\bibitem [{\citenamefont {Ljubotina}\ \emph {et~al.}(2017)\citenamefont
  {Ljubotina}, \citenamefont {{\v{Z}}nidari{\v{c}}},\ and\ \citenamefont
  {Prosen}}]{Prosen}%
  \BibitemOpen
\bibfield  {journal} {  }\bibfield  {author} {\bibinfo {author} {\bibfnamefont
  {M.}~\bibnamefont {Ljubotina}}, \bibinfo {author} {\bibfnamefont
  {M.}~\bibnamefont {{\v{Z}}nidari{\v{c}}}}, \ and\ \bibinfo {author}
  {\bibfnamefont {T.}~\bibnamefont {Prosen}},\ }\href {\doibase
  10.1038/ncomms16117} {\bibfield  {journal} {\bibinfo  {journal} {Nat.
  Commun.}\ }\textbf {\bibinfo {volume} {8}},\ \bibinfo {pages} {16117}
  (\bibinfo {year} {2017})}\BibitemShut {NoStop}%
\bibitem [{\citenamefont {Bulchandani}\ \emph {et~al.}(2018)\citenamefont
  {Bulchandani}, \citenamefont {Vasseur}, \citenamefont {Karrasch},\ and\
  \citenamefont {Moore}}]{Moore}%
  \BibitemOpen
  \bibfield  {author} {\bibinfo {author} {\bibfnamefont {V.~B.}\ \bibnamefont
  {Bulchandani}}, \bibinfo {author} {\bibfnamefont {R.}~\bibnamefont
  {Vasseur}}, \bibinfo {author} {\bibfnamefont {C.}~\bibnamefont {Karrasch}}, \
  and\ \bibinfo {author} {\bibfnamefont {J.~E.}\ \bibnamefont {Moore}},\ }\href
  {\doibase 10.1103/PhysRevB.97.045407} {\bibfield  {journal} {\bibinfo
  {journal} {Phys. Rev. B}\ }\textbf {\bibinfo {volume} {97}},\ \bibinfo
  {pages} {045407} (\bibinfo {year} {2018})}\BibitemShut {NoStop}%
\bibitem [{\citenamefont {Gopalakrishnan}\ and\ \citenamefont
  {Vasseur}(2019)}]{Gopalakrishnan}%
  \BibitemOpen
  \bibfield  {author} {\bibinfo {author} {\bibfnamefont {S.}~\bibnamefont
  {Gopalakrishnan}}\ and\ \bibinfo {author} {\bibfnamefont {R.}~\bibnamefont
  {Vasseur}},\ }\href {\doibase 10.1103/PhysRevLett.122.127202} {\bibfield
  {journal} {\bibinfo  {journal} {Phys. Rev. Lett.}\ }\textbf {\bibinfo
  {volume} {122}},\ \bibinfo {pages} {127202} (\bibinfo {year}
  {2019})}\BibitemShut {NoStop}%
\bibitem [{\citenamefont {Castro-Alvaredo}\ \emph {et~al.}(2016)\citenamefont
  {Castro-Alvaredo}, \citenamefont {Doyon},\ and\ \citenamefont
  {Yoshimura}}]{Yoshimura}%
  \BibitemOpen
  \bibfield  {author} {\bibinfo {author} {\bibfnamefont {O.~A.}\ \bibnamefont
  {Castro-Alvaredo}}, \bibinfo {author} {\bibfnamefont {B.}~\bibnamefont
  {Doyon}}, \ and\ \bibinfo {author} {\bibfnamefont {T.}~\bibnamefont
  {Yoshimura}},\ }\href {\doibase 10.1103/PhysRevX.6.041065} {\bibfield
  {journal} {\bibinfo  {journal} {Phys. Rev. X}\ }\textbf {\bibinfo {volume}
  {6}},\ \bibinfo {pages} {041065} (\bibinfo {year} {2016})}\BibitemShut
  {NoStop}%
\bibitem [{\citenamefont {Bertini}\ \emph {et~al.}(2016)\citenamefont
  {Bertini}, \citenamefont {Collura}, \citenamefont {De~Nardis},\ and\
  \citenamefont {Fagotti}}]{Fagotti}%
  \BibitemOpen
  \bibfield  {author} {\bibinfo {author} {\bibfnamefont {B.}~\bibnamefont
  {Bertini}}, \bibinfo {author} {\bibfnamefont {M.}~\bibnamefont {Collura}},
  \bibinfo {author} {\bibfnamefont {J.}~\bibnamefont {De~Nardis}}, \ and\
  \bibinfo {author} {\bibfnamefont {M.}~\bibnamefont {Fagotti}},\ }\href
  {\doibase 10.1103/PhysRevLett.117.207201} {\bibfield  {journal} {\bibinfo
  {journal} {Phys. Rev. Lett.}\ }\textbf {\bibinfo {volume} {117}},\ \bibinfo
  {pages} {207201} (\bibinfo {year} {2016})}\BibitemShut {NoStop}%
\bibitem [{\citenamefont {Misguich}\ \emph {et~al.}(2017)\citenamefont
  {Misguich}, \citenamefont {Mallick},\ and\ \citenamefont
  {Krapivsky}}]{Misguich}%
  \BibitemOpen
  \bibfield  {author} {\bibinfo {author} {\bibfnamefont {G.}~\bibnamefont
  {Misguich}}, \bibinfo {author} {\bibfnamefont {K.}~\bibnamefont {Mallick}}, \
  and\ \bibinfo {author} {\bibfnamefont {P.~L.}\ \bibnamefont {Krapivsky}},\
  }\href {\doibase 10.1103/PhysRevB.96.195151} {\bibfield  {journal} {\bibinfo
  {journal} {Phys. Rev. B}\ }\textbf {\bibinfo {volume} {96}},\ \bibinfo
  {pages} {195151} (\bibinfo {year} {2017})}\BibitemShut {NoStop}%
\bibitem [{\citenamefont {Gobert}\ \emph {et~al.}(2005)\citenamefont {Gobert},
  \citenamefont {Kollath}, \citenamefont {Schollw\"ock},\ and\ \citenamefont
  {Sch\"utz}}]{Gobert}%
  \BibitemOpen
  \bibfield  {author} {\bibinfo {author} {\bibfnamefont {D.}~\bibnamefont
  {Gobert}}, \bibinfo {author} {\bibfnamefont {C.}~\bibnamefont {Kollath}},
  \bibinfo {author} {\bibfnamefont {U.}~\bibnamefont {Schollw\"ock}}, \ and\
  \bibinfo {author} {\bibfnamefont {G.}~\bibnamefont {Sch\"utz}},\ }\href
  {\doibase 10.1103/PhysRevE.71.036102} {\bibfield  {journal} {\bibinfo
  {journal} {Phys. Rev. E}\ }\textbf {\bibinfo {volume} {71}},\ \bibinfo
  {pages} {036102} (\bibinfo {year} {2005})}\BibitemShut {NoStop}%
\bibitem [{fn7()}]{fn7}%
  \BibitemOpen
  \href@noop {} {\bibinfo  {journal} {To be specific, $g(x) = \frac{1}{2}+
  \frac{1}{2}\tanh [(x-L/2)/\xi]$ with $\xi = 5$. We note that our results are
  not sensitive to the particular choice of $g(x)$, as long as it resembles a
  smoothed out step function}\ }\BibitemShut {NoStop}%
\bibitem [{fn9()}]{fn9}%
  \BibitemOpen
\bibfield  {journal} {  }\href@noop {} {\bibinfo  {journal} {For more
  information visit \url{https://dynamite.readthedocs.io/en/latest/}}\
  }\BibitemShut {NoStop}%
\bibitem [{\citenamefont {Parker}\ \emph {et~al.}()\citenamefont {Parker},
  \citenamefont {Cao}, \citenamefont {Scaffidi},\ and\ \citenamefont
  {Altman}}]{Parker_1812}%
  \BibitemOpen
\bibfield  {journal} {  }\bibfield  {author} {\bibinfo {author} {\bibfnamefont
  {D.~E.}\ \bibnamefont {Parker}}, \bibinfo {author} {\bibfnamefont
  {X.}~\bibnamefont {Cao}}, \bibinfo {author} {\bibfnamefont {T.}~\bibnamefont
  {Scaffidi}}, \ and\ \bibinfo {author} {\bibfnamefont {E.}~\bibnamefont
  {Altman}},\ }\href {http://arxiv.org/abs/1812.08657} {\bibinfo  {journal}
  {arXiv:1812.08657}\ }\BibitemShut {NoStop}%
\end{thebibliography}%


\begin{thebibliography}{11}%
\makeatletter
\providecommand \@ifxundefined [1]{%
 \@ifx{#1\undefined}
}%
\providecommand \@ifnum [1]{%
 \ifnum #1\expandafter \@firstoftwo
 \else \expandafter \@secondoftwo
 \fi
}%
\providecommand \@ifx [1]{%
 \ifx #1\expandafter \@firstoftwo
 \else \expandafter \@secondoftwo
 \fi
}%
\providecommand \natexlab [1]{#1}%
\providecommand \enquote  [1]{``#1''}%
\providecommand \bibnamefont  [1]{#1}%
\providecommand \bibfnamefont [1]{#1}%
\providecommand \citenamefont [1]{#1}%
\providecommand \href@noop [0]{\@secondoftwo}%
\providecommand \href [0]{\begingroup \@sanitize@url \@href}%
\providecommand \@href[1]{\@@startlink{#1}\@@href}%
\providecommand \@@href[1]{\endgroup#1\@@endlink}%
\providecommand \@sanitize@url [0]{\catcode `\\12\catcode `\$12\catcode
  `\&12\catcode `\#12\catcode `\^12\catcode `\_12\catcode `\%12\relax}%
\providecommand \@@startlink[1]{}%
\providecommand \@@endlink[0]{}%
\providecommand \url  [0]{\begingroup\@sanitize@url \@url }%
\providecommand \@url [1]{\endgroup\@href {#1}{\urlprefix }}%
\providecommand \urlprefix  [0]{URL }%
\providecommand \Eprint [0]{\href }%
\providecommand \doibase [0]{http://dx.doi.org/}%
\providecommand \selectlanguage [0]{\@gobble}%
\providecommand \bibinfo  [0]{\@secondoftwo}%
\providecommand \bibfield  [0]{\@secondoftwo}%
\providecommand \translation [1]{[#1]}%
\providecommand \BibitemOpen [0]{}%
\providecommand \bibitemStop [0]{}%
\providecommand \bibitemNoStop [0]{.\EOS\space}%
\providecommand \EOS [0]{\spacefactor3000\relax}%
\providecommand \BibitemShut  [1]{\csname bibitem#1\endcsname}%
\let\auto@bib@innerbib\@empty
\bibitem [{\citenamefont {Machado}\ \emph {et~al.}(2019)\citenamefont
  {Machado}, \citenamefont {Kahanamoku-Meyer}, \citenamefont {Else},
  \citenamefont {Nayak},\ and\ \citenamefont
  {Yao}}]{machado_exponentially_2017}%
  \BibitemOpen
  \bibfield  {author} {\bibinfo {author} {\bibfnamefont {F.}~\bibnamefont
  {Machado}}, \bibinfo {author} {\bibfnamefont {G.~D.}\ \bibnamefont
  {Kahanamoku-Meyer}}, \bibinfo {author} {\bibfnamefont {D.~V.}\ \bibnamefont
  {Else}}, \bibinfo {author} {\bibfnamefont {C.}~\bibnamefont {Nayak}}, \ and\
  \bibinfo {author} {\bibfnamefont {N.~Y.}\ \bibnamefont {Yao}},\ }\href
  {\doibase 10.1103/PhysRevResearch.1.033202} {\bibfield  {journal} {\bibinfo
  {journal} {Phys. Rev. Research}\ }\textbf {\bibinfo {volume} {1}},\ \bibinfo
  {pages} {033202} (\bibinfo {year} {2019})}\BibitemShut {NoStop}%
\bibitem [{\citenamefont {Lubkin}\ and\ \citenamefont
  {Lubkin}(1993)}]{lubkin_entropycorrection_1993}%
  \BibitemOpen
  \bibfield  {author} {\bibinfo {author} {\bibfnamefont {E.}~\bibnamefont
  {Lubkin}}\ and\ \bibinfo {author} {\bibfnamefont {T.}~\bibnamefont
  {Lubkin}},\ }\href {https://doi.org/10.1007/BF01215300} {\bibfield  {journal}
  {\bibinfo  {journal} {Int. J. Theor. Phys.}\ }\textbf {\bibinfo {volume}
  {32}},\ \bibinfo {pages} {933} (\bibinfo {year} {1993})}\BibitemShut
  {NoStop}%
\bibitem [{\citenamefont {White}\ \emph {et~al.}(2018)\citenamefont {White},
  \citenamefont {Zaletel}, \citenamefont {Mong},\ and\ \citenamefont
  {Refael}}]{White_2017}%
  \BibitemOpen
  \bibfield  {author} {\bibinfo {author} {\bibfnamefont {C.~D.}\ \bibnamefont
  {White}}, \bibinfo {author} {\bibfnamefont {M.}~\bibnamefont {Zaletel}},
  \bibinfo {author} {\bibfnamefont {R.~S.~K.}\ \bibnamefont {Mong}}, \ and\
  \bibinfo {author} {\bibfnamefont {G.}~\bibnamefont {Refael}},\ }\href
  {\doibase 10.1103/PhysRevB.97.035127} {\bibfield  {journal} {\bibinfo
  {journal} {Phys. Rev. B}\ }\textbf {\bibinfo {volume} {97}},\ \bibinfo
  {pages} {035127} (\bibinfo {year} {2018})}\BibitemShut {NoStop}%
\bibitem [{\citenamefont {Mallayya}\ \emph {et~al.}(2019)\citenamefont
  {Mallayya}, \citenamefont {Rigol},\ and\ \citenamefont
  {De~Roeck}}]{Mallayya}%
  \BibitemOpen
  \bibfield  {author} {\bibinfo {author} {\bibfnamefont {K.}~\bibnamefont
  {Mallayya}}, \bibinfo {author} {\bibfnamefont {M.}~\bibnamefont {Rigol}}, \
  and\ \bibinfo {author} {\bibfnamefont {W.}~\bibnamefont {De~Roeck}},\ }\href
  {\doibase 10.1103/PhysRevX.9.021027} {\bibfield  {journal} {\bibinfo
  {journal} {Phys. Rev. X}\ }\textbf {\bibinfo {volume} {9}},\ \bibinfo {pages}
  {021027} (\bibinfo {year} {2019})}\BibitemShut {NoStop}%
\bibitem [{\citenamefont {Gritsev}\ and\ \citenamefont
  {Polkovnikov}(2017)}]{Gritsev}%
  \BibitemOpen
  \bibfield  {author} {\bibinfo {author} {\bibfnamefont {V.}~\bibnamefont
  {Gritsev}}\ and\ \bibinfo {author} {\bibfnamefont {A.}~\bibnamefont
  {Polkovnikov}},\ }\href {\doibase 10.21468/SciPostPhys.2.3.021} {\bibfield
  {journal} {\bibinfo  {journal} {SciPost Phys.}\ }\textbf {\bibinfo {volume}
  {2}},\ \bibinfo {pages} {021} (\bibinfo {year} {2017})}\BibitemShut {NoStop}%
\bibitem [{\citenamefont {Reimann}(2007)}]{Reimann_2007}%
  \BibitemOpen
  \bibfield  {author} {\bibinfo {author} {\bibfnamefont {P.}~\bibnamefont
  {Reimann}},\ }\href {\doibase 10.1103/PhysRevLett.99.160404} {\bibfield
  {journal} {\bibinfo  {journal} {Phys. Rev. Lett.}\ }\textbf {\bibinfo
  {volume} {99}},\ \bibinfo {pages} {160404} (\bibinfo {year}
  {2007})}\BibitemShut {NoStop}%
\bibitem [{\citenamefont {Kloss}\ \emph {et~al.}(2018)\citenamefont {Kloss},
  \citenamefont {Lev},\ and\ \citenamefont
  {Reichman}}]{kloss_time-dependent_2018}%
  \BibitemOpen
  \bibfield  {author} {\bibinfo {author} {\bibfnamefont {B.}~\bibnamefont
  {Kloss}}, \bibinfo {author} {\bibfnamefont {Y.~B.}\ \bibnamefont {Lev}}, \
  and\ \bibinfo {author} {\bibfnamefont {D.}~\bibnamefont {Reichman}},\ }\href
  {\doibase 10.1103/PhysRevB.97.024307} {\bibfield  {journal} {\bibinfo
  {journal} {Phys. Rev. B}\ }\textbf {\bibinfo {volume} {97}},\ \bibinfo
  {pages} {024307} (\bibinfo {year} {2018})}\BibitemShut {NoStop}%
\bibitem [{\citenamefont {Ljubotina}\ \emph {et~al.}(2017)\citenamefont
  {Ljubotina}, \citenamefont {{\v{Z}}nidari{\v{c}}},\ and\ \citenamefont
  {Prosen}}]{Prosen}%
  \BibitemOpen
  \bibfield  {author} {\bibinfo {author} {\bibfnamefont {M.}~\bibnamefont
  {Ljubotina}}, \bibinfo {author} {\bibfnamefont {M.}~\bibnamefont
  {{\v{Z}}nidari{\v{c}}}}, \ and\ \bibinfo {author} {\bibfnamefont
  {T.}~\bibnamefont {Prosen}},\ }\href {\doibase 10.1038/ncomms16117}
  {\bibfield  {journal} {\bibinfo  {journal} {Nat. Commun.}\ }\textbf {\bibinfo
  {volume} {8}},\ \bibinfo {pages} {16117} (\bibinfo {year}
  {2017})}\BibitemShut {NoStop}%
\bibitem [{\citenamefont {Gopalakrishnan}\ and\ \citenamefont
  {Vasseur}(2019)}]{Gopalakrishnan}%
  \BibitemOpen
  \bibfield  {author} {\bibinfo {author} {\bibfnamefont {S.}~\bibnamefont
  {Gopalakrishnan}}\ and\ \bibinfo {author} {\bibfnamefont {R.}~\bibnamefont
  {Vasseur}},\ }\href {\doibase 10.1103/PhysRevLett.122.127202} {\bibfield
  {journal} {\bibinfo  {journal} {Phys. Rev. Lett.}\ }\textbf {\bibinfo
  {volume} {122}},\ \bibinfo {pages} {127202} (\bibinfo {year}
  {2019})}\BibitemShut {NoStop}%
\bibitem [{\citenamefont {Misguich}\ \emph {et~al.}(2017)\citenamefont
  {Misguich}, \citenamefont {Mallick},\ and\ \citenamefont
  {Krapivsky}}]{Misguich}%
  \BibitemOpen
  \bibfield  {author} {\bibinfo {author} {\bibfnamefont {G.}~\bibnamefont
  {Misguich}}, \bibinfo {author} {\bibfnamefont {K.}~\bibnamefont {Mallick}}, \
  and\ \bibinfo {author} {\bibfnamefont {P.~L.}\ \bibnamefont {Krapivsky}},\
  }\href {\doibase 10.1103/PhysRevB.96.195151} {\bibfield  {journal} {\bibinfo
  {journal} {Phys. Rev. B}\ }\textbf {\bibinfo {volume} {96}},\ \bibinfo
  {pages} {195151} (\bibinfo {year} {2017})}\BibitemShut {NoStop}%
\bibitem [{\citenamefont {Gobert}\ \emph {et~al.}(2005)\citenamefont {Gobert},
  \citenamefont {Kollath}, \citenamefont {Schollw\"ock},\ and\ \citenamefont
  {Sch\"utz}}]{Gobert}%
  \BibitemOpen
  \bibfield  {author} {\bibinfo {author} {\bibfnamefont {D.}~\bibnamefont
  {Gobert}}, \bibinfo {author} {\bibfnamefont {C.}~\bibnamefont {Kollath}},
  \bibinfo {author} {\bibfnamefont {U.}~\bibnamefont {Schollw\"ock}}, \ and\
  \bibinfo {author} {\bibfnamefont {G.}~\bibnamefont {Sch\"utz}},\ }\href
  {\doibase 10.1103/PhysRevE.71.036102} {\bibfield  {journal} {\bibinfo
  {journal} {Phys. Rev. E}\ }\textbf {\bibinfo {volume} {71}},\ \bibinfo
  {pages} {036102} (\bibinfo {year} {2005})}\BibitemShut {NoStop}%
\end{thebibliography}%


\end{document}


\title{Supplementary Information:\\Emergent hydrodynamics in non-equilibrium quantum systems}
\author{Bingtian Ye}
\affiliation{Department of Physics, University of California, Berkeley, CA 94720, USA}
\author{Francisco Machado}
\affiliation{Department of Physics, University of California, Berkeley, CA 94720, USA}
\author{Christopher David White}
\affiliation{Institute for Quantum Information and Matter, Caltech, Pasadena, CA 91125, USA}
\author{Roger S. K. Mong}
\affiliation{Department of Physics and Astronomy, University of Pittsburgh, Pittsburgh, PA 15260, USA}
\affiliation{Pittsburgh Quantum Institute, Pittsburgh, PA 15260, USA}
\author{Norman Y. Yao}
\affiliation{Department of Physics, University of California, Berkeley, CA 94720, USA}
\affiliation{Materials Science Division, Lawrence Berkeley National Laboratory, Berkeley, CA 94720, USA}

\maketitle
\section{Benchmarking dynamics with exact diagnolization}
In this section, we benchmark both DMT and Krylov subspace methods against exact diagonalization (ED). 
While the former serves as a direct test of DMT's accuracy at small system sizes, the latter ensures that we can use Krylov subspace method as a reliable tool to further benchmark DMT at larger system sizes, which is shown in the main text. 
We hasten to emphasize that a benchmark at larger system sizes ($L \sim 20$) is important. 
In particular, in order to observe and benchmark prethermalization dynamics, one has to carefully tune the numerics to resolve the following energy scales: The frequency of the Floquet drive must be larger than the local energy scale of the Hamiltonian, but smaller than the many-body bandwidth of the system.
In previous numerical studies, this has required going to system sizes of at least $L \sim 20$ \cite{machado_exponentially_2017}.

\subsection{Benchmarking DMT with ED}
\begin{figure}[h!]
  \centering
  \includegraphics[width = 5.0in]{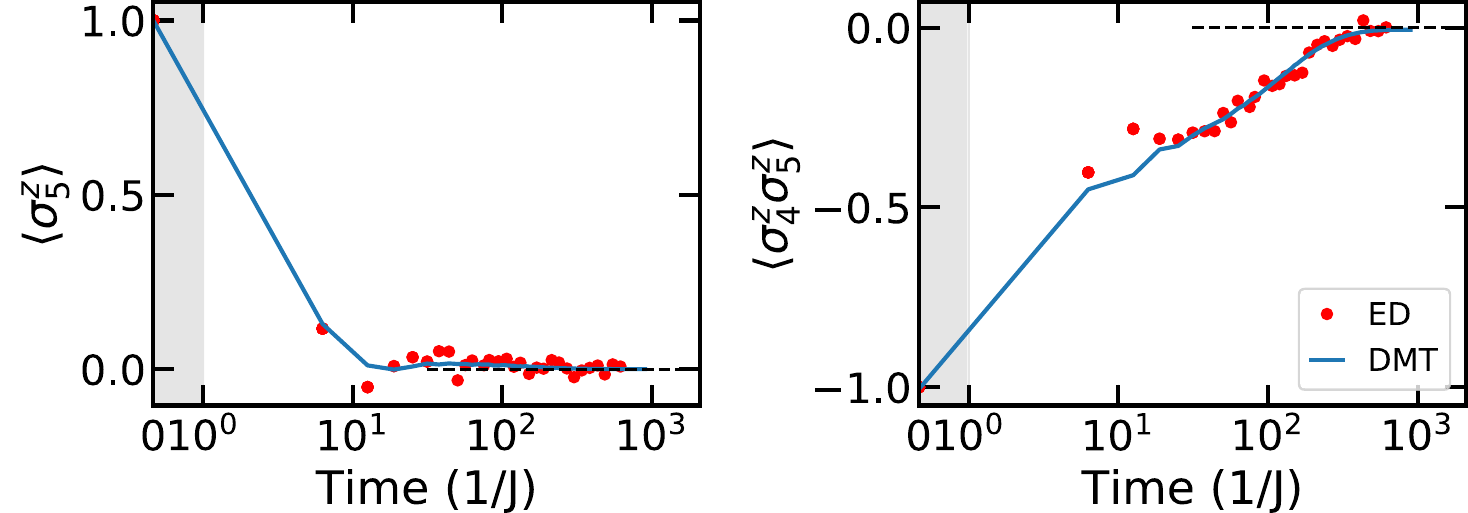}
  \caption{Benchmarking DMT against ED in a Floquet system with small driving frequency $\omega=1$  at system size $L=12$. DMT accurately captures the late-time dynamics of observables although it misses certain early-time fluctuations. 
  Shaded area is plotted linearly in time to emphasize the initial value of each quantity.
  }
  \label{benchmarkED}  
\end{figure}

Here we directly compare the local dynamics generated by DMT and ED for $L=12$ using the same Floquet Hamiltonian from the main text. 
As shown in Fig.~\ref{benchmarkED}, we find that DMT correctly captures the late-time dynamics of observables. 
As discussed in the main text, DMT disagrees with exact dynamics at early times, since it explicitly focuses on conserving observables within only the preservation diameter; indeed, obtaining the exact-early time dynamics would require knowledge of non-local operators.
Moreover, we note that the local observables exhibits smoother dynamics in DMT than in ED, since ED captures fluctuations due to the imperfect self-thermalization at small system sizes. 

\subsection{Benchmarking Krylov subspace method with ED}
By comparing it against exact diagonalization (ED) at small system size, we study the error of the Krylov subspace method.
For each system size $L \in \{4,6,8,10\}$, we consider a random initial product state of spins aligned in the $\hat{z}$ direction.
We then compute the evolution of the system under driving frequencies $\omega/J \in \{5,6,7,8,9,10\} $.

\begin{figure}
  \includegraphics[width = \textwidth]{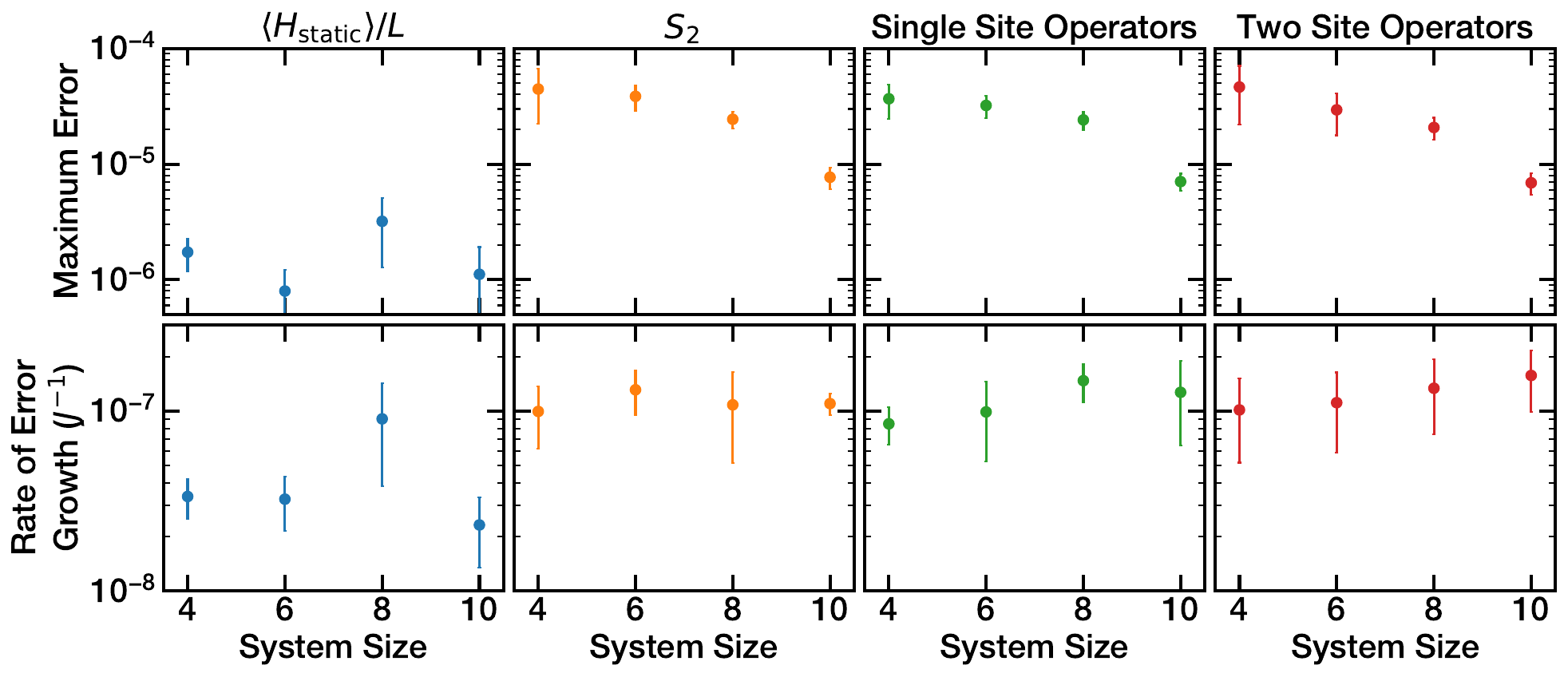} 
  \caption{
    Difference of measured quantities between exact diagonalization and Krylov subspace methods for $t\le 600/J$.
    In particular, we consider energy density $H_{\mathrm{static}}/L$, second R\'{e}niy Entropy $S_2$, single body operators ($\sigma^{\{x,y,z\}}_i$) and two body operators ($\sigma_i^x \sigma^x_{i+1}$ and $\sigma^z_i\sigma^z_{i+1}$), in their respective columns.
    In the top row we consider the largest difference observed within the elapsed time.
    In the lower row, we consider the largest rate of error growth, as defined in Eq.~\eqref{eq:rate_error}.
  }
  \label{fig:Krylov_bench}
\end{figure}

In Fig.~\ref{fig:Krylov_bench}, we show the difference between the two numerical methods for the quantities of interest in our study.
In particular we consider energy density $H_{\mathrm{static}}/L$, second R\'{e}nyi Entropy $S_2$, onsite operators ($\sigma^{\{x,y,z\}}_i$) and two-site operators ($\sigma_i^x \sigma^x_{i+1}$ and $\sigma^z_i\sigma^z_{i+1}$).
In the top row of plots of Fig.~\ref{fig:Krylov_bench}, we consider the maximum error observed in the first $600/J$ time units of the evolution.
Errorbars correspond to the standard deviation of the maximum error observed over 6 different initial states.
We note that the maximum error in this regime is not substantially affected by the system size.

It is perhaps more  enlightening to estimate the rate of error growth: for a given simulation parameters, Krylov subspace dynamics should induce a small constant error per timestep.
This observation is borne out by the data for $t<600/J$.
In the bottom row of Fig.~\ref{fig:Krylov_bench} we show the maximal rate of error growth for several quantities $O$:
\begin{align}\label{eq:rate_error}
  R = \max_{t\le 600/J} \frac{|O_{\mathrm{ED}}(t)-O_{\mathrm{Krylov}}(t)|}{t}~.
\end{align}
This provides an estimate of the error growth as a function of the simulation time.
Since the rate of growth is $\lesssim 2\times 10^{-7}$, we believe that Krylov subspace methods are suitable for benchmarking DMT, even at the times $t \sim 10^{3}$ we reach in this work.

\section{Effect of Trotter step size on DMT numerics}
In the main text, we considered the convergence of DMT with respect to the bond dimension $\chi$ and the size of preserved operators $\ell$. 
Here we complement that analysis by considering the convergence in the size of the Trotter step.
We quantify the error by measuring the average error $\delta_{\langle\hat{O_i}\rangle} \equiv\sqrt{L^{-1}\sum_i\big( \langle\hat{O_i}\rangle_\text{DMT}-\langle\hat{O_i}\rangle_\text{Krylov} \big)^2}$,
i.e., the error of a local observable averaged over all sites.

\begin{figure}[h!]
  \includegraphics[width=3.2in]{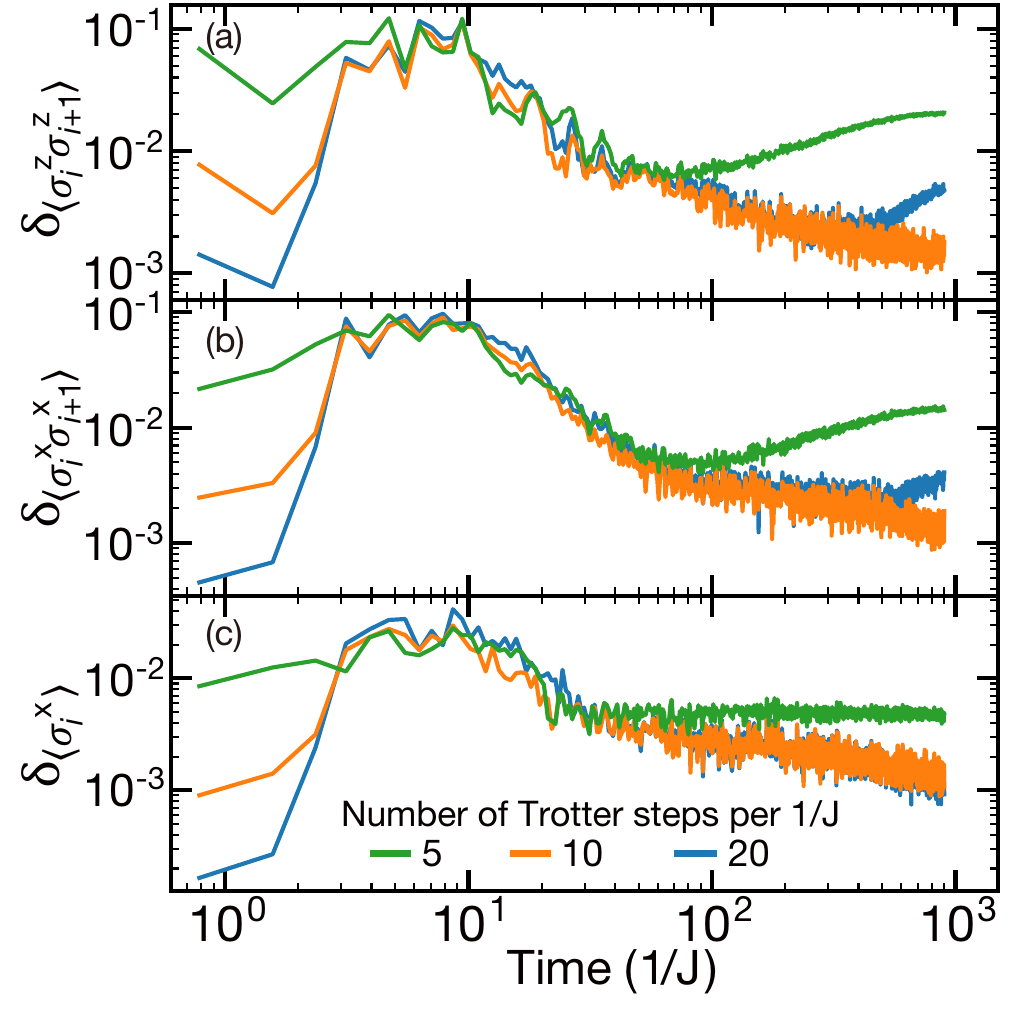}
  \centering
  \caption{Convergence of local observables with respect to Trotter steps ($L=20$, $\omega=8$, $\chi=128$, $\ell=3$).
	(a) $\langle\sigma^z_i\sigma^z_{i+1}\rangle$, (b) $\langle\sigma^x_i\sigma^x_{i+1}\rangle$, and (c) $\langle\sigma^x_i\rangle$ are the three terms that have non-zero contribution to energy density.}
  \label{SuppFigB}
\end{figure}

In Fig.~\ref{SuppFigB}, we take $\hat{O}_i \in \{\sigma^z_i \sigma^z_{i+1},\sigma^x_i \sigma^x_{i+1}, \sigma^x_i\} $, the three local observables that contribute to energy $H_{\mathrm{static}}$.
By decreasing the Trotter step size from $\frac{1}{4J}$ to $\frac{1}{10J}$, we observe an improvement of results, especially during the late-time heating. 
However, the simulation does not benefit from further decreasing the step size from $\frac{1}{10J}$ to $\frac{1}{20J}$.
While the very early-time dynamics shows an improvement, the late-time discrepancy increases; when we apply the truncation too frequently, the long-range correlators are destroyed more severely and the heating process is suppressed.
Since we are interested in the late-time dynamical properties of the system, we use Trotter step $10/J$ throughout this work.

\section{Entropy in DMT}
\label{app:entropy}
In the main text, we show the evolution of the entropy of the leftmost three sites. 
Here we motivate that choice with additional details on the evolution of the entropy of different subsystems. 

\subsection{Page-like correction at late time}
\begin{figure}[h!]
  \includegraphics[width=3.2in]{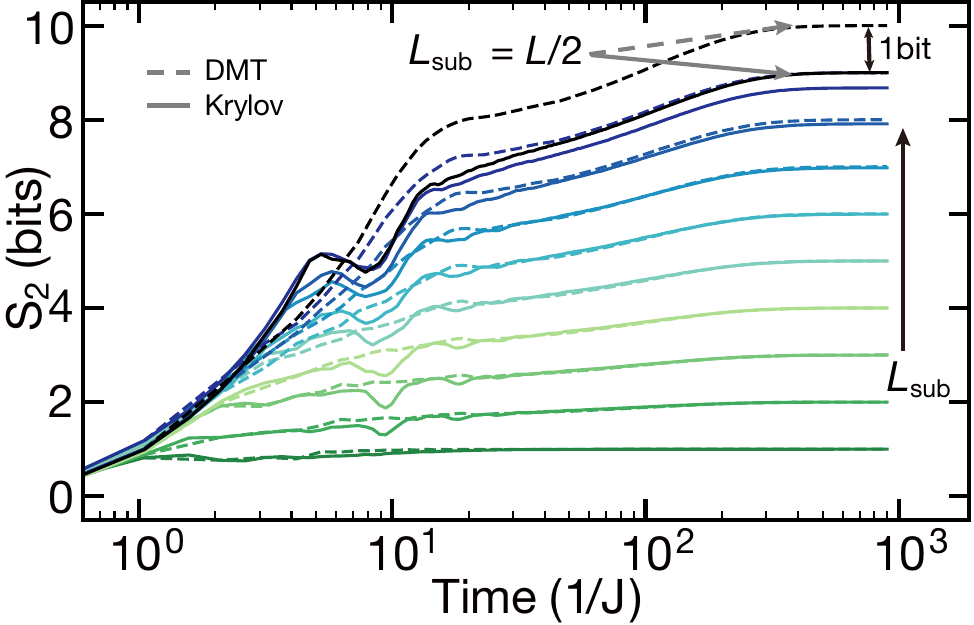}
  \centering
  \caption{Late-time second R\'{e}nyi entropy of subsystems with different sizes ($L=20, \omega=6$).
	The subsystem sizes $L_\text{sub}$ for the curves from the bottom to top are \{1,2,3,4,5,6,7,8,9,10\}. }
  \label{SuppFigCLate}
\end{figure}

For a subsystem with size $L_\text{sub} \sim L/2$, the bipartite entanglement entropy approaches the maximal value of $L_{\text{sub}}$ bits in DMT simulations, but approaches a smaller value in Krylov subspace simulations (Fig.~\ref{SuppFigCLate}). 
This discrepancy arises because the Krylov method considers a pure state, while DMT considers an MPDO, which can be entangled with a notional bath via the truncation procedure.
The difference between the two corresponds to a Page-like correction: at $L_\text{sub}=L/2$ and infinite temperature, this correction is exactly $\log 2$ (1 bit), in agreement with the theoretical prediction~\cite{lubkin_entropycorrection_1993}.
As one decreases $L_{\text{sub}}$, the correction decreases exponentially, and the two methods agree in their late-time entanglement.

\begin{figure}[h!]
  \includegraphics[width=3.2in]{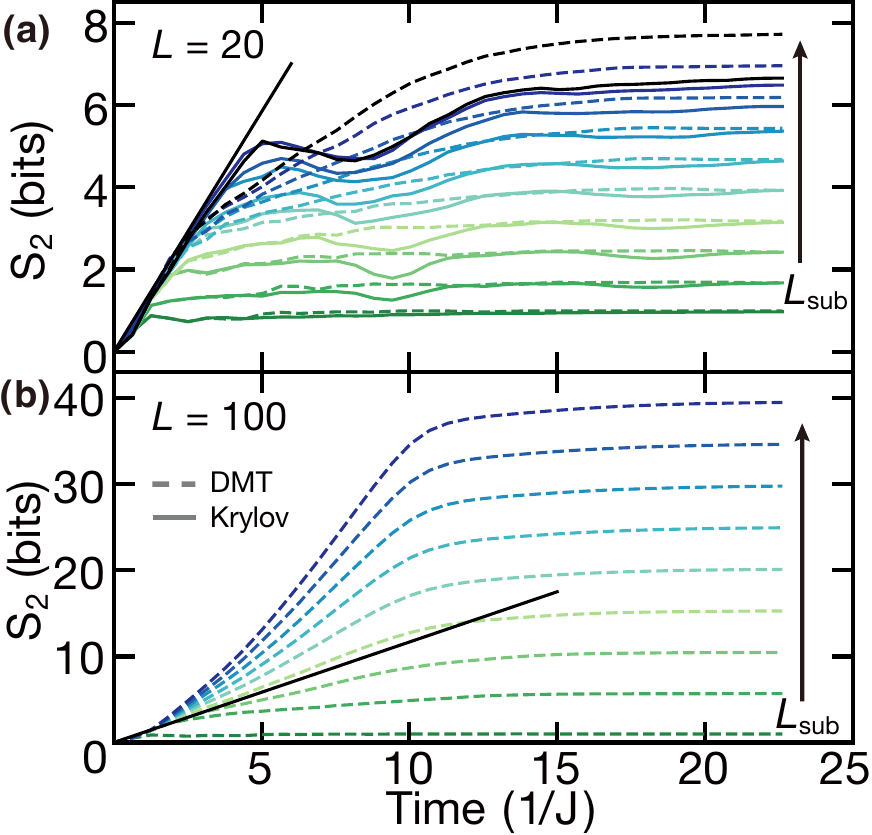}
  \centering
  \caption{Early-time second R\'{e}nyi entropy of subsystems with different sizes ($\omega=10$).
	(a) Small system with $L=20$, and the subsystem sizes $L_\text{sub}$ for the curves from the bottom to top are \{1,2,3,4,5,6,7,8,9,10\}.
	(b) Large system with $L=100$, and the subsystem sizes $L_\text{sub}$ for the curves from the bottom to top are \{1,9,17,25,33,41,49\}.
	The black straight line sets the upper bound of the entropy, of which the rate is extracted from Krylov data with $L=20$. }
  \label{SuppFigCEarly}
\end{figure}
\subsection{Early-time behaviors}
For $t<\tau_{D_\eff}$, the system is well described by the time-independent interacting Hamiltonian $D_{\eff}$.
For an initial product state, the entropy of a subsystem is expected to increase linearly with time.
Using Krylov subspace methods, we indeed observe the linear increase of entropy at early time, and extract the early-time entropy production rate $\Gamma^S_{early}$.
Importantly, $t\cdot\Gamma^S_{early}$ gives an upper bound on the entropy (black line in Fig.~\ref{SuppFigCEarly}).
Nevertheless, $S_2$ can exhibit some non-monotonic behaviors before it approaches the prethermal value at $\tau_{D_\eff}$ (Fig.~\ref{SuppFigCEarly}a), indicating some resonance-like effects in the system. 
In particular, the dips emerging in $S_2$ (marked by arrows in Fig.~2c in the main text) reflect the coherent revival of local operators. 

However, DMT fails to capture both the effect of the many-body coherences, as well as the aforementioned bound in entropy growth.
On the one hand, the truncation destroys the long-range correlations necessary to capture the coherent revivals, resulting in a smoother entropy curve.
On the other hand, at subsystem sizes greater than the preservation diameter $\ell$, the entropy in DMT can exceed the upper bound $t\cdot\Gamma^S_{early}$ (Fig.~\ref{SuppFigCEarly}b).
This is because the truncation via DMT can convert some entanglement entropy to thermal entropy, which does not care about the subsystem boundary: the entropy of a subsystem can increase with the number of bonds truncated and, thus, in larger subsystems we observe a higher rate of entropy growth, Fig.~\ref{SuppFigCEarly}b.

However, if the subsystem size is at most $\ell$, the truncation will preserve the reduced density matrix. 
Thus, the truncation will not affect the $\ell$-site entropy.
Nevertheless, errors in $\ell$-site entropy can still occur via the propagation of errors in longer range operators.
DMT fails to accurately capture the $\ell$-site entropy to the extent that errors in longer ranged operators propagate down to the three-site density operators via the system's dynamics.

\section{Approach to Gibbs Ensemble}

\begin{figure}[h!]
  \includegraphics[width=3.2in]{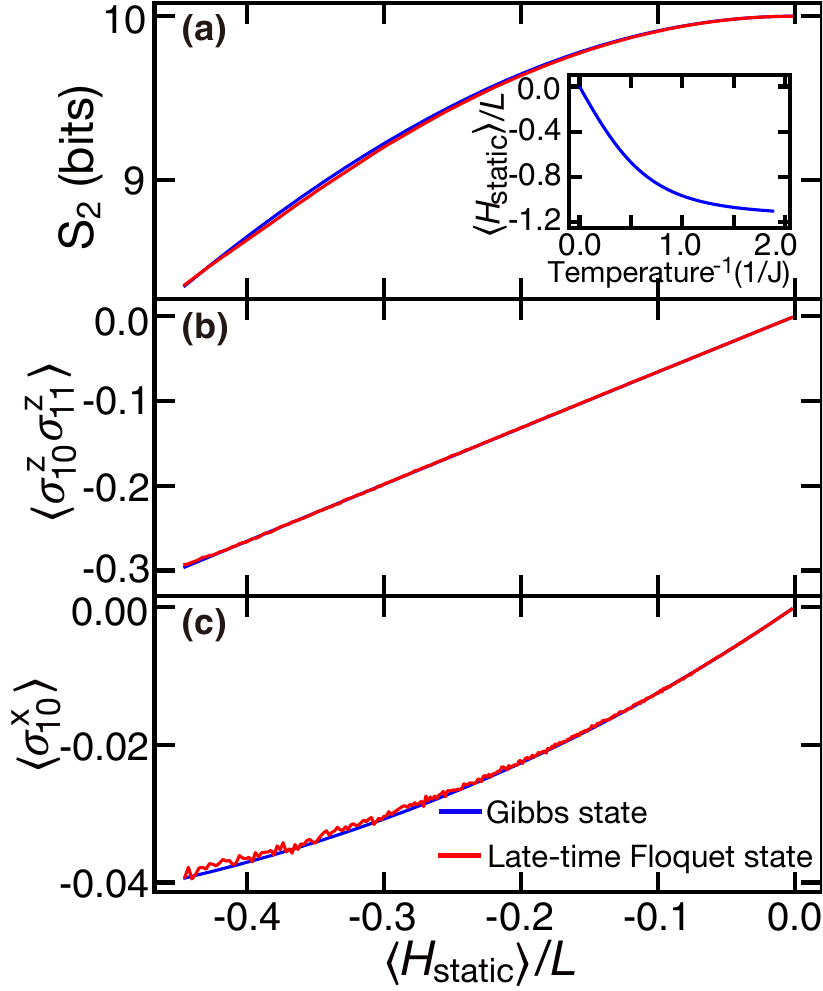}
  \centering
  \caption{Physical quantities as a function of the average energy density for imaginary-time evolved Gibbs states and late-time Floquet evolution ($L=20,$ $\omega=6$). (a) second R\'{e}nyi entropy of the half chain, (b) a two-site local observable, and (c) a one-site local observable. Inset: The average energy density as a function of inverse temperature. }
  \label{SuppFigD}
\end{figure}

To show that the system approaches a Gibbs ensemble at late time, we compare the expectation value of local observables of late-time Floquet evolution and the Gibbs state of the static Hamiltonian $H_\static$ (such Hamiltonian being the zeroth order approximation to $D_\eff$). 
By performing imaginary-time evolution with DMT, we obtain the Gibbs states of the  $H_{\text{static}}/L$ at different temperatures. 
The inset of Fig.~\ref{SuppFigD} shows the average energy density $\langle H_{\text{static}}\rangle$ at different temperatures; similarly we can obtain other physical quantities as a function of temperature. 
This provides a map between energy density and the expectation value of other observables in the Gibbs state.
Using this map, we can directly compare the late-time Floquet evolution (where we lack a direct measure of the temperature) to the Gibbs state \emph{at the same averaged energy density}.  
As shown in Fig.~\ref{SuppFigD}, given the same energy density, the two states exhibit the same entropy and local observables, indicating that the Floquet system can be described by a Gibbs ensemble with respect to the prethermal Hamitonian (for local observables). 

How can these results be consistent with our claim in the main text that the relationship between effective temperature and (e.g.) the energy density of $H_\static$ is itself frequency dependent?
Essentially, consistency requires that the particular local observables of Fig.~\ref{SuppFigD} depend weakly on frequency.
We find that they depend on frequency only at second order or higher in $\omega^{-1}$, because those observables and $H_\static$ are even under global spin flip, while our drive is odd.

To see how these symmetry considerations play out, 
suppose the system is in a Gibbs state
\begin{equation}
  \rho(t; \omega) = \frac 1 Z e^{-\beta(t) D_\eff(\omega^{-1})}
\end{equation}
where the effective temperature $ T(t) = 1/\beta(t)$ is determined by the energy density as
\begin{equation}
  \expct{D_\eff(t)} \equiv Z^{-1} \tr [ D_\eff (\omega^{-1})~ e^{-\beta(t) D_\eff(\omega^{-1})} ]\;
\end{equation}
with $Z = \tr [ e^{- \beta(t) D_\eff(\omega^{-1})} ]$.
For compactness we drop the time dependence of the temperature.
Expanding $D_\eff$ in powers of $\omega^{-1}$:
\begin{equation}
  D_\eff = H_\static + \omega^{-1} D'\; + O(h^2\omega^{-2}).
\end{equation}
As per Ref.~\cite{machado_exponentially_2017}:
\begin{align}
  \begin{split}
    \omega^{-1} D' &= \frac 1 T \int_0^T dt\, i \int_0^t dt' [H_\drive(t'), H_\static]\\
    &= \frac 1 T \int_0^T dt\, i \int_0^t dt' v(t) \sum_{jk}[h_z \sigma^z_j + h_y \sigma^y_j, J\sigma^z_k \sigma^z_{k+1} + J_x \sigma^x_k \sigma^x_{k+1} + h_x \sigma^x_k] \\
    &= i \frac{\pi\omega^{-1}}{2}  \sum_{jk}[h_z \sigma^z_j + h_y \sigma^y_j, J\sigma^z_k \sigma^z_{k+1} + J_x \sigma^x_k \sigma^x_{k+1} + h_x \sigma^x_k]\\
    &= \pi\omega^{-1} \sum_{k}\left[-h_z J_x (\sigma^y_k\sigma^x_{k+1} + \sigma^x_k \sigma^y_{k+1}) - h_zh_x \sigma^y_k -h_y J (\sigma^x_k\sigma^z_{k+1}  +\sigma^z_k\sigma^x_{k+1})\right.\\
      &\hspace{3.5cm}\left.+ h_yJ_x(\sigma^z_k \sigma^x_{k+1} + \sigma^x_k \sigma^z_{k+1}) + h_yh_x \sigma^z_k\right]~.
    \end{split}
\end{align}
It is immediately apparent that $D'$ is odd under a $\pi$ rotation about the $x$ axis, while $H_\static$ is even---more specifically, if
\begin{equation}
  X = \prod_j \sigma^x_j
\end{equation}
then
\begin{align}
  \begin{split}
    X H_\static X &= H_\static\\
    X D' X &= - D'\;,
  \end{split}
\end{align}
so
\begin{equation}
  0 = \tr [D' H_\static^n]\;.
\end{equation}
Then the partition function is 
\begin{align}
  \begin{split}
    Z &= \tr \left[e^{-\beta (H_\static + \omega^{-1} D')}\right] + O(\omega^{-2})\\
    &\approx \tr \left[e^{-\frac{\beta \omega^{-1}}{2}D'}e^{-\beta H_\static} e^{-\frac{\beta \omega^{-1}}{2}D'} \right] + O(\omega^{-2})\\
    &\approx \tr \left[\left(1 - \frac{\beta \omega^{-1}}{2} D'\right)e^{-\beta H_\static} \left(1 - \frac{\beta \omega^{-1}}{2} D'\right)\right] + O(\omega^{-2})\\
    &\approx \tr \left[e^{-\beta H_\static} (1 - \beta \omega^{-1}D') \right] + O(\omega^{-2})\\
    &= \tr \left[e^{-\beta H_\static}\right] + O(\omega^{-2})\\
    &= Z_0 + O(h^2\omega^{-2})
  \end{split}
\end{align}
with $Z_0 = \tr\left[e^{-\beta H_\static}\right]$ the partition function of the static Hamiltonian.

Consider now some (local) operator $O_j$. Its expectation value in the Gibbs state $\rho$ is
\begin{align}
  \begin{split}
    \tr  O_j \rho &= Z_0^{-1} \tr\left[ e^{-\frac{\beta \omega^{-1}}{2}D'}e^{-\beta H_\static} e^{-\frac{\beta \omega^{-1}}{2}D'} \times O_j \right] + O(\omega^{-2})\\
    &= Z_0^{-1} \tr\left[ \left(1 - \frac{\beta \omega^{-1}}{2}D'\right)e^{-\beta H_\static} \left(1 - \frac{\beta \omega^{-1}}{2}D'\right) \times O_j \right] + O(\omega^{-2})\\
    &= Z_0^{-1} \tr\left[e^{-\beta H_\static} O_j\right] - \frac{\beta \omega^{-1}}{2} Z_0^{-1}\tr\left[e^{-\beta H_\static}\{D',O_j\} \right] + O(\omega^{-2})\;,
  \end{split}
\end{align}
where $\{\cdot, \cdot\}$ corresponds to the anti-commutator.

\begin{figure}
  \centering
  \includegraphics[width=0.5\textwidth]{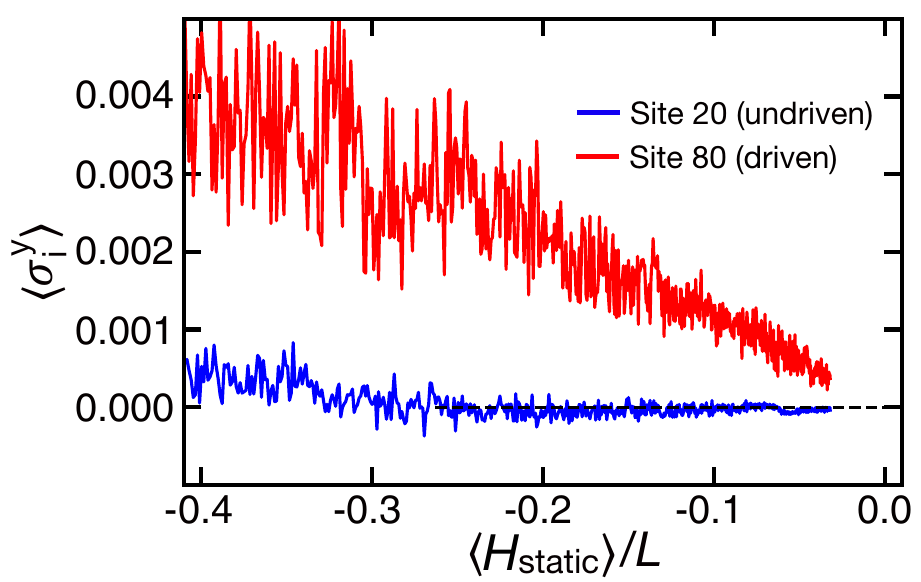}
  \caption{
    Comparison of the expectation value of $\sigma^y$ for different sites on the half-driven chain, at frequency $\omega=6$ and using the parameters of the main text.
    In red, the site is located within the driven side of the chain, where the effective Hamiltonian is modified by the drive.
    In blue, the site is located within the undriven side of the chain, where the effective Hamiltonian is only modified by the drive at very large order.
  }
  \label{fig:sigmay}
\end{figure}

If $O_j$ is even under $X$, as the operators of Fig.~\ref{SuppFigD} are, $\tr\left[e^{-\beta H_\static}D'O_j \right] = 0$ and, to first order in $\omega^{-1}$, $O_j$ takes the same expectation value as in the $H_\static$ Gibbs state:
\begin{equation}
  \tr \left[ \rho~ O_j^{\mathrm{even}}\right] = Z_0^{-1} \tr\left[e^{-\beta H_\static} O_j^{\mathrm{even}}\right] + O(\omega^{-2})\;.
\end{equation}
If $O_j$ is odd under $X$, then
\begin{equation}
  \tr\left[ \rho~ O_j^{\mathrm{odd}}\right]  = - Z_0^{-1} \frac{\beta \omega^{-1}}{2} \tr\left[e^{-\beta H_\static} \{D',O_j^{\mathrm{odd}}\} \right] + O(\omega^{-2})\;.
\end{equation}
Here we note that $Z_0^{-1} \tr\left[e^{-\beta H_\static}\{D',O_j^{\mathrm{odd}}\} \right]$ corresponds to the sum of expectation values of one and two-body operators.
Taking $O_j^{\mathrm{odd}} = \sigma^y_k$, the significant terms in $D'$ are $h_zJ_x(\sigma^x_{k+1}\sigma^y_{k}+\sigma^y_k\sigma^x_{k+1}) + h_zh_x \sigma^y_k$.
Using, $\beta \sim 0.2, h_z =0.13, h_x=0.21, J_x=0.75,\omega = 6$ and the data from Fig.~\ref{SuppFigD}, we estimate $\tr \rho \sigma^y_k \sim \beta \pi \omega^{-1} (2 h_z J_x \langle \sigma^x_k\rangle+h_zh_x) \sim 2.2\times 10^{-3}$.
Indeed, when we consider our half-driven chain and compare $\sigma^y$ for a site in the undriven part ($D_\eff = H_{\mathrm{static}}$) to a site in the driven part ($D_\eff = H_{\mathrm{static}} + \omega^{-1}D' + \mathcal{O}(\omega^{-2})$), we find that in the driven part $\expct{\sigma^y} \sim 3 \times 10^{-3}$ whereas in the undriven region $\expct{\sigma^y} \sim 0$ (Fig.~\ref{fig:sigmay}), in agreement with our estimate.

\section{Preservation of \texorpdfstring{$l$}{l}-site operators}
Here we present the preservation of $\ell$-site operators within DMT for any $\ell$ as a simple generalization of the preservation of $3$-site operators focused in the original DMT paper \cite{White_2017}.
Suppose we are making a truncation at the bond between site $i$ and $i+1$.
Then the reduced density matrix of the whole system can be written in the following form:
\begin{equation}
  \rho = \sum^{\chi-1}_{\alpha=0} \hat{x}_{L\alpha} s_{\alpha} \hat{x}_{R\alpha}
\end{equation}
in which
\begin{equation}
  \begin{split}
    \hat{x}_{L\alpha}&=\sum_{\{\mu\}} [A_1^{\mu_1}\dots A_i^{\mu_i}]_\alpha \hat{\sigma}_1^{\mu_1}\dots \hat{\sigma}_i^{\mu_i}\\
    \hat{x}_{R\alpha}&=\sum_{\{\mu\}} [B_{i+1}^{\mu_{i+1}}\dots B_L^{\mu_L}]_\alpha \hat{\sigma}_{i+1}^{\mu_{i+1}}\dots \hat{\sigma}_L^{\mu_L}.
  \end{split}
\end{equation}
As in the preservation of $3$-site operators, we perform a basis transformation before SVD decomposition and truncation:
\begin{equation}
  \begin{split}
    \hat{y}_{L\beta}&=\sum^{\chi_1}_{\alpha=0}\hat{x}_{L\alpha}Q^*_{L\alpha\beta}\\
    \hat{y}_{R\beta}&=\sum^{\chi_1}_{\alpha=0}Q^*_{R\alpha\beta}\hat{x}_{R\alpha}.
  \end{split}
\end{equation}
However, the tranformations $Q_{L,R}$ in our method are given by
\begin{equation}
  \begin{split}
    Q_{L\alpha\beta}R_{L\beta}^\lambda&= \tr[\hat{x}_{L\alpha}\hat{O}^\lambda_{i+1-n,i}]\in \mathbf{C}^{\chi\times 4^n}\\
    Q_{R\alpha\beta}R_{R\beta}^\lambda&= \tr[\hat{x}_{R\alpha}\hat{O}^\lambda_{i+1,i+n}]\in \mathbf{C}^{\chi\times 4^n}.
  \end{split}
\end{equation}
where $n$ is an integer controlling the size of the preserved operators, and the $\hat{O}^\lambda_{j,k}$ form a basis for operators on the subsystem $[i+1-n, i]$ indexed by $\lambda$.
After the transformation, we follow the same method as the original DMT, which means we do not change the first $4^n$ rows and $4^n$ columns during the truncation.
This procedure can preserve the reduced density matrices of the subsystem $[1,i+n]$ and the subsystem $[i+1-n,L]$ (the proof is almost the same as in \cite{White_2017}).
We remark that to guarantee this requires the bond dimension $\chi\ge\chi^{preserve}=2\times4^n$, where $4^n$ is the number of all possible $\hat{O}^\lambda_{i+1-n,i}$ (or $\hat{O}^\lambda_{i+1,i+n}$), i.e. the number of operators living in the subsystem $[1,i+n]$ (or the subsystem $[i+1-n,L]$).

\begin{figure}[h!]
  \includegraphics[width=4.8in]{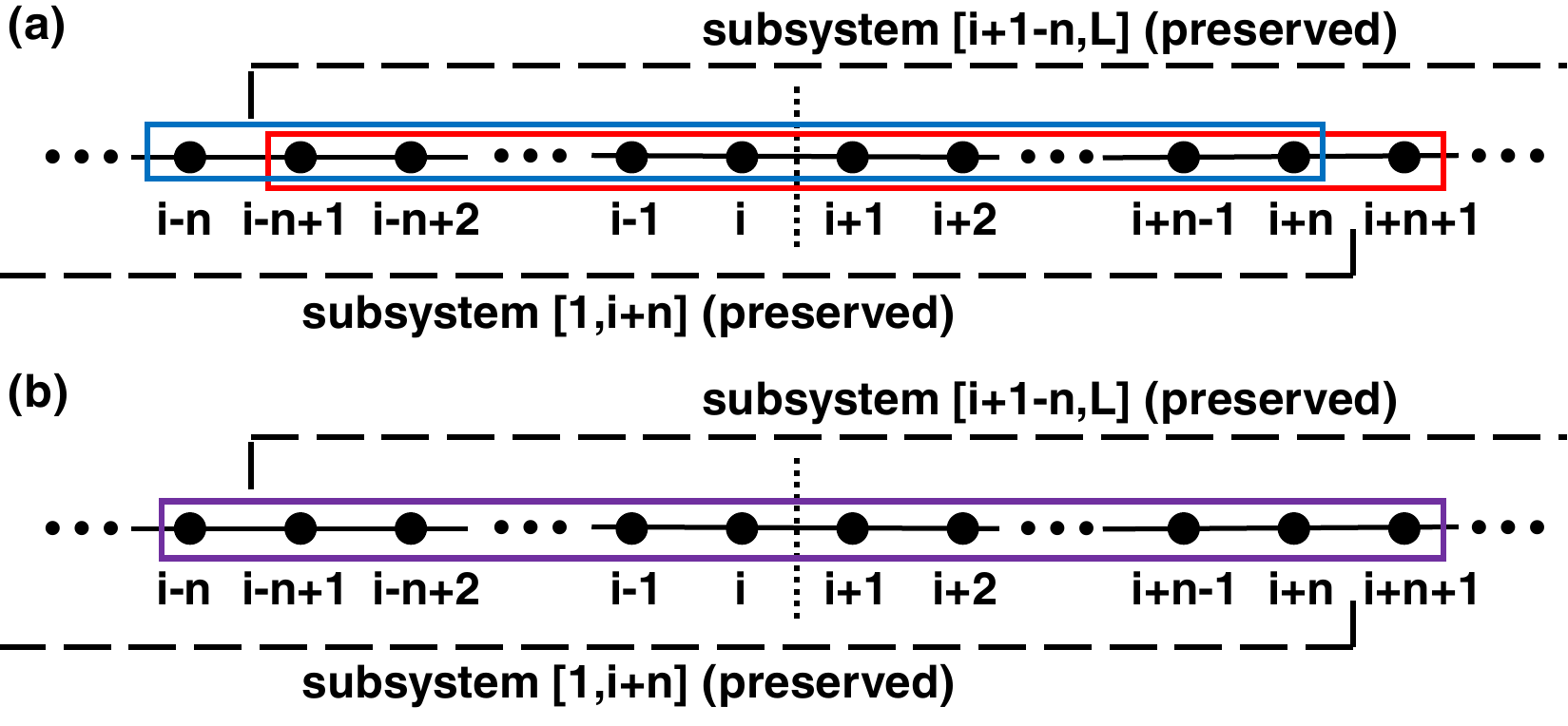}
  \centering
  \caption{The operators preserved during the truncation. (a) $2n+1$ is the maximum size of the operators that can be preserved. (b) A truncation can change the expectation of a $(2n+2)$-site operator.}
  \label{SuppFigE}
\end{figure}
For an operator on $\ell$ consecutive sites, we only need to consider the case when the truncation seperates it  into two parts. Let $\ell_{left}$ and $\ell_{right}$ denote the size of the left and the right parts (relative to the truncation) of this operator respectively.
Since $\ell_{left}+\ell_{right}=\ell$, we have $\mathrm{min}\{\ell_{left},\ell_{right}\}\le \lceil \ell/2\rceil$.
If $\ell_{right}\le n$, the $\ell$-site operator will live in the subsystem $[1,i+n]$, and is thus preserved (the marginal case is the blue frame in Fig.~\ref{SuppFigE}a).
Similarly, if $\ell_{left}\le n$, the $\ell$-site operator will also be preserved (the marginal case is shown by the red frame in Fig.~\ref{SuppFigE}a).
Therefore, any $l$-site operator with $\lceil \ell/2\rceil \le n$ is preserved during a truncation on any bond $i$, which means for a given $n$, we can preserve all $(2n+1)$-site operators.
However, a $(2n+2)$-site operator can be changed by the truncation at the middle of it (Fig.~\ref{SuppFigE}b).
Combining the previous expression for $\chi^{preserve}$ and $\ell=2n+1$, we prove that to preserve all $\ell$-site operators requires bond dimension $\chi^{preserve}=2^\ell$.

\section{Details for Heating Timescales}
\subsection{Comparison of growth rates of energy and entropy}
\label{sec:DiffInTauStar}
Before describing in detail the extraction of $\tau^*$ from the dynamics of $\langle H_{\text{static}}\rangle$ and $S_2$, we would like to highlight the slightly different definition of heating timescales for these two quantities.
To be specific, we define the heating timescales $\tau^*$ as
\begin{equation}
  \begin{split}
    \frac{d}{dt}\langle H_{\text{static}}\rangle &= -\frac{1}{\tau^*_E}\langle H_{\text{static}}\rangle\\
    \frac{d}{dt} \Delta S_2 &= -\frac{2}{\tau^*_S} \Delta S_2
  \end{split},
\end{equation}
where $\Delta S_2=S_2^{T=\infty}-S_2$. 
Here, we justify that with this factor of 2 difference, $\tau^*_E$ and $\tau^*_S$ measure the same heating timescale in the high-temperature regime.

For a Gibbs ensemble of $H$ at temperature $T$, the probability $p_i$ assigned to the $i$'th eigenstate (with $\epsilon_i$ being its eigenenergy) can be approximated to the first order as
\begin{equation}
  p_i =\frac{e^{-\beta\epsilon_i}}{\tr[e^{-\beta H}]}
  \approx  \frac{1-\beta\epsilon_i}{\tr[1-\beta H]}
  = 2^{-L}(1-\beta\epsilon_i),
\end{equation}
where $\beta=1/T$, $2^{L}$ is the dimension of the Hilbert space, and we use the fact that $\tr H = 0$.
In the high-temperature regime, the energy is then written as:
\begin{equation}
  E=\sum_i p_i\epsilon_i\approx 2^{-L}\sum_i (1-\beta\epsilon_i)\epsilon_i
  = -2^{-L}\sum_i\beta \epsilon_i^2 \propto \beta.
\end{equation}
A similar estimate can be made for the second R\'{e}nyi entropy of the entire system:
\begin{gather}
\begin{split}
  S_{\mathrm{entire}} &=-\log_2\sum_i p_i^2\\
  &\approx -\log_2 \left[2^{-2L} \sum_i (1-\beta\epsilon_i)^2\right]\\
  &= L -\beta^2 2^{-L}\sum_i\epsilon^2_i.
\end{split}
\end{gather}
Hence, 
\begin{equation}
  \Delta S_{\mathrm{entire}}\equiv L-S_{\mathrm{entire}} \simeq \beta^2 2^{-L}\sum_i\epsilon^2_i\propto \beta^2. 
\end{equation}
Since the entropy is an extensive quantity for a Gibbs state, one expects this behavior to hold for any subsystem; thus $\Delta S\propto \beta^2$.
Therefore, using our definition for $\tau^*_E$ and $\tau^*_S$:
\begin{gather}
\begin{split}
  \frac{1/\tau^*_E}{1/\tau^*_S}&=\left(\frac{1}{E}\frac{dE}{dt}\right) /\left(\frac{1}{2}\frac{1}{\Delta S}\frac{d\Delta S}{dt}\right)\\
  &=\left(\frac{d\log |E|}{dt}\right)/\left(\frac{1}{2}\frac{d\log \Delta S}{dt}\right)\\
  &=\left(\frac{d\log |E|}{d\log \beta}\right)/\left(\frac 1 2 \frac{d\log \Delta S}{d\log \beta}\right)\\
  &= 1 \,.
\end{split}
\end{gather}

\subsection{Extracting the heating timescale \texorpdfstring{$\tau^*$}{tau*}}
\label{sec:taustarExt}

\begin{figure}[h!]
  \includegraphics[width=3.2in]{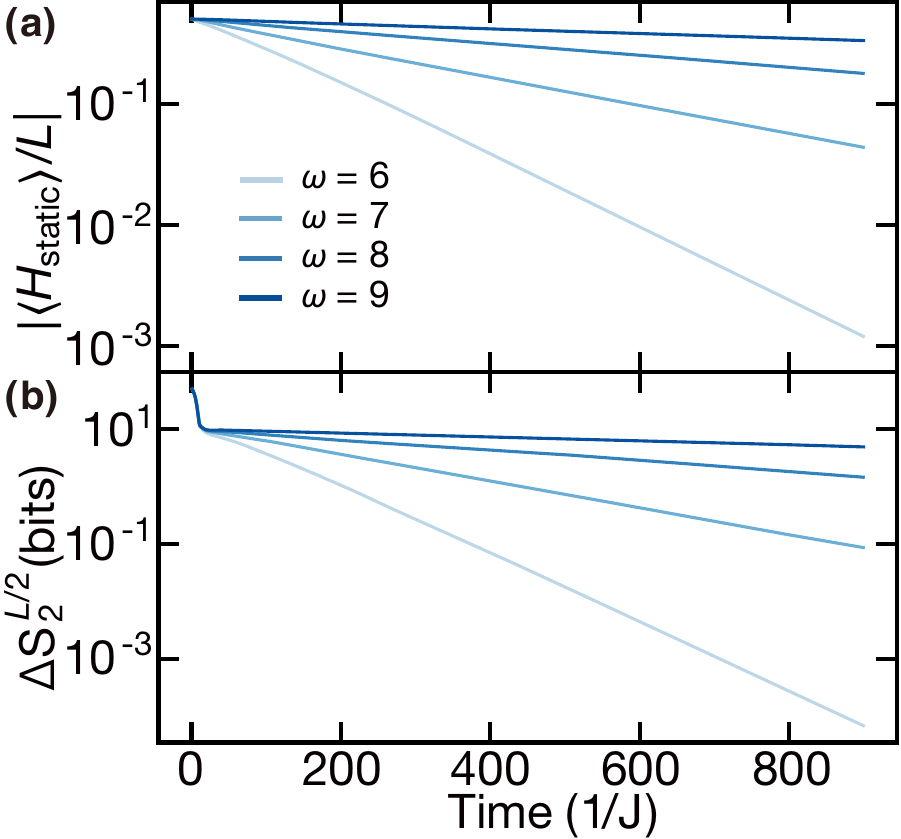}
  \centering
  \caption{The average energy density $\braket{ H_{\text{static}} }/L$ and the entropy $\Delta S_2 = S_2^{T=\infty}-S_2(t)$ decay exponentially zero. }
  \label{SuppFigF}
\end{figure}

We observed that both the average energy density $\langle H_{\text{static}}\rangle$ and the entropy $S_2$ decay exponentially to their infinite-temperature values (Fig.~\ref{SuppFigF}). 
Therefore, $\tau^*_E$ and $\tau^*_S$ can be naturally defined by the equations $|\langle H_{\text{static}}\rangle|\propto e^{-t/\tau^*_E}$ and $(S_2^{T=\infty} - S_2) \propto e^{-2t/\tau^*_S}$ respectively.
The existence of the factor of $2$ ensures that $\tau^*_E$ and $\tau^*_S$ are consistent, as described in the previous section.
We obtain their values by fitting to the inverse of the slope of the logarithm of the two quantities, as plotted in Fig.~\ref{SuppFigF}.
We estimate the errors by fitting to different ranges of time (after prethermalization), and taking the standard deviation of the obtained rates. 

\subsection{Finite-size effects analysis}
\begin{figure}
  \centering
  \includegraphics[width = 5.6in]{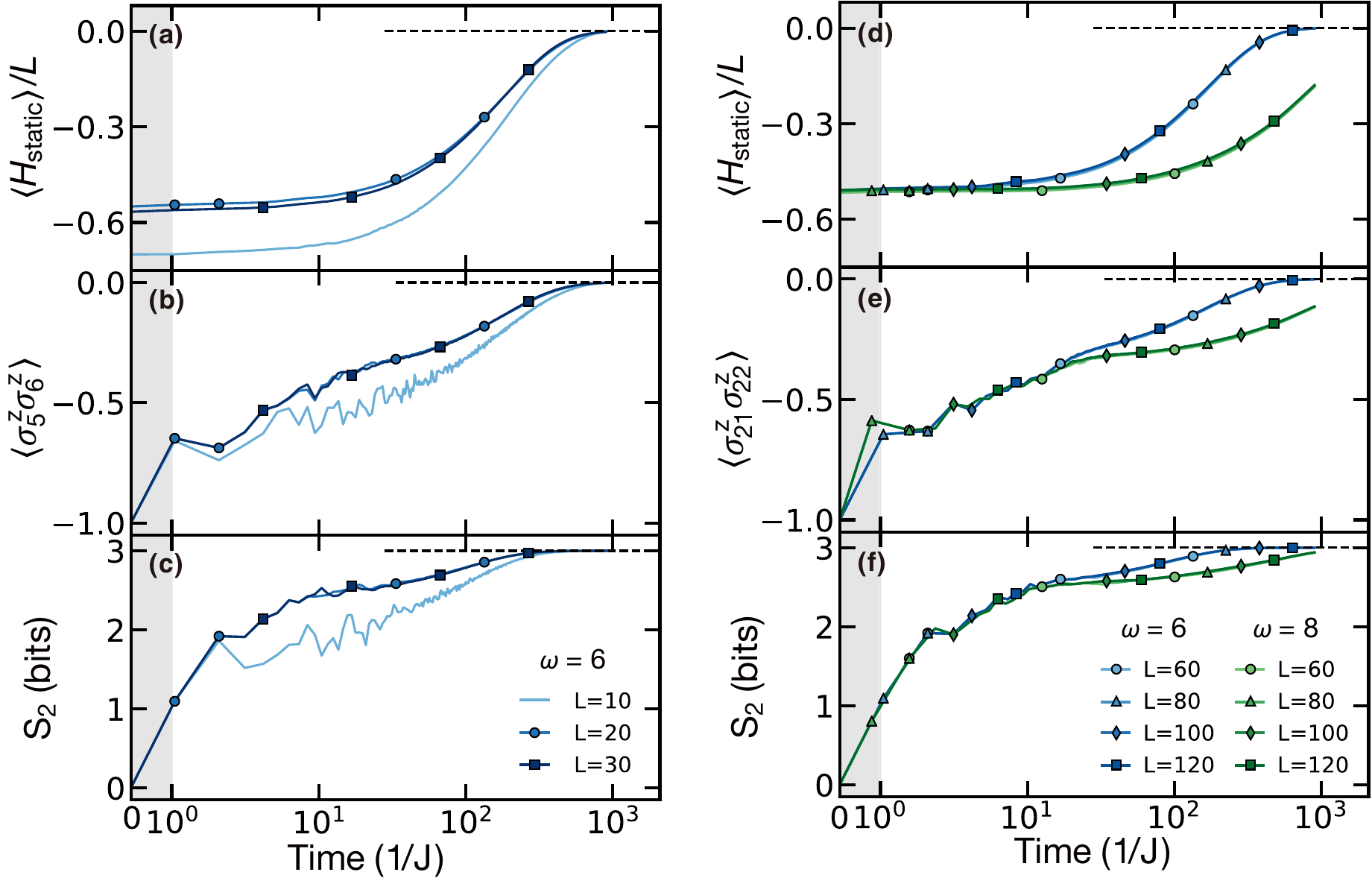}
  \caption{Finite-size scaling analysis of the Floquet dynamics. We simulate the evolution of systems with different sizes. In particular, we study (a)(d) energy density, (b)(e) local two-body operator, (c)(f) second R\'enyi entropy. Left panel ($L=10,20,30$): for very small system sizes, we observed finite-size effects. Right panel ($L=60,80,100,120$): for large system sizes, we observed the collapse of all quantities considered. This ensures that the results in the main text are not affected by finite-size effects.
    Shaded area is plotted linearly in time to emphasize the initial value of each quantity.
  }
  \label{finite-size}  
\end{figure}

\begin{figure}[h!]
  \centering
  \includegraphics[width = 3.2in]{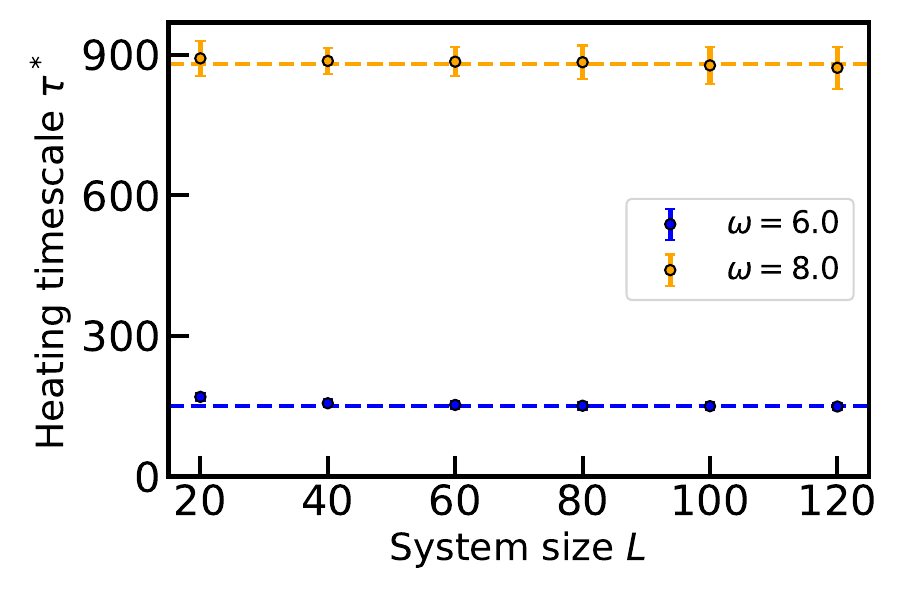}
  \caption{Heating timescale $\tau^*$ for different system sizes as measured via energy density. $\tau^*$'s extracted for different system sizes agree with each other. 
  }
  \label{finite-size1}  
\end{figure}

In this subsection, we numerically confirm that our observation is not limited by finite-size effects. 
In particular, we perform extensive additional numerics on different system sizes ($L=10,20,30,40,60,80,100,120$) as shown in Fig.~\ref{finite-size}. 
Although we observe some finite-size effects for small size $L\lesssim 30$ (Fig.~\ref{finite-size} left panels), above a certain system size, $L=60$, we observe the collapse of all the time traces of different quantities including energy density, entropy and local observables (Fig.~\ref{finite-size} right panels).
  We believe this discrepancy does not arise from the details of the numerical method itself, but rather from the impact a nearby edge and of slightly modifying the energy density.
  Intriguingly, we find that the heating time scale, which we studied extensively in the manuscript, is even more robust.
  In particular, already for $L=20$, it exhibits the correct large $L$ value (Fig.~\ref{finite-size1}). 

\section{Thermalization dynamics of different Floquet models}
\subsection{Integrable models}
Throughout this work, we have considered the dynamics in non-integrable models. 
In this subsection, we comment on how integrability can play a role in the thermalization dynamics of the system. 
Crucially, integrability can emerge in two contexts: 1) the time averaged Hamiltonian $H_\static$ is integrable; 2) the entire effective Hamiltonian $D_\eff$ is integrable. 

\subsubsection{Case 1: $H_\static$ is integrable}
	
	While the Floquet evolution can lead to an integrable $H_\static$, in general, higher order corrections to $D_\eff$ will break the integrability of the effective evolution.
	As a result, the system does not simply relax to a Generalized Gibbs Ensemble (GGE), but rather approaches the thermal Gibbs Ensemble on a time scale $\mathcal{O}(\omega^2)$ \cite{Mallayya}.
	At this point, the system, much like a non-integrable model, will remain in the prethermal Gibbs state for an exponentially long time $\mathcal{O}(e^{\omega/J_\local})$ before absorbing energy from the drive and approaching the late time infinite temperature state. 
	We remark that in this case, owing to the parametric difference between the two time scales, the  deformation away from GGE arises solely from the integrability breaking terms of $D_\eff$ and not from the heating dynamics.

\subsubsection{Case 2: $D_\eff$ is integrable}
  If $D_\eff$ is integrable to all orders, then the entire Floquet dynamics will be integrable (as explored in detail in Ref.~\cite{Gritsev}).
In fact, the only known cases where this occurs are when $D_\eff$ is non-interacting.
  %
  %
  Crucially, for a fully non-interacting Floquet integrable system, one does \emph{not} actually expect heating to occur, which means that such a system will never deform into the infinite-temperature state.
  This is precisely the same reason that one does not need to worry about heating issues for single particle Floquet bandstructures. 
  Thus, a completely integrable Floquet system will, in general, not heat up to the fully mixed state, but instead approach a GGE state at intermediate times and remain in such state for all subsequent times.

\begin{figure}[h]
  \centering
  \includegraphics[width = 2.8in]{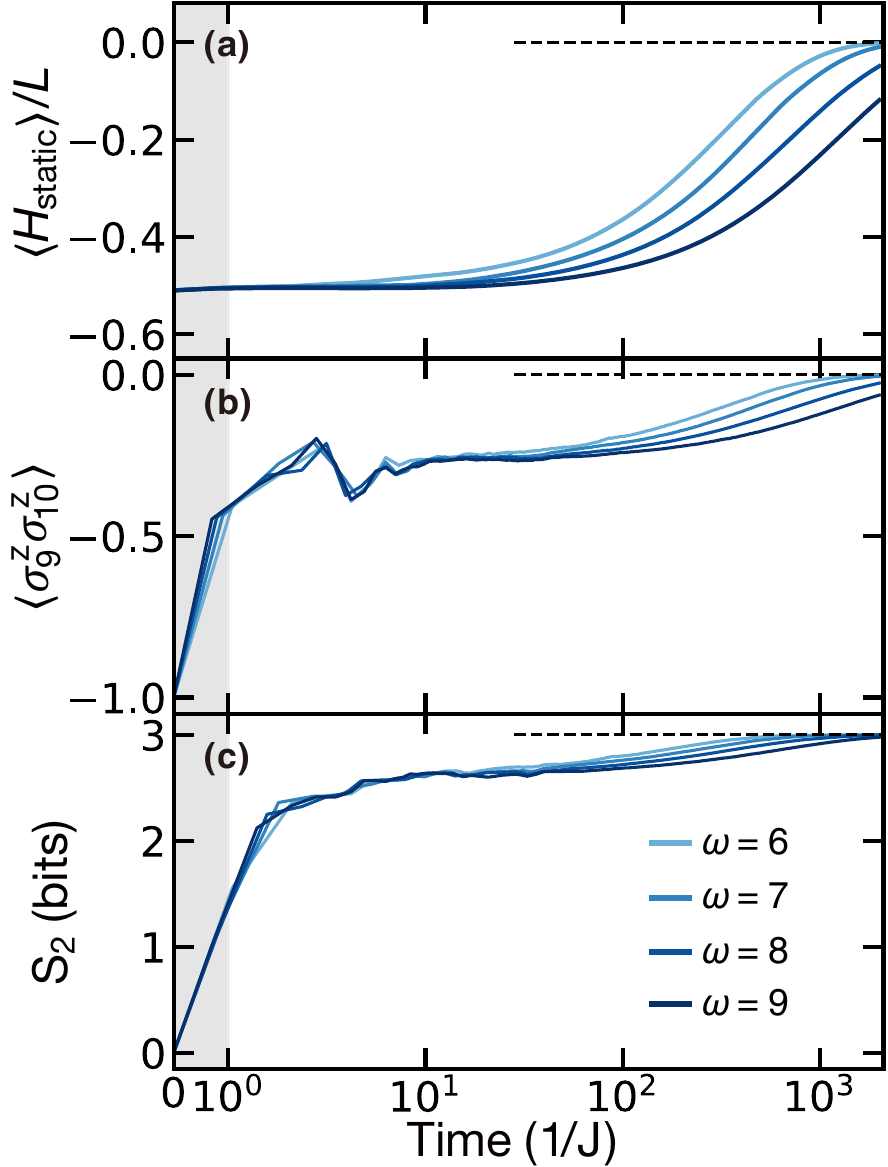}
  \caption{Thermalization dynamics of the driven XXZ model with parameters $\{J_z, J_x, h_x, h_y, h_z\} = \{1, 0.63, 0.29, 0.23, 0.17\}$ and system size $L=100$.
    The observed phenomenology is in excellent agreement with that of the main text.
    In particular, the panels depict  (a) the average energy density, (b) a typical local observable $\sigma^z_{49}\sigma^z_{50}$, and (c) the second R\'enyi entropy.
    Shaded area is plotted linearly in time to emphasize the initial value of each quantity.
  }
  \label{figXXZ}  
\end{figure}

\subsection{Moving away from integrability}
In this subsection, we investigate the dynamics of a different Floquet model via DMT, showing the generality of the observations found in in the main text. 
In particular, we study a driven XXZ spin chain with integrability breaking terms: 
  \begin{equation}
    H_{\text{XXZ,driven}}(t) = \sum_{i} J_x(\sigma^x_i \sigma^x_{i+1} + \sigma^y_i \sigma^y_{i+1}) + J_z \sigma^z_i \sigma^z_{i+1} + h_x \sigma^x_i
    + \begin{cases} 
      \sum_i h_y \sigma^y_i + h_z \sigma^z_i & 0\le t<T/2\\
      -\sum_i h_y \sigma^y_i + h_z \sigma^z_i & T/2\le t<T\\     
    \end{cases}
  \end{equation}
  where we choose the parameters to be $\{J_z, J_x, h_x, h_y, h_z\} = \{1, 0.63, 0.29, 0.23, 0.17\}$. As shown in Fig.~\ref{figXXZ}, we observe the same phenomenology (Floquet heating) as the model presented in the manuscript. 
  In particular, the energy density remains constant until an exponentially long heating timescale, after which it begins to approach its infinite temperature value. 
  For the local observables and the second R\'{e}nyi entropy, they first reflect the system's equilibration process to the prethermal plateaus, and then exhibit an exponentially slow increase to their infinite temperature values. 
  While the heating timescales and the local equilibration process are model-dependent, the dynamics of this Floquet XXZ model is qualitatively the same as the model used in our manuscript.

\section{Classical diffusion equation}
\label{sec:heat_eq}
\subsection{Derivation and approximation}
In general, the equations governing the heat transport in classical systems are:
\begin{align}
  \partial_t \epsilon(x,t) &=  \partial_x j(x,t) + q(x,t) \\
  j(x,t) &\propto \partial_x T(x,t)\;.
\end{align}
The first equation encodes energy conservation, while the second equation reflects that the energy density current $j(x,t)$ arises from an inhomogeneity in temperature $T(x,t)$ across the chain. 

Since we drive only the right half of the chain, the mapping from energy density $\epsilon(x,t)$ to temperature $T(x,t)$, as well as the energy absorption $q(x,t)$, varies explicitly in position along the chain.
In particular, the lowest order $\epsilon(x,t)$ correction to $T(x,t)$ yields a heat current $j(x,t)\propto \partial_x T(x,t) \propto \partial [(1+\eta g_1(x))\epsilon(x,t)]$, in which $g_1(x)$ captures this inhomogeneity in the mapping from $\epsilon$ to $T$, and the small parameter $\eta$ characterizes its magnitude.
Motivated by the heating term in the global drive case, we expect the local heating $q(x,t) = -g_2(x) \epsilon /\tau^*$, where $g_2(x)$ is a spatial function charaterizing where the heating happens. 
Combining all above, the diffusion equation then reads:
\begin{equation}
  \partial_t \epsilon = D \partial_x^2 [(1+\eta g_1(x))\epsilon] - q(x,t) =  D \partial_x^2 [(1+\eta g_1(x))\epsilon] - \frac{\epsilon}{\tau^*} g_2(x). 
\end{equation}
In our model, both $g_1(x)$ and $g_2(x)$ should resemble the step function $\Theta(x-L/2)$. 
In practice, we use the same smoothed-out step function $g(x) = \frac{1}{2}+ \frac{1}{2}\tanh [(x-L/2)/\xi]$ with $\xi=5$ to approximate the two, and our results are not sensitive to the exact form of $g(x)$. 

A few remarks about the diffusion equation are in order. 
First, the diffusion coefficient $D$ depends on the temperature of the system, or equivalently, depends on the energy density $\epsilon$. 
As the system is heated up, $\epsilon$ increases globally, leading to a temporally varying $D$, and we take this effect into account when numerically solving the diffusion equation. 

\begin{figure}[h!]
  \includegraphics[width=3.2in]{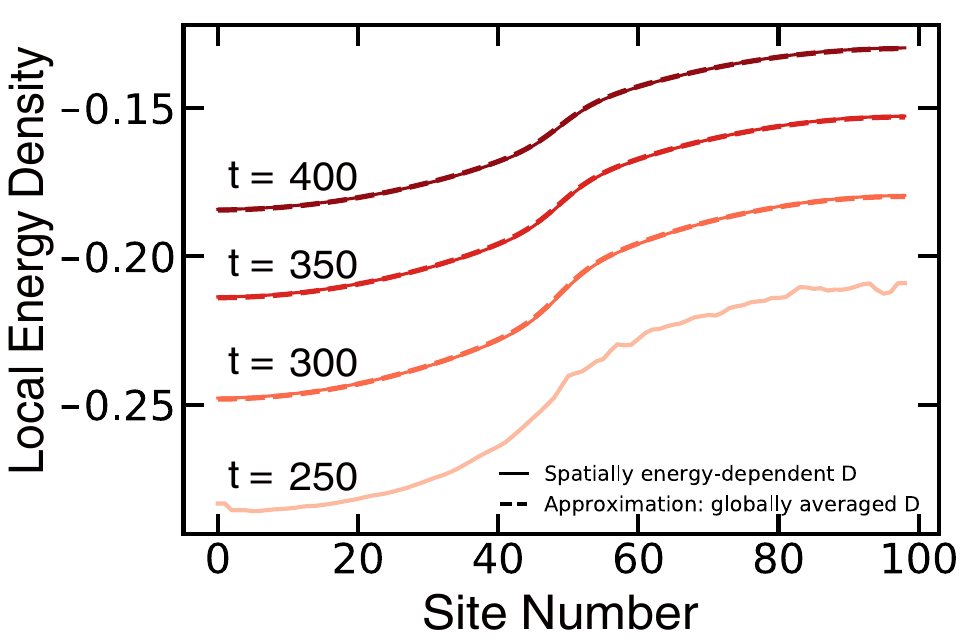}
  \centering
  \caption{Evolution of the energy density solved by diffusion equation, with the diffusion coefficient $D$ varying across the spin chain or approximated by its instantaneous global average. We set the lowest curve as the initial state, and observed no significant difference between the two cases during the following evolution. }
  \label{SuppFigApproxD}
\end{figure}

Second, at any time slice, $\epsilon$ changes across the spin chain, leading to a spatially varying $D$. 
However, this effect is less important, because the spatial inhomogeneity of $\epsilon$ is relatively small, compared to the change of $\epsilon$ over the entire evolution. 
Indeed in Fig.~\ref{SuppFigApproxD}, no significant error is observed when we replace the spatially varying $D$ with a globally averaged value. 
Thus in practice, we only consider the former effect, making it easier to solve the diffusion equation (see the formal solution in the next section). 
Being very clear, in the main text and the following discussion, we treat $D$ as a spatially homogeneous constant, depending on the instantaneous average energy density of the system. 

Third, a spatially varying effective Hamiltonian $D_\eff$ may also lead to another two modifications in the heat equation: a spatially dependent definition of $\epsilon$, and spatially dependent diffusion constant $D$. 
A heat equation including all these modifications can be written as:
\begin{equation}
  \partial_t [(1+\lambda g(x))\epsilon] = \partial_x \{ [D+\delta D g(x)]\partial_x [(1+\eta g(x))\epsilon]\} - \frac{\epsilon}{\tau^*} g(x)
\end{equation}
where $\lambda$, $\delta D$ and $\eta$ are all small numbers.
We find that the dynamics in our experiment is not sensitive to the inclusion of $\lambda$ and $\delta D$, as they do not lead to qualitatively different terms in the equation.
In fact, we find that the $\eta$ correction is the most meaningful.
At early times, before the heating occurs ($\tau_{D_\eff}<t<\tau^*$), we already observe a non-homogenous spatial profile of energy density, due to the temperature inhomogeneity induced by $\eta$.
Moreover, since the spatial profile $g(x)$ is close to the step function $\Theta(x-L/2)$, a higher derivative of it will contribute to larger correction. 
These considerations are both captured by the $\eta$ term. 

\subsection{Solving the Heat Equation}
Since no energy can flow out of the boundary of the system, we require $0 = j(t)|_{x=0,L} \approx\left.\partial_x [(1+\eta g(x)) \epsilon] \right|_{x = 0,L}$.
Considering that $g(x)$ remains constant deeply inside the driven and undriven parts, we can simplify the boundary condition as $\left.\partial_x \epsilon \right|_{x = 0,L} = 0$.
This can be immediately achieved by considering the cosine series of the problem:
\begin{align}
  \epsilon(x,t) =\sum_{n=0}^\infty f_n(t)\cos\frac{n\pi x}{L}
\end{align}

The differential equation then becomes:
\begin{equation}
  \sum_{n=0}^\infty \cos\frac{n\pi x}{L} \partial_t f_n(t) =   -D\sum_{n=0}^\infty \left[\frac{n\pi}{L}\right]^2 \cos\frac{n\pi x}{L} f_n(t) + D\eta\sum_{n=0}^\infty \partial_x^2 \left[ g(x)\cos\frac{n\pi x}{L}\right] f_n(t) - \frac{1}{\tau^*}\sum_{n=0}^\infty g(x)\cos\frac{n\pi x}{L} f_n(t).
\end{equation}

Integrating both sides with the kernel of $\cos \frac{k\pi x}{L}$ for $k\in \mathbb{N}$ yields (the second term on the right-hand side is integrated by parts):
\begin{align}
   \partial_t f_k(t) =  - f_k(t) D\left[\frac{k\pi}{L}\right]^2 -  \frac{2}{L(1+\delta_{k0})}\sum_{n=0}^\infty f_n(t) \left\{ D\eta \left[\frac{k\pi}{L}\right]^2 + \frac{1}{\tau^*}\right\} \int_{0}^{L}dx~g(x) \cos \frac{k \pi x}{L} \cos \frac{n \pi x}{L}  \label{eq:diff_fn}
\end{align}

The resulting equations can be cast in a vectorial form as:
\begin{align}
  \partial_t \vec{f}(t) = M \vec{f}(t)
\end{align}
where $\vec{f}$ is the vector of the Fourier components and $M$ describes the coupling between the modes in the right-hand side of Eq.~\eqref{eq:diff_fn}.
In general, $M$ is time dependent, since it contains $D$.
Hence, the formal solution for $\vec{f}$ becomes:
\begin{equation}\label{eq:solution}
  \vec{f}(t) = \mathcal{T}\exp\left\{ \int_{t_0}^t M(t') dt'\right\} \vec{f}(t_0),
\end{equation}
where $\mathcal{T}$ indicates that the exponential is time-ordered. 

Here, we also remark that although in principle $\vec{f}_n(t)$ is an infinite-dimentional vector, the magnitude of these Fourier modes decays very quickly with $n$.
Therefor, in practice we can consider only the first $n=40$ and not incur significant error.

\subsection{Dynamics of the energy density}

We now describe the procedure by which we can obtain the dynamics of the energy density at late times in the $L=100$ system.

The hydrodynamical description holds only for systems near a local equilibrium.
As such to ensure we have a meaningful initial state, we choose some initial time $t_0$ and Fourier transform the energy density profile at that time. 

\begin{figure}[h!]
  \includegraphics[width = \linewidth]{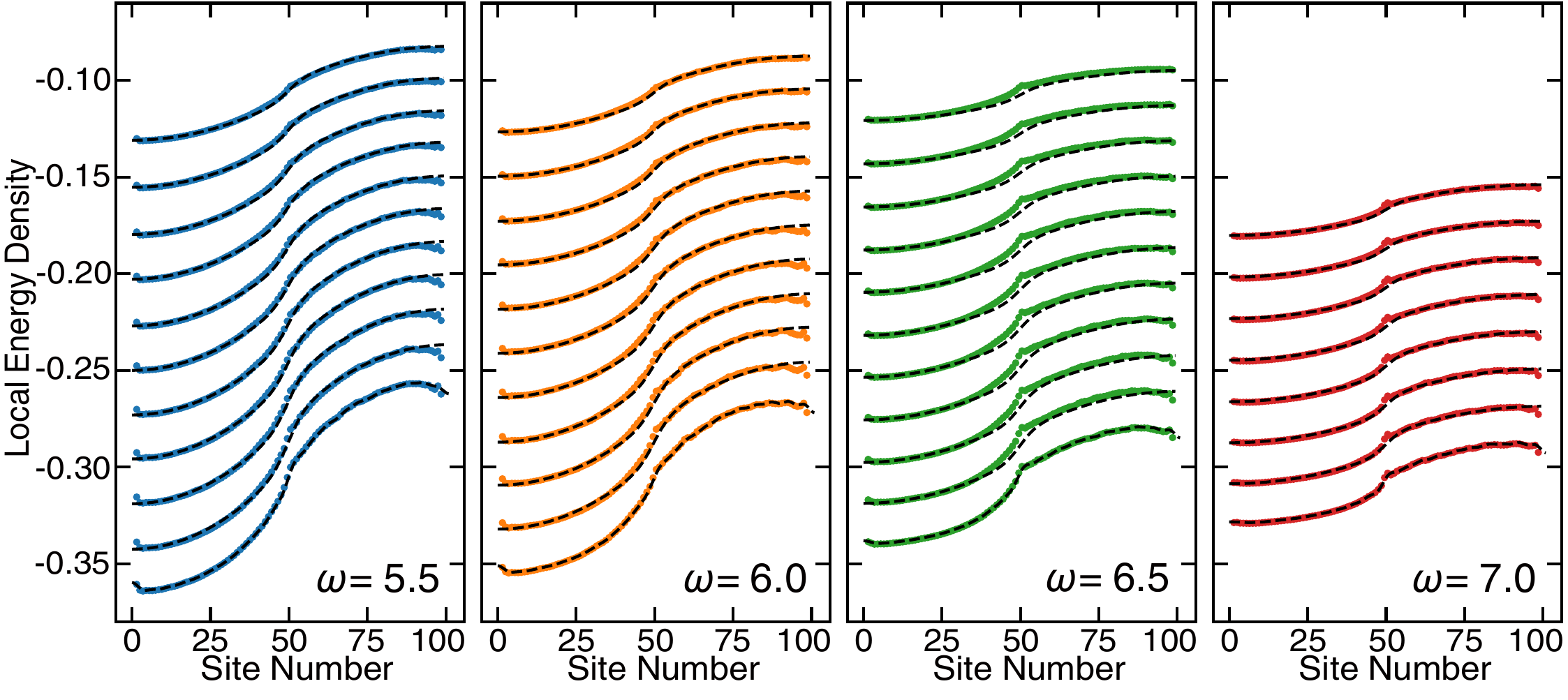}
  \centering
  \caption{
    Time evolution of the energy density profile starting with the lowest energy density state as the initial state.
    We observe great agreement with the DMT results.
    The extracted values of $\eta$ are shown in Fig.~\ref{fig:eta}.
  }
  \label{fig:energydensityevolution}
\end{figure}

We chose a set of times $\{t_n\}$ to compare the hydrodynamical evolution with DMT. 
Here, the heating timescale $\tau^*$ and the diffusion coefficient $D(\epsilon)$ are obtained from our previous analysis of the global heating rate and the diffusion under static Hamiltonian, respectively. 
For the energy-independent parameter $\eta$, we optimize its value to minimize the discrepency between the two evolutions (DMT and hydrodynamical model).
In particular, we charaterize the discrepency by the standard deviation averaged over all time slices, namely, $\sum_{t_n}\sqrt{\sum_x |\epsilon(x,t_n)_\text{DMT}-\epsilon(x,t_n)_\text{Hydro}|^2/L}$. 

\begin{figure}[h!]
  \includegraphics[width = 0.32\linewidth]{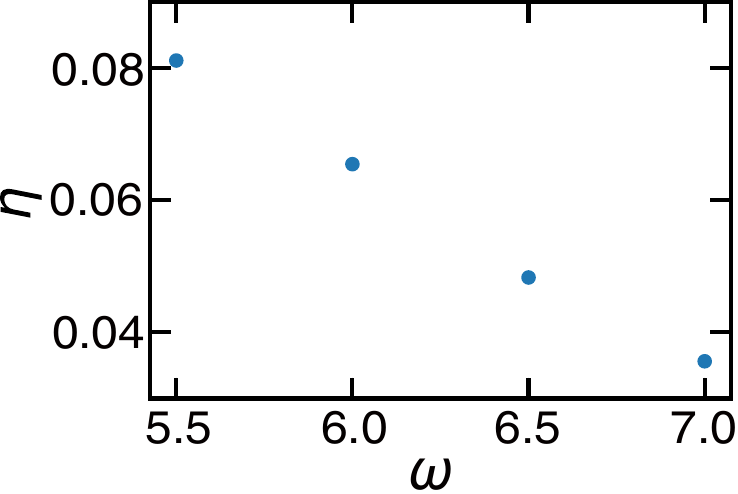}
  \centering
  \caption{Extracted $\eta$ decreases as the driving frequency increases. }
  \label{fig:eta}
\end{figure}

\begin{figure}[h!]
  \includegraphics[width = 0.7\linewidth]{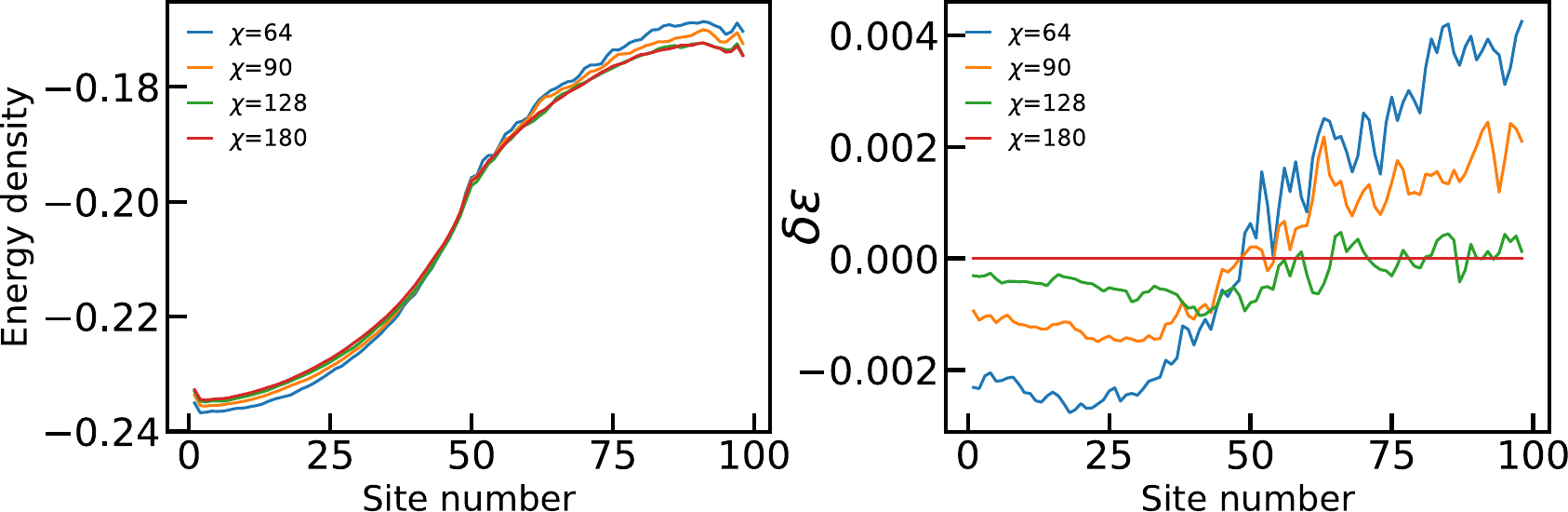}
  \centering
  \caption{Convergence of energy density. (a) Direct comparison between different bond dimension $\chi$. (b) The error of local energy density $\delta\epsilon=\epsilon_{\chi}-\epsilon_{\chi=180}$. We chose $\omega=6$ and averaged energy density $\bar{\epsilon}=-0.2$ as an illustration. For other choice of parameters, we observe similar trends. }
  \label{HydroConvergence1}
\end{figure}

In Fig.~\ref{fig:energydensityevolution}, we compare the energy density profile from the simulation with that arising from Eq.~\eqref{eq:solution}.
Using the lowest energy density state as the initial state, we can then apply the above procedure to obtain the energy density dynamics for later times across a large frequency range.
We also observe that $\eta$ has a negative dependence on frequency $\omega$ as expected (Fig.~\ref{fig:eta}), since its value is determined by the higher order corrections to $D_\eff$ in $\omega^{-1}$, which decreases when the driving frequency increases.
Moreover, by increasing bond dimension $\chi$ in DMT, we check the convergence of the energy density (Fig.~\ref{HydroConvergence1}). 

\section{Extracting diffusion coefficients of a spatially uniform static Hamiltonian}

In this section we discuss the method used to extract the diffusion constant for a spatially uniform Hamiltonian $H$, which is a sum of local terms (with wave length $\frac{2(L-1)}{k}$):
\begin{align}
  H = \sum_{\text{site}~i=0}^{L-1} h_i
\end{align}
We present the numerical experiment performed using DMT and Krylov subspace methods, and how the diffusion constant can be extracted from the obtained data.

\subsection{Numerical Experiment}

To properly probe diffusive behavior, it is imperative that the system is perturbed around an equilibrium (i.e. thermal) state of $H$.
More specifically, we want to initialize the system in a thermal state of $H + \eta H_{\mathrm{perturb}}$, where the form of $H_{\mathrm{perturb}}$ controls the type of perturbation imposed, while $\eta$ controls its strength.

Since we are interested in studying the diffusion of energy when evolving under $H$, we want the perturbations to correspond to the eigenmodes of the diffusion equation, spatial oscilations of the energy density.
We then consider a family of $H_{\mathrm{perturb}}^{[k]}$ which generate the k-th mode:
\begin{align}
  H_{\mathrm{perturb}}^{[k]} = \sum_{\text{site}~i=0}^{L-1} h_i \cos \frac{k i \pi}{L-1}
\end{align}
In DMT, the thermal state can be straighforwardly generated by performing imaginary time evolution on the infinite temperature state $\rho_{T=\infty} \propto \mathds{1}$:
\begin{equation}
  \rho_{\beta} = Z^{-1} \exp\left\{-\beta \left[H + \eta H_{\mathrm{perturb}}\right]\right\}\;.
\end{equation}
In contrast, because Krylov subspace methods can only treat pure states, it is impossible to directly compute expectations of the thermal state.
Nevertheless, expectation values over the thermal density matrix $\rho_\beta$ can be obtained by averaging over initial states, which are then imaginary time evolved:
\begin{equation}
  \tr{O \rho_{\beta}} = \frac{1}{D} \sum_{i=1}^D \left\langle \psi_i \left|\rho_\beta^{1/2} ~ O~ \rho_{\beta}^{1/2}\right| \psi_i\right \rangle\approx \frac{1}{N_{ave}} \sum_{i=1}^{N_{ave}} \left[\left\langle \psi_i \left|\rho_\beta^{1/2}\right] O \left[\rho^{1/2}_\beta \right| \psi_i  \right\rangle\right]~.
\end{equation}
Due to the large size $D$ of the Hilbert space, we cannot perform the entire calculation.
Instead we approximate it by averaging over $N_{ave}$ number of \emph{random} initial states $\ket{\psi_i}$:
\begin{equation}
  \ket{\psi_i} \propto \sum_{i=1}^D c_i \ket{i}, \quad c_i~\text{normal distributed complex variables}
\end{equation}
Due to quantum typicality, such random states behave as infinite temperature states (for local operators) \cite{Reimann_2007}, and so the number of $N_{ave}$ need not be very large (we use $N_{ave} = 50$).

Once the initial state is generated, the system is time evolved with $H$, and the local energy $\langle h_i\rangle$ is calculated as a function of time evolved.
We observed that the initial spatial profile of the local energy quickly decays and the system becomes spatially uniform due to the diffusion of the energy density.

\subsection{Extraction of the diffusion constant}

Consider a system with some conserved quantity $S = \sum_j s_j$ such that $s_j$ are local operators.
Moreover let $\dot{s}_j = i [H,s_j]$ be also local (as is guaranteed in for a local Hamiltonian).
We call $s_j$ local conserved quantities.
In our case, $S = H$ and $s_j=h_j$, the local energy. 

When $H$ is spatially uniform and the system is at equilibrium, $s_j(t)$ will be constant for all sites (up to edge effects).
As a result, we can measure the distance from equilibrium by:
\begin{equation}
  \label{eq:dist}
  \mathcal P(t) = \sqrt{\sum_j (\expct{s_j(t)} - \bar s)^2}
\end{equation}
where $\bar s\equiv S/L $ is independent of time.
The decay of this quantity provides a proxy for the diffusion coefficient:
if the system is diffusive with diffusion coefficient $D$,
then the decay rate of this quantity is given by the decay rate of the slowest non-zero diffusive mode.
For a generic initial state, this corresponds to:
\begin{equation}
  \mathcal P(t) \propto \exp[-\pi^2 t D / (L-1)^2 ]
\end{equation}
for $t \gg  L^2/(4\pi^2 D)$ (the decay rate of the second-slowest mode).
The diffusion coefficient is extracted by fitting this long-time behavior of $\mathcal{P}(t)$.

Alternatively, we can probe these slowest modes directly, by exciting a particular diffusive mode and measuring its magnitude.
This is most straightforwardly implemented by preparing the lowest diffusive mode $k=1$, $s_j(t=0) \approx \cos(j\pi/L) + C$ as the initial state and measuring the amplitude of the corresponding Fourier mode:
\begin{equation}
  s_{q = \pi/L}(t) = \frac{2}{L} \sum_{j=0}^{L-1} \cos\frac{j\pi}{L-1} s_j(t)~.
\end{equation}

In this scenario, the decay of $s_{q=\pi/L}(t)$ will be $\propto e^{-\pi^2 t D /L^2}$, from where $D$ can be extracted.
We note that the profile of $s_j(t)$ can also be fitted, with a least-square method, to the lowest Fourier mode.
Both methods yield the same results.

\begin{figure}[t]
  \includegraphics[width = 1.0\linewidth]{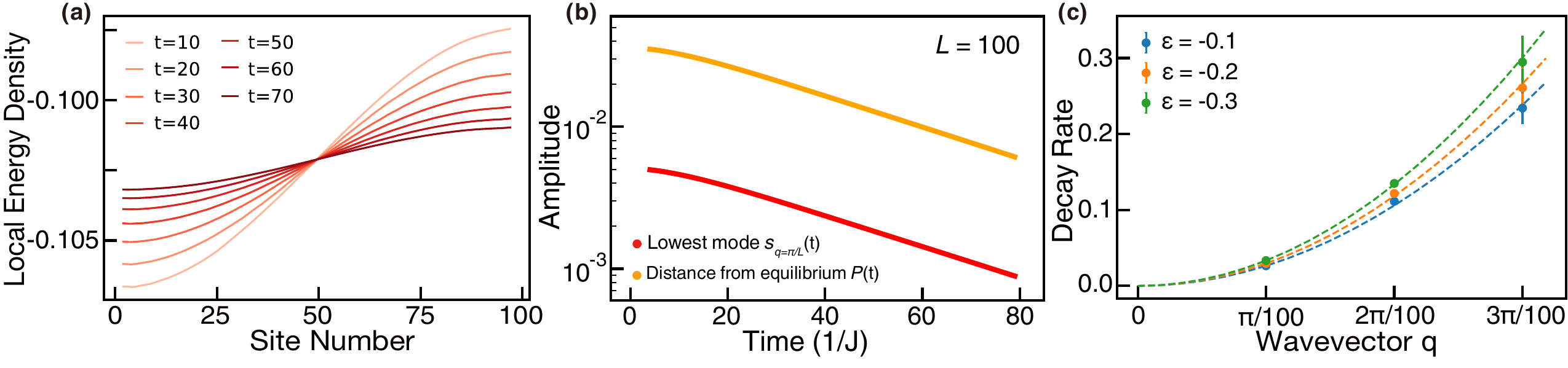}
  \centering
  \caption{(a) The evolution of the first Fourier mode under $H_\static$.
    (b) Decay of Fourier modes at large system size.
    $\beta$ is chosen such that the average energy density is set to be $\bar{\epsilon}=-0.1$. 
    (c) The decay rate of Fourier modes depends quadratically on the wavevector. The system size considered is $L=100$.}
  \label{relaxation}
\end{figure}

Fig.~\ref{relaxation}a,b illustrate both methods, investigating $\mathcal{P}(t)$ and $s_{q=\pi/L}(t)$ for $H = H_\static$.
We see both the slowest mode and the sum over all modes decay exponentially with time at the same decay rate. Here we note that, at early times, we observe non-exponential behavior in the decay of $s_{q=\pi/L}(t)$ until a timescale $\sim 1/h_x$, the integrability breaking term of our system.
It is at this timescale that we expect the interactions to induce the appropriate diffusive behavior.

Moreover, we can study the decay of higher Fourier modes by using the same methodology. 
In Fig.~\ref{relaxation}c, we observe the quadratic dependence of the decay rate on the wavevector, supporting that the dynamics of local energy density is diffusive in our system. 

\subsection{Accuracy of extracted diffusion coefficients}

The work of Kloss, Bar Lev and Reichman~\cite{kloss_time-dependent_2018} and ongoing (unpublished) work of Leviatan~et~al.\ find that TDVP shows ``false convergence'': it can converge very quickly in bond dimension---but to dynamics with an unphysical diffusion coefficient.
We must therefore check that DMT shows the \emph{correct} diffusive dynamics. 
In this section we compare the diffusion constant extracted between DMT and Krylov-space dynamics.

\begin{figure}[t]
  \includegraphics[width = 0.7\linewidth]{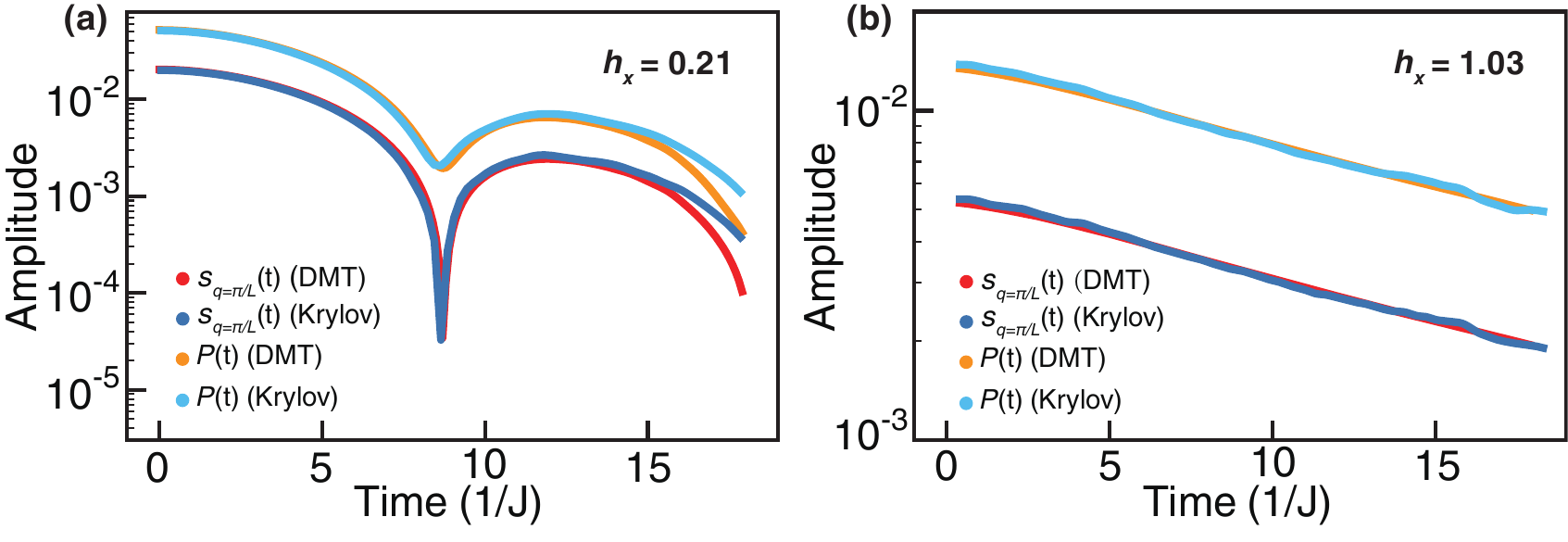}
  \centering
  \caption{Decay of Fourier modes at small system size ($L=20$) for (a) our model $H_\static$ near integrability, with the parameters used in the bulk of the paper, and for (b) our model far from integrability (b). The average energy density is set to be $\bar{\epsilon}=-0.1$. }
  \label{relaxation_benchmark}
\end{figure}

Consider our static Hamiltonian
\begin{equation}
    H_\static = \sum^{L-1}_{i=1}[J\sigma^z_i \sigma^z_{i+1}+J_x \sigma^x_i \sigma^x_{i+1}] + h_x\sum_{i=1}^L\sigma^x_i~.
\end{equation}

In the main text we used $\{J,J_x, h_x\} = \{1, 0.75, 0.21\}$, similar to previous work \cite{machado_exponentially_2017}.
Because we are considering nearest neighbor interactions, the only integrability breaking term is $h_x$, leading to a  na\"ive estimate for the scattering length of $\lambda \sim J/h_x \simeq 5$.
As a result, observing diffusion at small system sizes is difficult.
Fig.~\ref{relaxation_benchmark} (left) highlights this difficulty.
Nevertheless, we observe good agreement between DMT and Krylov in the dynamics.
We note that we expect DMT to artificially increase the dephasing rate for the model's quasiparticles; this explains the gradually increasing discrepancy between the DMT and Krylov simulations.

To check how well DMT can capture diffusion, we increase $h_x = 1.03$, decreasing the scattering lengthscale and making small system sizes more amenable to studies of diffusion.
Indeed, Fig.~\ref{relaxation_benchmark} clearly demonstrates the agreement between the two methods, and as a result, the ability of DMT to probe the diffusive physics.  

\section{Hydrodynamics in integrable models}
In the main text, we focus on late-time hydrodynamics in non-integrable models. 
However, in integrable systems, one may expect qualitatively distinct late-time dynamics, since integrable systems in general relax to generalized Gibbs ensembles (GGE) rather than simple Gibbs ensembles. 
From this perspective, the ability to capture the dynamics of integrable models is another important test for DMT. 
To this end, in this section, we study hydrodynamics of integrable models as a complement to the investigation of non-integrable systems considered in the main text.

In particular, we consider a generic XXZ model, which is known to host different dynamical transport properties in different parameter regimes. 
To be specific, the Hamiltonian of XXZ model is written as: 
\begin{equation} \label{eq:HXXZ_integrable}
  H_{\text{XXZ}}=\sum_i J_x(\sigma^x_i \sigma^x_{i+1}+\sigma^y_i \sigma^y_{i+1})+J_z\sigma^z_i \sigma^z_{i+1}. 
\end{equation}
At high temperature, by tuning $J_z$ to be $<J_x$, $=J_x$ or $>J_x$ (here we set $J_x=1$), the spin transport becomes ballistic, super-diffusive and diffusive, respectively (see Table~\ref{Table1}). 
To demonstrate that DMT can accurately capture these different dynamical regimes, we study two different setups: 1) polarization transfer across a domain wall; 2) decay of Fourier modes. 
Here we hasten to emphasize that while the polarization transfer (Setup 1) can be studied via other numerical methods \cite{Prosen}, as far as we are aware, the decay of Fourier modes (Setup 2) can only be simulated with DMT via our methodology. 
This opens the door to probing emergent hydrodynamics and to extracting quantities like the diffusion coefficient as a function of temperature for this first time. 

\begin{table}[h!]
  \centering
 \begin{tabular}{|c|c|c|c|c|}
    \hline
    Setup & Probe & $J_z<1$ & $J_z=1$ & $J_z>1$ \\
    \colrule
    Domain wall & $P(t)\propto t^\alpha$ & \;$\alpha = 1$\; & \;$\alpha=2/3$~\cite{Prosen}\; &  \;$\alpha=1/2$\; \\
    \;Fourier mode\; & \;$A_k(t) = \tilde{A}(kt^\beta)$\; & $\beta = 1$ & $\beta = 3/2$ & $\beta = 2$ \\
    \hline
  \end{tabular}
    \caption{
    Different types of dynamics for different parameter regimes of $H_{\text{XXZ}}$ at high temperature.
  }
  \label{Table1}
\end{table}
\begin{figure}
  \centering
  \includegraphics[width = 6.5in]{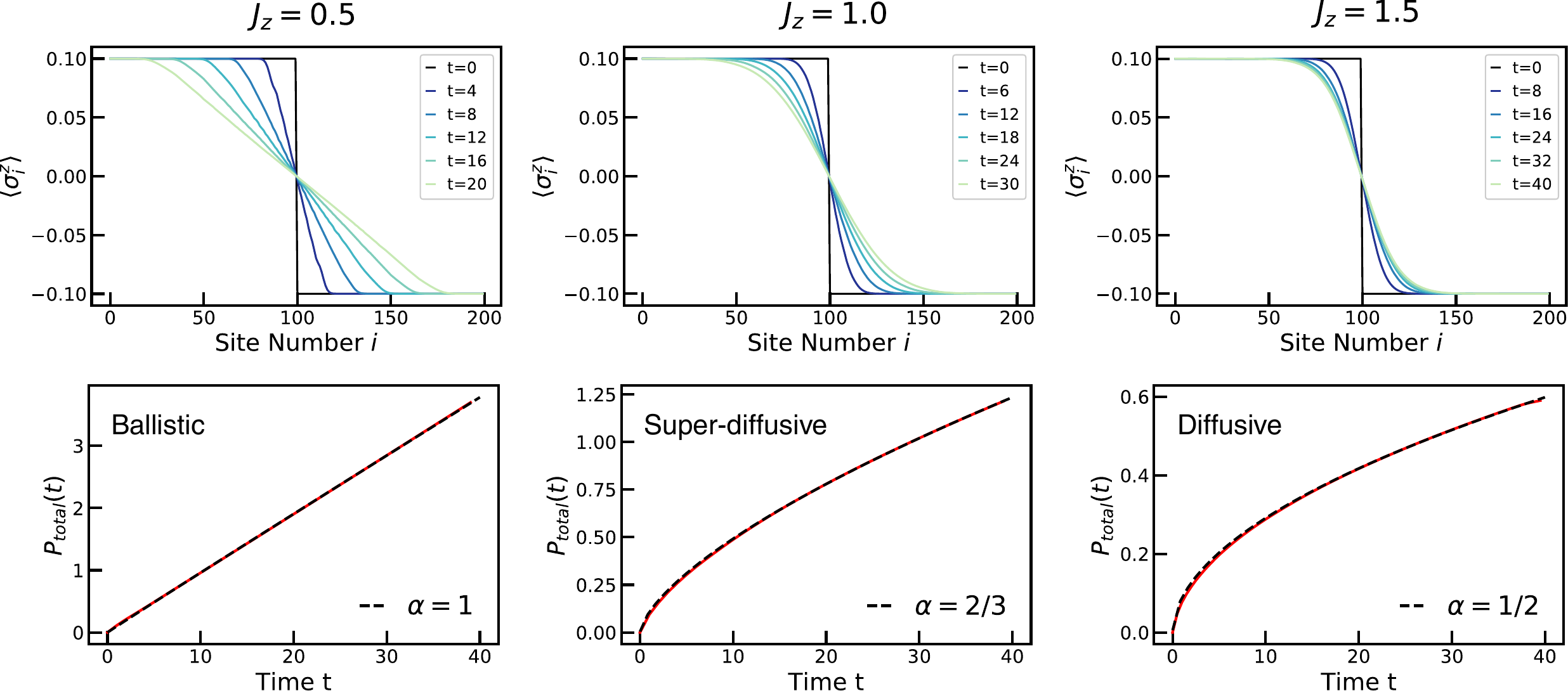}
  \caption{Dynamics of the integrable XXZ model initialized with a domain-wall state at high temperature. Upper panel: spatial profiles of local polarization. Lower panel: the total spin polarization transferred across the domain wall.
  For  different values of $J_z$, $P(t)$ has a power-law dependence on time $t$ with distinct exponents, evincing ballistic, super-diffusive and diffusive transport in agreement with Table~\ref{Table1}.
  }
  \label{XXZ_transport1}  
\end{figure}

\begin{figure}
  \centering
  \includegraphics[width = 6.5in]{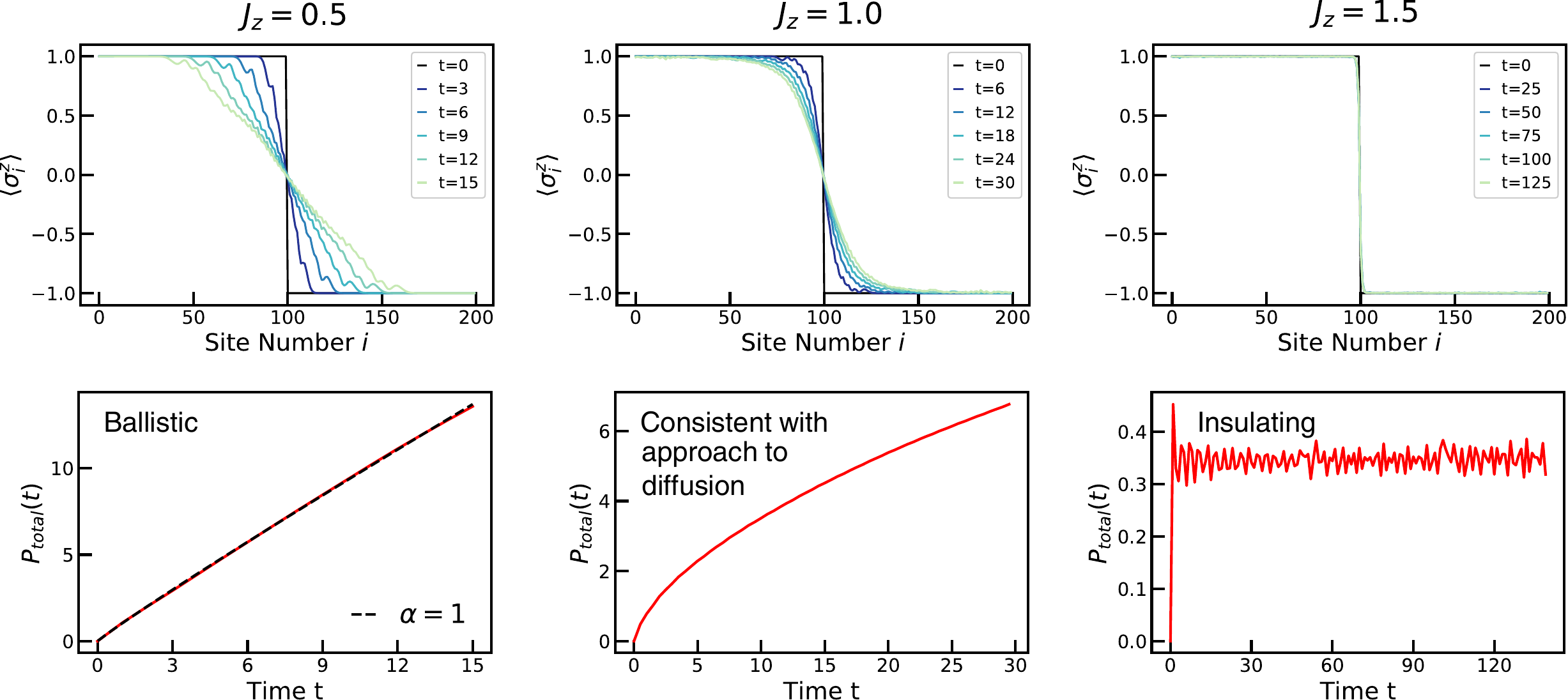}
  \caption{Dynamics of the integrable XXZ model initialized with a pure domain-wall state. Upper panel: spatial profiles of local polarization. Lower panel: the total spin polarization transferred across the domain wall.
  }
  \label{XXZ_transport1_lowT}  
\end{figure}

\subsection{Setup 1: polarization transfer across a domain wall}
 In this setup, we start with a domain-wall state with a single domain wall at the center of the chain:
\begin{equation}
\rho_{ini} = \left( \bigotimes_{i\le L/2}\frac{I+\mu\sigma^z_i}{2}\right)\otimes\left( \bigotimes_{i>L/2}\frac{I-\mu\sigma^z_i}{2}\right), 
\end{equation}
where $\mu$ controls the purity of the state. 
 A conventional measure of spin transport is $P_{total}(t)=\sum_{i\le L/2} (1-\langle \sigma^z_i \rangle)$, i.e.~the total polarization transferred across the center of the chain as a function of time. 
 In general, we expect $P_{total}(t)\propto t^{\alpha}$, where the power-law identifies the type of underlying spin transport ($\alpha=1$, $1/2<\alpha<1$, and $\alpha=1/2$ for ballistic, super-diffusive and diffusive transport, respectively). 
 
 At high temperature ($\eta\ll 1$), we demonstrate that DMT can indeed capture these three distinct forms of emergent late time hydrodynamics (Fig.~\ref{XXZ_transport1}). Our observations are in precise agreement with previous work \cite{Prosen, Gopalakrishnan}.
 %
 This demonstrates DMT's ability to accurately capture the dynamics of integrable systems and to probe emergent hydrodynamics, even when the system relaxes to a GGE state rather than a thermal Gibbs state. 

\begin{figure}
  \centering
  \includegraphics[width = 3.1in]{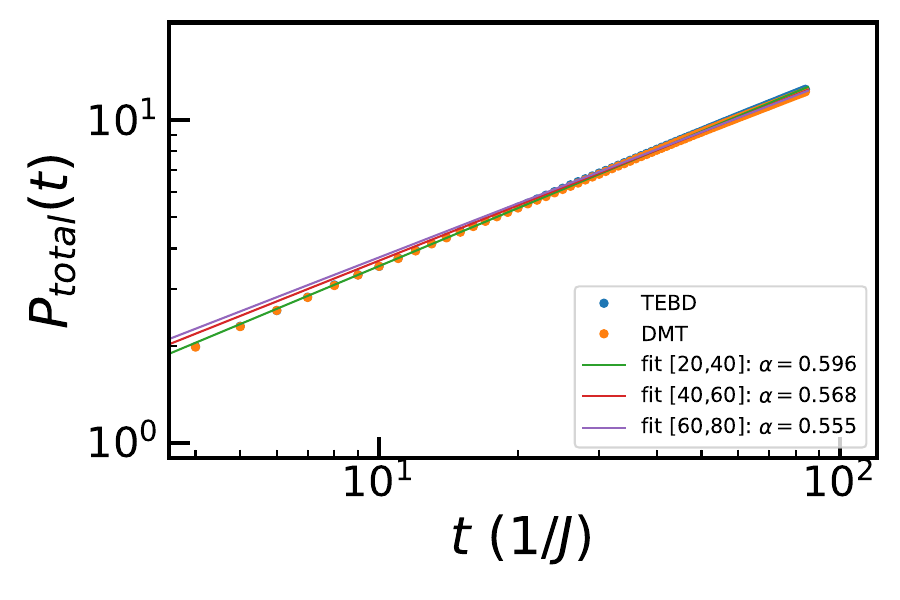}
  \caption{Total polarization transfer $P_{total}(t)$ as a function of time $t$ in the isotropic case ($J_z=1.0$). DMT simulation exhibit quantitative agreement with TEBD numerics. Consistent with previous work \cite{Misguich}, fitting the data with $P_{total}\propto t^\alpha$ in different time windows, we observe a decreasing exponent $\alpha$, which seems to tend toward diffusion (with a possible logarithmic correction). 
  }
  \label{diff_correc}
\end{figure}

\begin{figure}[h]
  \centering
  \includegraphics[width = 3.2in]{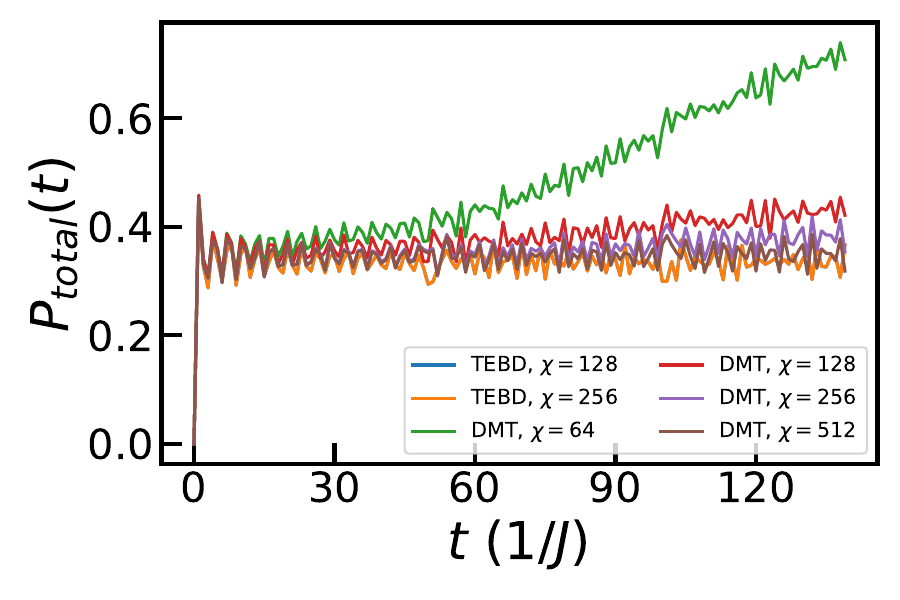}
  \caption{Convergence of the frozen dynamics in the isotropic XXZ model ($J_z=1$) at zero temperature. As we increase the bond dimension $\chi$, DMT converges to capture the finite polarization transfer. 
  }
  \label{frozendyn}  
\end{figure}

For pure initial states ($\eta = 1$) the spin dynamics can be markedly distinct (Fig.~\ref{XXZ_transport1_lowT}).
While for $J_z < 1$ the dynamics remain ballistic, when $J_z \ge 1$ the initial state is very close to the ground state of the system altering its dynamics.
When $J_z = 1$, we observe dynamics consistent with an approach to diffusive transport with possible logarithmic corrections (Fig.~\ref{diff_correc}), in agreement with previous works \cite{Prosen,Misguich}.
We note, however, that a definite conclusion on the precise nature of the dynamics requires a more careful calculation and analysis of longer-time evolution in larger-size systems.
By contrast, when $J_z > 1$, the spin dynamics become frozen and $P_{total}(t)$ remains bounded even for $t\rightarrow\infty$ \cite{Gobert}. 
Here we note the importance of bond dimension scaling in the analysis of the dynamics;
when considering small bond dimensions (e.g. $\chi = 64$), we observe a very slow diffusive transport, which arises from DMT artificially pushing the system towards a mixed state and destroying important coherences of the frozen dynamics. 
Nevertheless, when the bond dimension is increased, the DMT simulation converges and correctly captures the arrest of transport behavior (Fig.~\ref{frozendyn}). 

%
%
%
%

\begin{figure}[h]
  \centering
  \includegraphics[width = 6.5in]{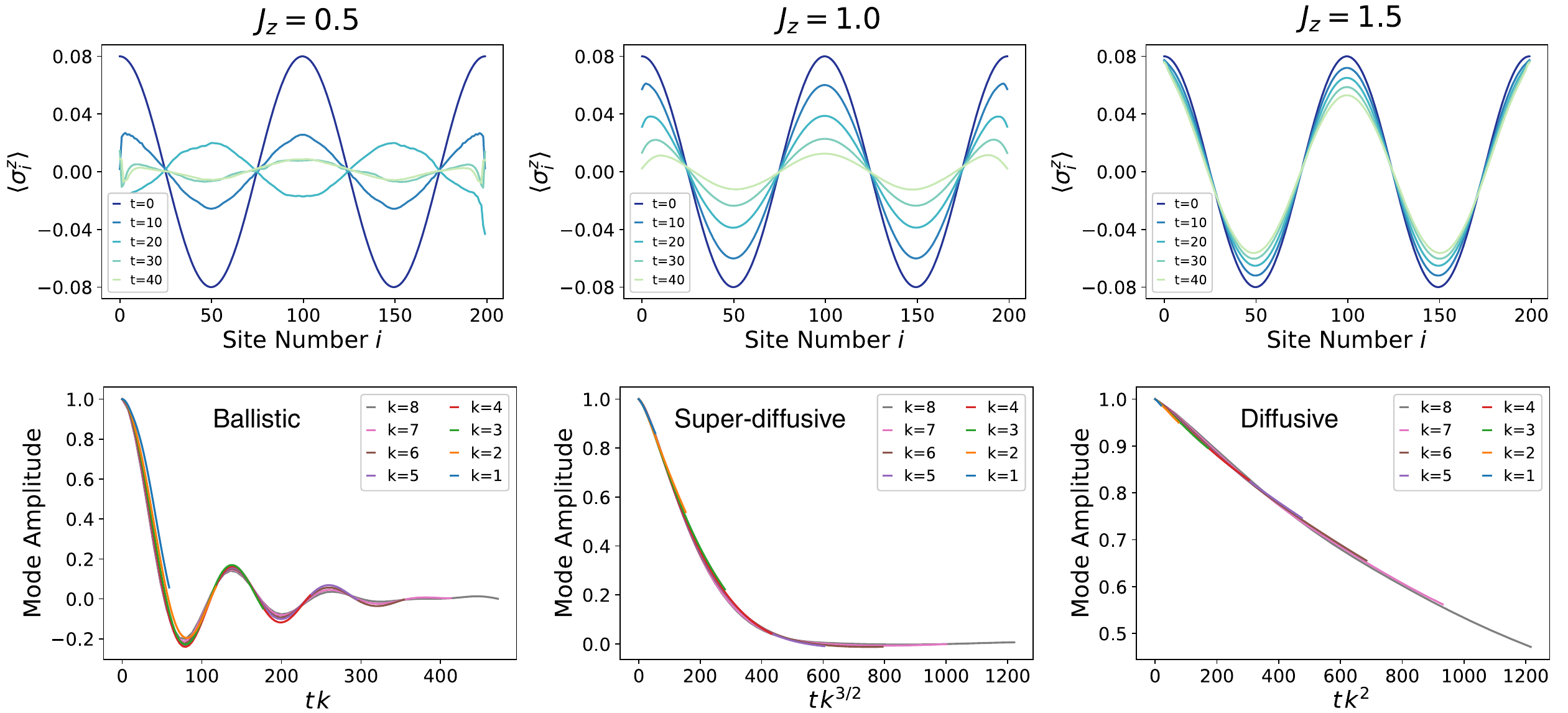}
  \caption{Dynamics of the integrable XXZ model initialized with Fourier modes. Upper panel: spatial profiles of local polarization. Lower panel: the amplitude of the Fourier modes as a function of time. In different parameter regimes, we utilize different scaling exponents to collapse the curves for different wavevectors $k$, again clearly demonstrating ballistic, super-diffusive and diffusive transport in agreement with Table~\ref{Table1}.
  }
  \label{XXZ_transport2}  
\end{figure}
\subsection{Setup 2: Decay of Fourier modes}
 In this case, we follow the methodology introduced in our main text. 
 In particular, we initialize the system in a completely mixed state with a small spatial inhomogeneity in the local magnetization (taken to be a Fourier mode with wavevector $\pi k/L$). 
 Then we evolve the system with $H_{\text{XXZ}}$, and extract the amplitude of the Fourier mode $A_k(t)$ (Fig.~\ref{XXZ_transport2}). 
 For different transport behaviors, the decay of $A_k(t)$ exhibits a distinct scaling with $k$.
 In particular, the Fourier amplitude dynamics should exhibit a collapse to a simple scaling function $A_k(t) = \tilde{A}(tk^\beta)$, where $\beta=1$, $1<\beta<2$, and $\beta=2$ correspond to ballistic, super-diffusive and diffusive transport, respectively.  
 Indeed, in the bottom panels of Fig.~\ref{XXZ_transport2}, we observe the expected collapse for the three regimes considered!

\bibliography{references}